\documentclass[aps,twocolumn,nofootinbib]{revtex4-1}
\usepackage{graphicx}
\usepackage{color}
\usepackage{epsfig}
\usepackage[caption=false]{subfig}
\usepackage{float}

\begin{document}

\title{Glass Transition in Supercooled Liquids with Medium Range 
Crystalline Order}
\author{Indrajit Tah$^{1}$}
\author{Shiladitya Sengupta$^{2}$}
\author{Srikanth Sastry$^{3}$}
\author{Chandan Dasgupta$^{4}$}
\author{Smarajit Karmakar$^{1}$}
%\email{smarajit@tifrh.res.in}
\affiliation{$^1$ Centre for Interdisciplinary Sciences,
Tata Institute of Fundamental Research, 
21 Brundavan Colony, Narisingi, Hyderabad, 500075, India,\\
$^2$ Department of Chemical Physics, Weizmann Institute of Science, Israel,\\
$^3$ Jawaharlal Nehru Centre for Advanced 
Scientific Research, Bangalore 560064, India,\\
$^4$ Centre for Condensed Matter Theory, 
Department of Physics, Indian Institute of Science, Bangalore, 560012, 
India}
%\contributor{Submitted to Proceedings of the National Academy of Sciences
%of the United States of America}

%% Significant statement 
%% word limit is 120; now 118  
%\significancetext{A comprehensive understanding of the origins
%of dynamical slow down in glass forming liquids remains a challenge, 
%and has been sought to be understood in terms of growing static and 
%dynamical length scales. In systems with medium range crystalline order 
%(MRCO), it has been claimed that the so-called point-to-set method for 
%estimating static length scales fails. We show that glass formers 
%exhibiting MRCO are fundamentally different in that the structural order
%length scale governs all other length scales, which is not true generically. 
%Thus behaviour of glass formers with MRCO cannot be applied broadly in 
%understanding glassy behaviour.  We also show that when appropriately 
%implemented the point-to-set method is a robust, order-agnostic 
%method for estimating static length scales. }
%\maketitle

%% word limit 250; now 234 
\begin{abstract}
The origins of rapid dynamical slow down in glass forming liquids 
in the growth of static length scales, possibly associated with identifiable 
structural ordering, is a much debated issue. Growth of medium range crystalline 
order (MRCO) has been observed in various model systems to be associated with 
glassy behaviour. Such observations raise the question about the eventual state 
reached by a glass former, if allowed to relax for sufficiently long times. Is 
a slowly growing crystalline order responsible for slow dynamics?  Are the 
molecular mechanisms for glass transition in liquids with and without MRCO the same? 
If yes, glass formers with MRCO provide a paradigm for understanding 
glassy behaviour generically. If not, systems with MRCO form a new class 
of glass forming materials whose molecular mechanism for slow dynamics 
may be easier to understand in terms of growing crystalline order, and 
should be approached in that manner,  even while they will not provide 
generic insights. In this study we perform extensive molecular dynamics 
simulations of a number of  glass forming liquids in two dimensions and 
show that the static and dynamic properties of glasses with MRCO are different from
other glass forming liquids with no predominant local order. We also resolve
an important issue regarding the so-called Point-to-set method for determining 
static length scales,  and demonstrate it to be a robust, order agnostic, 
method for determining static correlation lengths in glass formers. 
\end{abstract}

\maketitle
%SS: General comments: 
%SS: For all the figures, (i) use spline or fit lines to connect data points. Very hard to tell the different sets apart.  (ii) The legends have to be much bigger, and use keys (eg. BC instead of typing out Binder Cumulant) which is explained in the figure caption. 

\begin{figure}[h] 
\begin{center}
\includegraphics[scale=0.28]{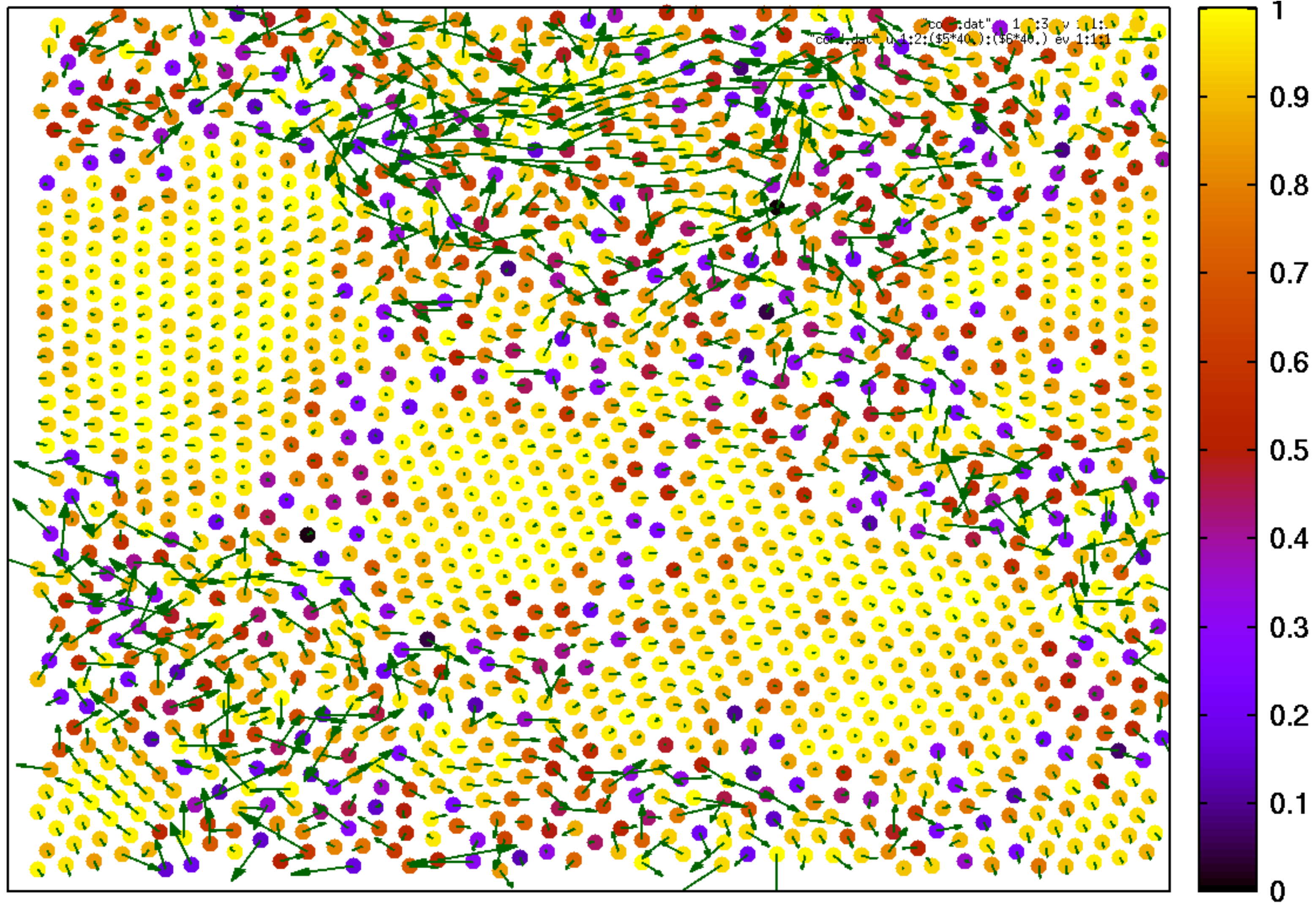}
%\includegraphics[scale=0.30]{hopdp1.eps}
%\hspace{0.9em}
%\includegraphics[width=2.7in,keepaspectratio]{ALLFIGURE/2dKA/probhop2dKA.eps}
\caption{\textbf{Dynamic heterogeneity and MRCO for 2dKA syatem}. A snapshot 
%CDQ: how does this plot show dynamic heterogeneity?
of the system with $N = 2000$ at $T = 0.930$ and $\rho = 1.20$ with particles 
coloured according to their local hexatic order parameter $|\psi_6^i|$, as indicated 
in the right side colour bar of the figure. The arrows are the displacements of particles 
calculated over $\alpha$ relaxation time starting from this configuration.} 
\label{fig:snapshot2dKA}
\end{center}
\end{figure}

The existence of growing length scales and their connection to 
the rapid slowing down of dynamics of glass forming liquids approaching 
glass transition are still a subject of active research. Based on extensive 
research on the existence of various static and dynamical length scales 
in glass formers, it is now understood that multiple growing length scales 
exist and correlate with increasing relaxation time scales as the glass 
transition is approached, but whether a single or a small number of such length 
scales satisfactorily describe the essential physics of dynamical slow down 
and the glass transition is not fully resolved \cite{11BB,arcmp,ROPPSMA,
annurev.physchem.51.1.99,05Berthier,RFOT2}. For quite sometime in the 
study of glass formers, there has been a focus on identifying structural 
motifs which are favoured locally but which can not span the whole system 
due to frustration arising from the incommensurate nature of the structural 
motifs. There are many experimental and simulation studies which claim the 
existence of such structural motifs in a variety of glass forming liquids
\cite{paddyRoyal1,paddyRoyal2}. 
A strong connection between the growth of such locally favoured structures 
(also widely known as medium range crystalline order, MRCO) and slow 
relaxation has also been proposed in some glass formers, with a fair degree 
of supporting evidence \cite{PhysRevLett.99.215701, 
Nat2010,J.Phys.Condens.Matter23-194121}. In Figure \ref{fig:snapshot2dKA},  snapshot of such MRCO 
(red regions in the figure) is shown for one of the model glass formers 
studied in this work. While the dynamics of different glass forming liquids 
are very similar to each other in essential aspects, only some glass formers 
manifest growing MRCO, whereas such structural motifs can
not be clearly identified in many others. The failure to identify growing structural order 
may either be due to their absence or due to the subtlety of the relevant 
structural order that may not be captured by the metrics employed. It is thus 
important to understand whether the physics that govern the slowing down 
of dynamics in different glass forming liquids are generically the same 
irrespective of the presence of well characterised MRCO or whether essential 
differences can be found between those systems 
with MRCO and those without. If it turns out that indeed the physics 
governing the slowing down in different glass forming liquids are same and 
all glass forming liquids eventually will start to have locally preferred 
structurally ordered domains, then the complexity of glass transition will be 
reduced to the formation of such domains and efforts must be directed at 
detecting through suitable metrics the growing structural order.  On the 
other hand if it turns out that the behaviour of glass formers with MRCO 
is not generic, then they form a distinguishable special class of glass 
forming systems, whose behaviour is perhaps less complex, and must be 
analysed differently.

Recently, in a series of papers Tanaka et. al. \cite{PhysRevLett.99.215701, 
Nat2010,J.Phys.Condens.Matter23-194121} showed that the glass transition in 
systems with MRCO is determined by the length scale associated with that of 
locally preferred structures. For example, for two dimensional systems, the 
correlation length associated with hexatic order is the important length 
scale and the glass transition in these systems are completely controlled 
by the hexatic correlation length. In \cite{PNASUSA2015}, a polydisperse two 
dimensional glass forming liquid was studied in which the hexatic correlation 
grows rapidly with increasing packing fraction. The dynamic heterogeneity 
\cite{PhysRevLett.79.2827, PhysRevE.58.3515, science.1166665} length scale 
obtained from the scaling analysis of the wave-vector ($q$) dependence of 
the four-point structure factor, $S_4(q,t)$ \cite{JCP119-14} was compared with 
the hexatic correlation length obtained from the spatial correlation function 
of the hexatic order parameter, $\psi_6$ \cite{J.Phys.Condens.Matter23-194121, 
PhysRevE.75.041503, JCM14} defined as
\begin{equation}
\psi_6 = \frac{1}{N}\sum_{i=1}^{N}\psi_6^i, \quad \mbox{where} \quad
\psi_6^j = \frac{1}{n_j}\sum_{k=1}^{n_j}\exp\left({\imath 6\theta_{jk}}
\right)
\label{psi6} 
\end{equation}
where $\theta_{jk}$ is the angle made by the vector from particle $j$ to its %CD
neughbour $k$  with the $x$-axis and $n_j$ is the number of neighbours
of particle $j$ (see \textbf{SI} for further details). These two length scales 
were found to be proportional to each other over the studied density 
range. 

The static length scale obtained from few different methods including the 
Point-to-set (PTS) \cite{Nature2008, PhysRevLett.108.225506} 
method  was also 
compared with the hexatic correlation length. The PTS length scale 
was found to be very different from the hexatic correlation length in 
the studied density window. In this system with local hexatic 
order, the behaviour is completely determined by the hexatic correlation 
length. So, if the PTS method is ``order agnostic'', then it should pick
up the same hexatic correlation length. From the large difference between the
calculated values of the PTS and hexatic length scales, 
it was concluded that the PTS method
fails to pick up the relevant static length scale for systems with MRCO. 
These results cast a serious doubt about the 
usefulness of the PTS method in measuring  important static 
length scales relevant for slow dynamics  in all glass forming liquids.   

%CD: changed the wording in this paragraph
In this work we have addressed the important issue of whether a deep connection 
between structure and dynamics exists for all glass forming liquids.
As noted above, a number of theoretical approaches \cite{ ROPPSMA, RevModPhys.83.587, PhysRevLett.97.195701, 
ARPC58, AIP2003} have explored the relevance of a growing length scale 
to dynamical slow down. A specific goal in some of these approaches were to 
study the connection between heterogeneous dynamics (also referred to as dynamic 
heterogeneity) in glass formers \cite{annurev.physchem.51.1.99, J.Phys.Soc.Jpn, 
PhysRevE.52.1694, PhysRevLett.80.2338} with rapid change in viscosity.  
We show that the slowing down in the
dynamics while approaching the glass transition in liquids which exhibit a pronounced 
increase in structural order is strongly coupled to the growth of dynamic 
heterogeneity length scales, as well as to the growth of the length scale that captures the  
local structural order. In two-dimensional glass formers with MRCO, the static length scale 
determined by growing hexatic order is the same as both the dynamic heterogeneity 
length scale and a static amorphous length scale which we independently 
evaluate using the PTS method. On the other hand, for liquids in which such 
structural ordering is not prominent, the temperature dependence of the dynamic 
heterogeneity length scale is very different from that of the static amorphous length scale. 
This suggests that the glass transition in liquids which 
show a strong tendency to the growth of locally favoured structural order is 
fundamentally different from the transition in liquids in which such 
local crystalline structures do not grow very prominently.

We address these issues by performing extensive computer simulations of four 
different glass forming liquids in two dimensions and measuring all the relevant 
length scales in each system using a wide range of procedures. The model systems 
studied are: (i) the two-dimensional Kob-Andersen binary Lennard-Jones mixture 
(referred here as 2dKA) \cite{PhysRevE.51.4626} at fixed density, with the 
temperature as the variable that determines changes in dynamics; (ii) a two-
dimensional system characterized by a repulsive inverse power-law potential, 
referred to as 2dIPL \cite{PhysRevLett.105.157801}, studied with density as 
the relevant variable. For the above two systems, structural order or MRCO 
grows significantly while approaching the putative 
glass transition. We study another system (iii) with a repulsive 
interaction potential (referred to as 2dR10) \cite{PhysRevE.82.031301}. 
No pronounced crystalline order is found in the studied temperature 
range for this model. This is consistent with previous work \cite{PhysicaA} that showed that 
the 2dR10 model has almost no tendency to exhibit locally favored crystalline 
order. We also study (iv) the polydisperse system studied by Tanaka et al 
\cite{PNASUSA2015}, for reasons which we discuss later. The details of the 
models and simulations are given in the SI.

%SS Should we not also mention the polydisperse model here? 
%SK Yes, it is good to mention it here as you have done to complete the 
%detials of the models studied.

Rest of the paper is organized as follows: first, we briefly discuss the systems 
with local hexatic order  (2dKA,  2dIPL) and compare various length scales in those 
model systems. We then discuss the model (2dR10) for which there is no prominent 
hexatic order and compare the findings with those for the previous 
set of model systems. Then we will discuss the controversy regarding the 
PTS method and describe results that resolve that controversy. Finally, 
we conclude with a summary of the perspective regarding the relationship 
between structural order and dynamics in different classes of glass formers. 
 
\begin{figure*}
%\vskip -0.5cm                                                           
\hskip -0.9cm
\includegraphics[scale=0.35]{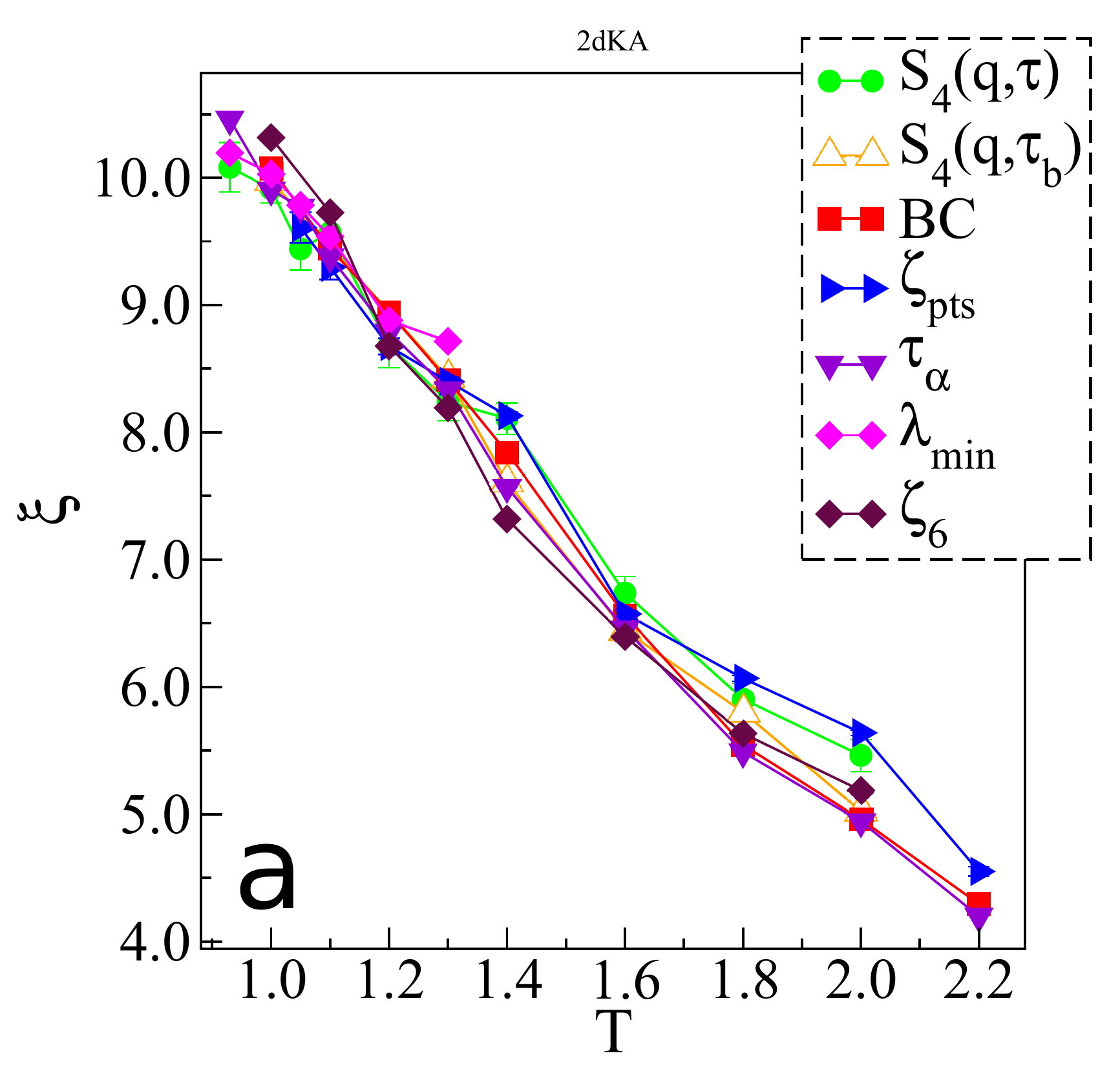}%
%\hskip -0.5cm                 
\includegraphics[scale=0.23]{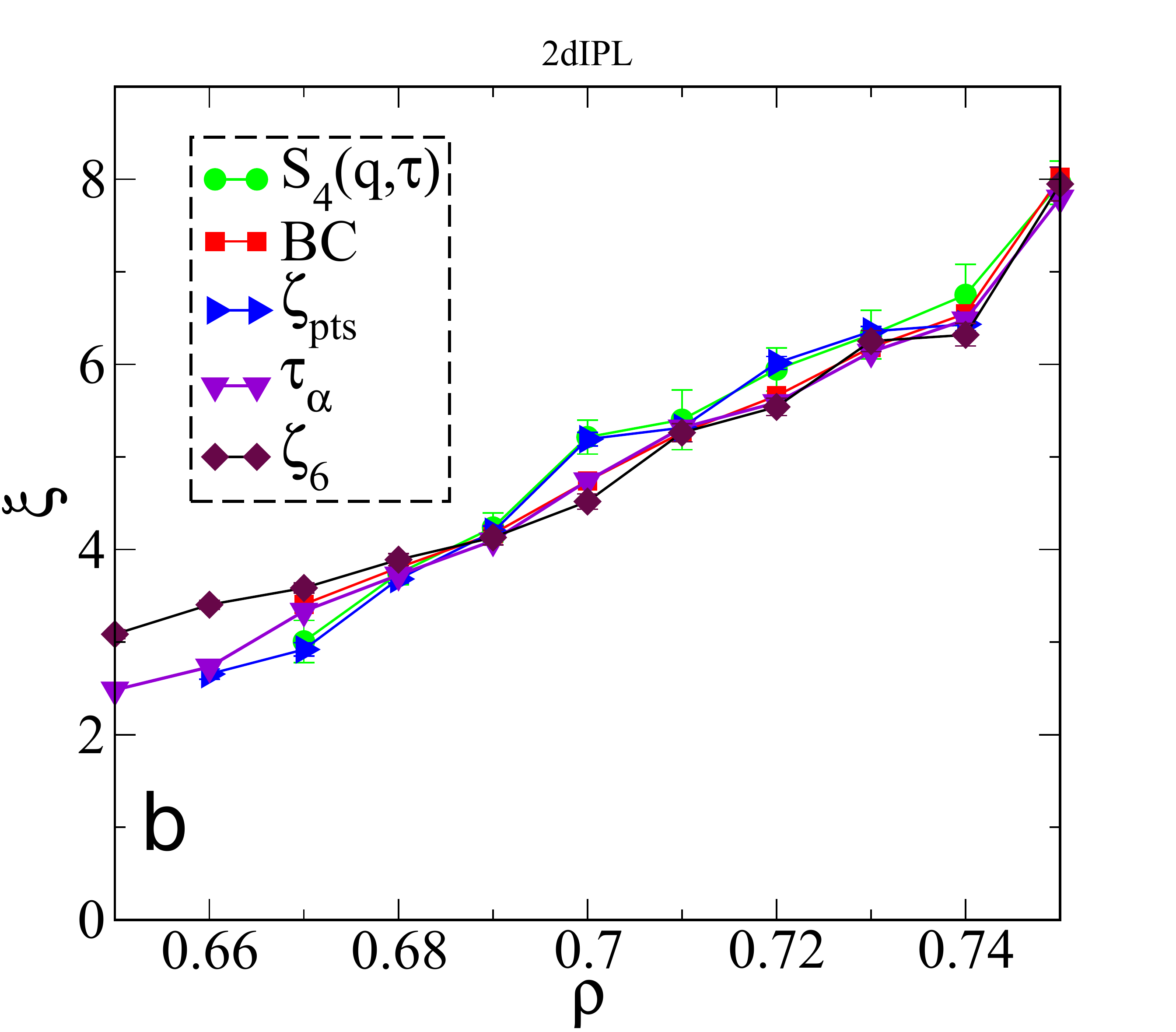}% 
\hskip -0.5cm             
\includegraphics[scale=0.32]{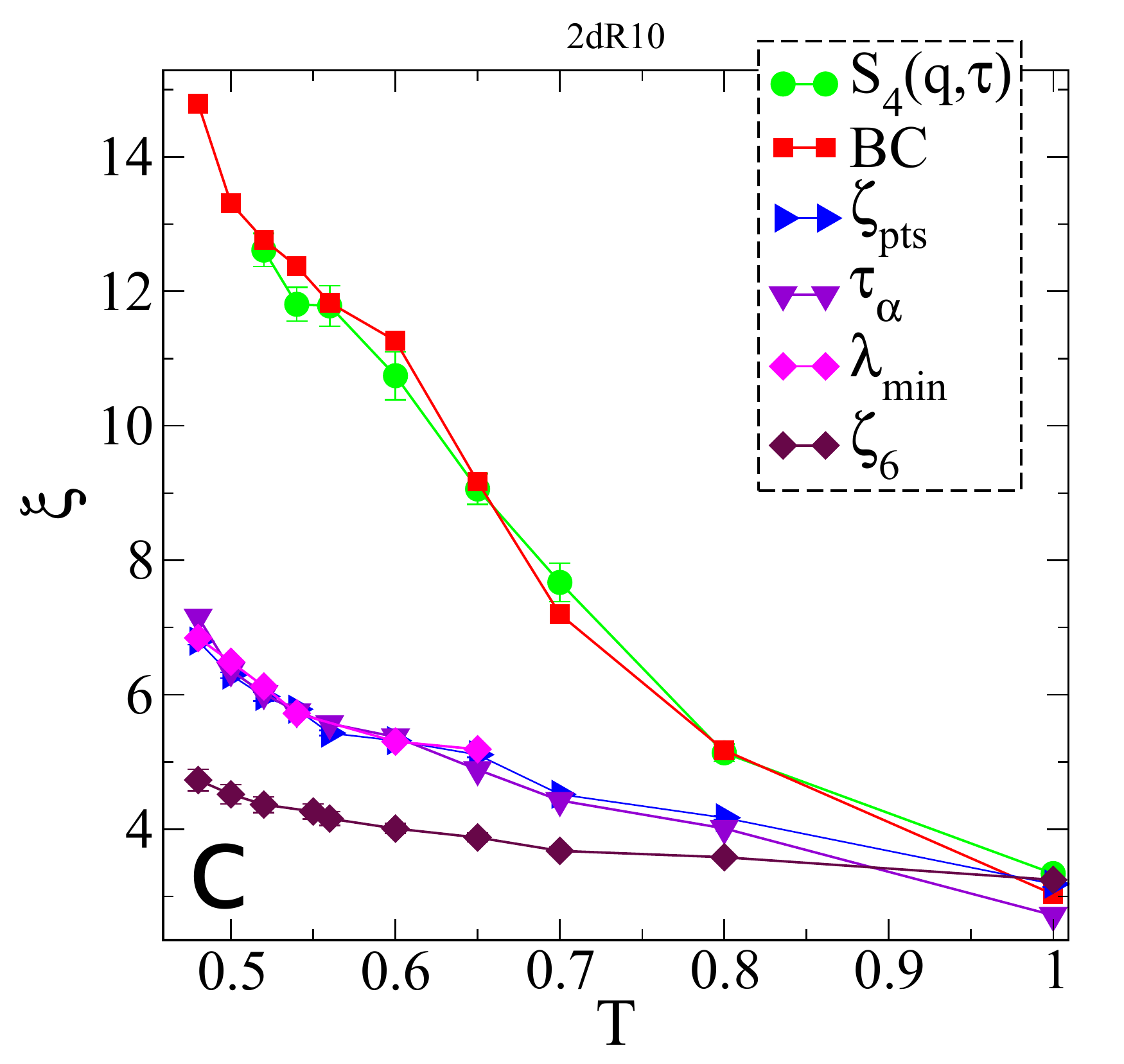}                               
 \vskip +0.5cm     
%{\textbf{d}}\includegraphics[scale=0.15]{ALLFIGURE/cavity.png} 
%\hspace{2mm}%                
\includegraphics[scale=0.22]{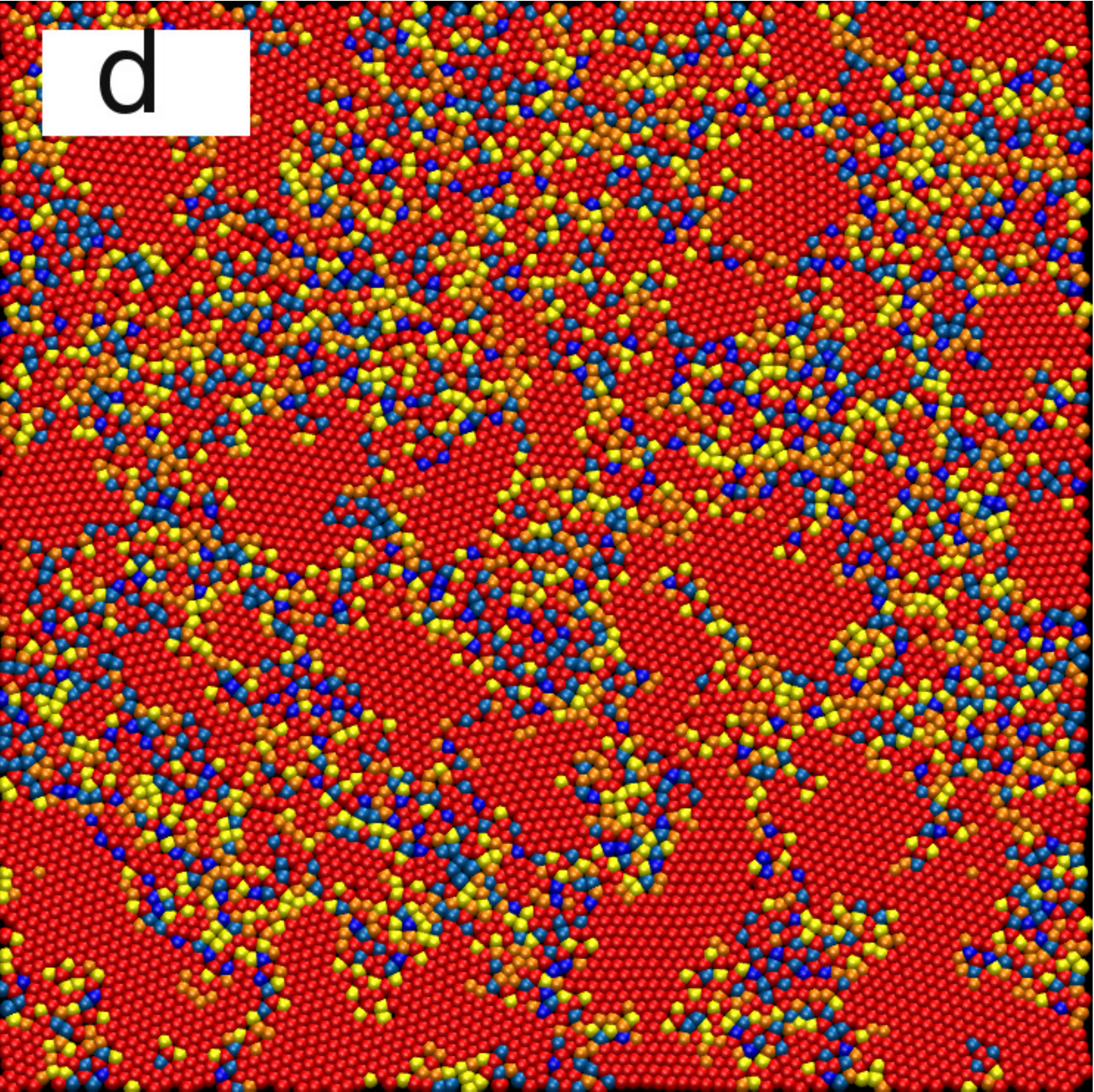}% 
\hspace{3mm}%             
\includegraphics[scale=0.20]{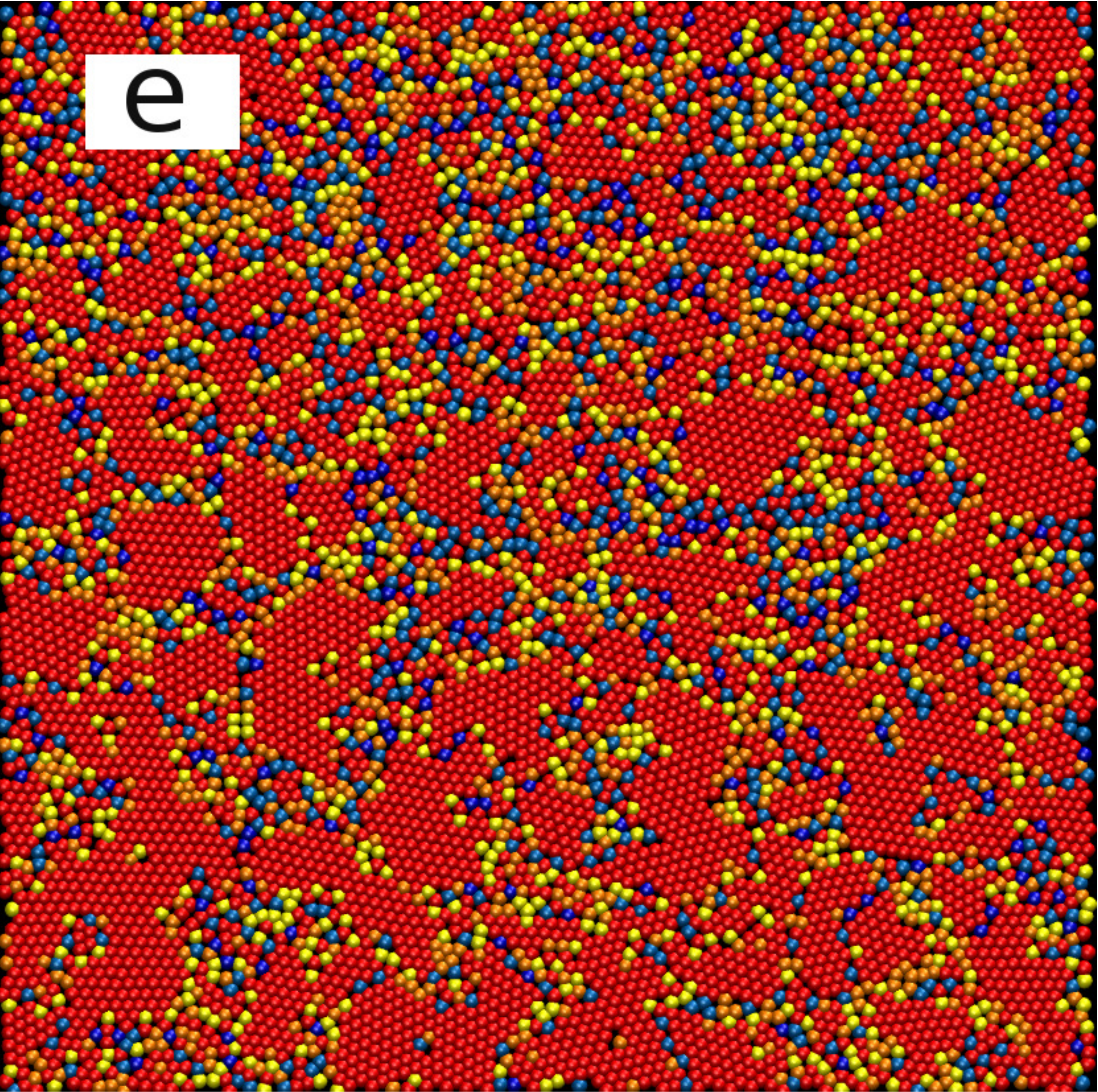}   
%\hspace{-0.05cm}%             
\includegraphics[scale=0.265]{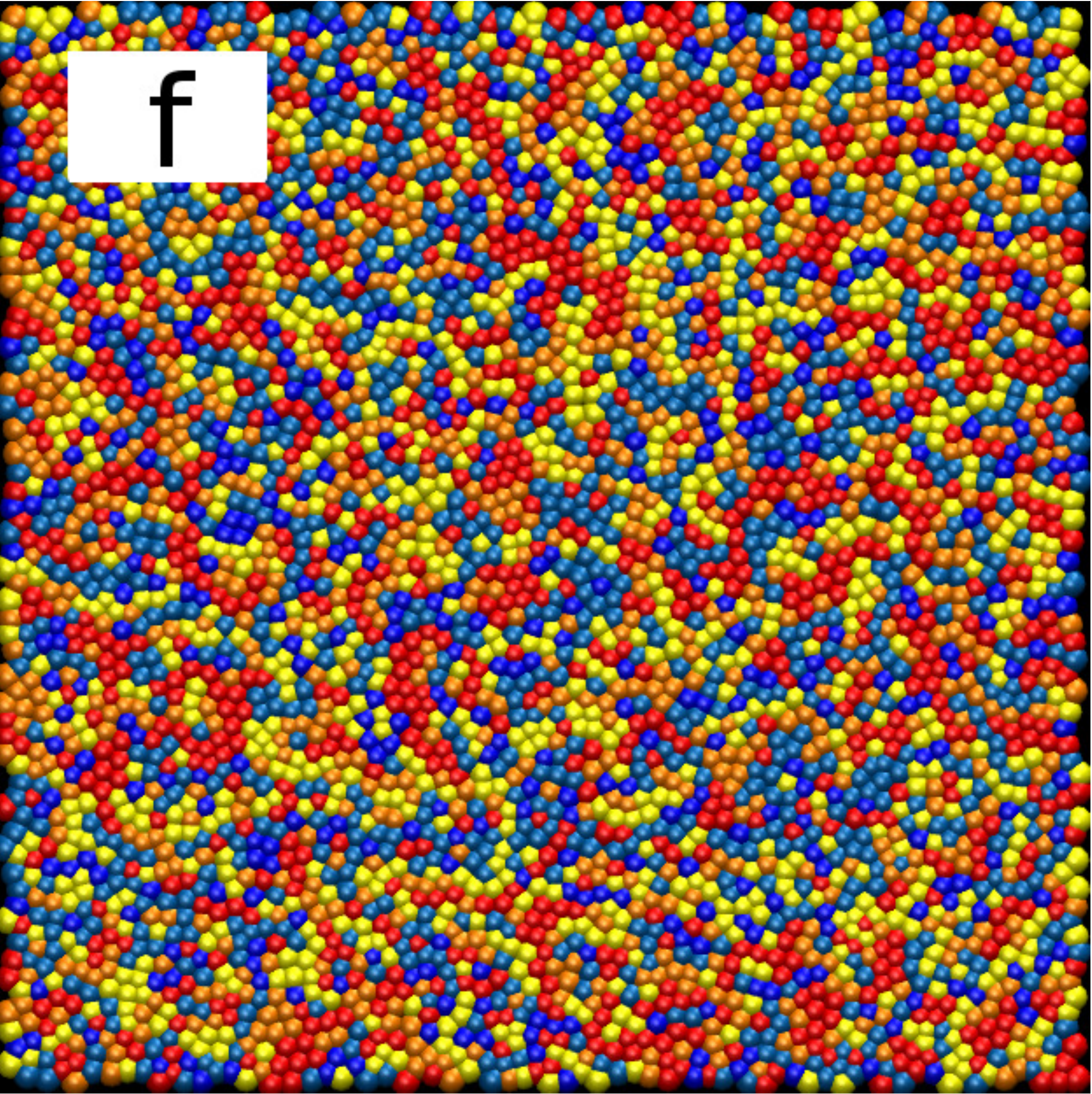}                                      
\caption{Growth of dynamical and static length scales for three different 
systems 2dKA(a), 2dIPL(b), and 2dR10 (c). Snapshots show hexatic order 
for the same systems (d, e, f). (a) Correlation lengths as a function 
of temperature $T$ for the 2dKA system. Symbols for different quantities 
are as follows: Green circle 
for the dynamical length scale $\xi_d$ are obtained from $S_4(q,\tau)$; 
Orange triangles for the dynamical length scale obtained from $S_4(q,\tau_b)$ 
(details in SI); 
Red squares for dynamical length scale $\xi_d$ calculated 
by Binder cumulant scaling; Blue right 
triangle for PTS length scale $\xi_{pts}$; Purple lower triangle for 
static length 
scales obtained from finite size scaling of $\tau_{\alpha}$; 
Pink diamonds for the static length scale 
obtained from the smallest eigenvalue of the Hessian matrix;  
Brown diamond for hexatic correlation
length $\xi_6$. Same colours and symbols are used in the other two panels. 
(b) Correlation lengths as a function of density $\rho$ for the 2dIPL system.  
(c) Correlation lengths as a function of temperature $T$ for the 2dR10 system. 
(d) Snapshot of a configuration at T=1.0 (2dKA) in which the particles are 
coloured according to the following criteria: Red for very high hexatic order 
(($|\psi_6^i|$) $\geq 0.80$); Orange for high hexatic order 
(($|\psi_6^i|$) $\geq 0.60$ and ($|\psi_6^i|$) $< 0.80$);  Yellow for medium hexatic order (($|\psi_6^i|$) $\geq 
0.40$ and ($|\psi_6^i|$) $< 0.60$); Light blue for low hexatic order (($|\psi_6^i|$) $\geq 0.20$ and 
($|\psi_6^i|$) $< 0.40$); Dark blue for very low hexatic order (($|\psi_6^i|$) $\geq 0.0$ and ($|\psi_6^i|$) 
$< 0.20$). Same colouring is used in the following two panels. (e) Snapshot of 
a configuration at $\rho$ = 0.74 (2dIPL), (f) Snapshot of a configuration at 
$T = 0.500$ (2dR10).}
%CDQ: why use "hop"? Isn't it the same as |\psi_6^i|? How is panel (d) different from Fig.1?
 \label{fig:lengthscale}
\end{figure*}

For the 2dKA model, hexatic order grows with decreasing temperature as 
illustrated in Figure \ref{fig:snapshot2dKA} where particles are coloured 
according to the value of the local hexatic order parameter, $\psi_6^i$. Similarly,  
hexatic order grows for the 2dIPL model with increasing density. We have 
extracted the correlation length associated with the hexatic order by 
fitting the decay profile of the peaks of the hexatic correlation function, 
$g_6(r)$ \cite{JOPCMEPP} defined as
\begin{equation}
g_6(r) = \left< \psi_6^*(\vec{r})\psi_6(0)\right>, %CD
\end{equation}
where $\psi_6(\vec{r}) = \sum_{i=1}^{N}\delta (\vec{r}-\vec{r}_i)\psi_6^i$.
We have normalized it by pair correlation function $g(r)$ (see SI for 
further details). We have used exponential function and the two dimensional version of
the Ornstein-Zernike function, to fit the decay of the peaks of the correlation
function to get the correlation length. The correlation length $\xi_6$ is 
plotted in Figure \ref{fig:lengthscale} (a), \ref{fig:lengthscale} (b) and 
\ref{fig:lengthscale} (c) together with other dynamical and static 
correlation lengths which are described below. 

Previous studies have shown \cite{Nat2010, J.Phys.Condens.Matter23-194121} 
a one-to-one correspondence between dynamic heterogeneities and regions of 
high/low hexatic order. We also find the same correspondence for the  model 
systems studied here where medium range hexatic correlations are prominent. 
We have calculated the dynamical length scale $\xi_4$ from the four 
point structure factor $S_4(q,t)$ \cite{JCP119-14, PhysRevLett.97.195701}, 
which is defined as
%in terms of fluctuations of the Fourier transform of the overlap

\begin{equation} 
S_4(q,t) = N\left[\langle \tilde{Q}(q,t)\tilde{Q}(-q,t)\rangle - \langle \tilde{Q}(q,t)\rangle^2\right].
\end{equation}
with $\tilde{Q}(q,t)$ is defined as 
\begin{equation}
\tilde{Q}(q, t) = \frac{1}{N}\sum_{i=1}^{N} \exp(i {\bf q}.{\bf r}_i(0)) w\left(|\vec{r}_i(t) - \vec{r}_i(0)|\right),
\end{equation}
where the window function, $w(x) = 1$ if $x < 0.3$ and $0$ otherwise.  

%CDQ: shouldn't there be averages \langle ...\rangle here? Also, Q(t) doesn't
%have any r-dependence. So, what is the meaning of the Fourier transform of Q? 
%SK text is rewritten to remove this ambiguity

A dynamical susceptibility $\chi_4(t)$ is defined in terms of the fluctuations of 
two point overlap correlation function $Q(t)$ as  
\begin{equation}
\chi_4(t) = N\left[\langle Q^2(t)\rangle - \langle Q(t)\rangle^2 \right].
\end{equation}
The function $Q(t)$ (see SI for further details) defined as
\begin{equation}
Q(t) = \frac{1}{N}\sum_{i=1}^{N}w\left(|\vec{r}_i(t) - \vec{r}_i(0)|\right),
\end{equation}
This function measures the overlap of a configuration of particles at 
a given initial time $(t=0)$ with the configuration at a later time $t$.
Moreover $\chi_4(t) \equiv \lim_{q \to 0} S_4(q,t)$. The $\alpha$ 
relaxation time $\tau_\alpha$ is defined as $Q(t=\tau_\alpha)=1/e$. 
We define another time scale $\tau$ as the time at which $\chi_4(t)$ peaks - 
it is proportional to $\tau_\alpha$ with a proportionality constant close to 
1. We consider $S_4(q,t)$ evaluated as $t=\tau$ in our analysis using the 
Ornstein-Zernike form,
\begin{equation}
S_4(q,\tau) = \chi_0 \mathcal{F}(q\xi_4),
\end{equation}

to obtain the dynamical correlation length, $\xi_4$. Here, 
$\chi_0 = S_4(q=0,\tau)$ is the $q=0$ value of the structure factor 
and is same as $\chi_4(\tau)$ only in the Grand canonical Ensemble 
\cite{ROPPSMA, PhysRevLett.105.015701, PhysRevLett.105.019801, 
PhysRevLett.105.217801}, and $\mathcal{F}(x)$ is a
scaling function (see SI for further details of the scaling analysis).

We have also estimated the dynamical correlation length $\xi_4$ from 
detailed finite size scaling analysis of the Binder Cumulant 
\cite{ROPPSMA, J.Phys.Condens.Matter23-194121,  PNASUSA2009, Zeitschrift}, 
defined as
\begin{equation} 
B(N,T) = 1 - \frac{<Q(N,T)_{\tau}^4>}{3<Q(N,T)_{\tau}^2>^2},
\end{equation} 
where $Q(N,T)_{\tau} \equiv Q(t = \tau)$ for system size $N$ at 
temperature $T$. The Binder cumulant is defined in terms of $4^{th}$ 
and $2^{nd}$ moments of the distribution of $Q$ so that it becomes zero 
for a Gaussian distribution and acquires positive values for bimodal 
distributions. It is a scaling function of only $L/\xi_4$ where $L$ is 
the linear dimension of the system (further details can be found in SI).
From Figures~\ref{fig:lengthscale}.(a), ~\ref{fig:lengthscale}.(b) 
and ~\ref{fig:lengthscale}.(c) it is clear that this length scale 
obtained from finite size scaling of $B(N,T)$ is in good agreement 
with the dynamical length scale calculated from the four-point structure 
factor $S_4(q,\tau)$.  For one model (2dKA) we also compute the dynamical 
length scale from the four-point structure factor of bond breakage 
correlation function (see SI for details), and as seen in 
Figure.~\ref{fig:lengthscale} (a), this length scale also compares very 
well with the lengths obtained by the other methods. 

The growth of the calculated dynamical correlation length, $\xi_4$ is 
found to be in good agreement with the hexatic correlation length, $\xi_6$ 
for models 2dKA and 2dIPL. This is in complete agreement with the 
observation made in Ref.\cite{PNASUSA2015} where it was shown that for 
systems with growing hexatic order, the hexatic correlation length is the 
same as the dynamical heterogeneity length scale.

To understand possible connections between the growing hexatic order 
and static length scales associated with amorphous order, we have 
calculated the static length scale using three different methods: 
(a) PTS correlation method, (b) Finite size scaling of 
the relaxation time $\tau_\alpha$ and (c) Finite size scaling of the minimum
eigenvalue of the Hessian matrix \cite{PhysicaA} calculated at local minima 
of the potential energy  (known as Inherent structures) visited by 
the system at a given temperature. Computing PTS correlations 
\cite{Nature2008, PhysRevLett.108.225506} is an elegant, general 
method to capture the multi-point static structural correlations in 
viscous liquids. The main idea is to freeze the majority of particles 
from an equilibrated configuration of the liquid, outside a specified 
volume, and to measure how the structure of the unfrozen particles in the inner 
volume are affected. It has been argued that in 
the geometry in which particles outside a spherical cavity are frozen, 
the PTS correlation should detect the typical domain size predicted 
by the random first order transition theory (RFOT) 
\cite{JChemPhys.2013Mar28, JCP121-7347}. Here we calculate
the point to set length scale (referred to as $\xi_{PTS}$) in the cavity 
geometry. Bulk equilibrium configurations are generated at desired 
temperature and density using NVT Molecular dynamics simulations and 
then cavities are constructed by freezing the particles outside a 
circular region of radius $R$. Static overlap correlations are 
calculated as a function of cavity radius $R$ and from the decay profile 
we obtain the PTS length scale (see \textbf{SI} for further details).

The static length scale obtained from the finite size scaling of the
relaxation time is referred to here as $\xi_{\tau}$ \cite{arcmp,PNASUSA2009}. 
The details of the analysis and the scaling plots are given in the SI. 
The length scale $\xi_{\lambda}$ obtained from the scaling of the minimum 
eigenvalue of the Hessian matrix
is calculated only at low temperatures due to the harmonic nature of the 
analysis which will not be a good approximation at higher temperatures
(see the SI for further details of the analysis and scaling plots for 
different model systems).

The comparison of different length scales are summarized in 
Figures ~\ref{fig:lengthscale}.(a), ~\ref{fig:lengthscale}.(b) and 
~\ref{fig:lengthscale}.(c). The data for 2dKA and 2dIPL in which MRCO 
grows significantly, both dynamic and static length scales
obtained using different methods grow very similarly as the hexatic  
correlation length. On the other hand for the 2dR10 model where no significant 
hexatic order is observed in the studied temperature range, the dynamic length 
scales grow much faster than both the static length scale and the hexatic length scale.
Note  that for the 2dR10 model, the growth of the hexatic 
correlation length is even smaller than the modest growth of the static 
length scale.

These results clearly show that the relationship between dynamical, static and 
hexatic order length scales is significantly different for glass formers with 
MRCO (for which all length scales are comparable) and 
those without (for which different length scales are significantly different). 

\begin{figure}[!h]
\begin{center}
\vskip -0.8cm
\hspace{-0.5cm}                                                          
%{\textbf{a}} 
\includegraphics[scale=0.32]{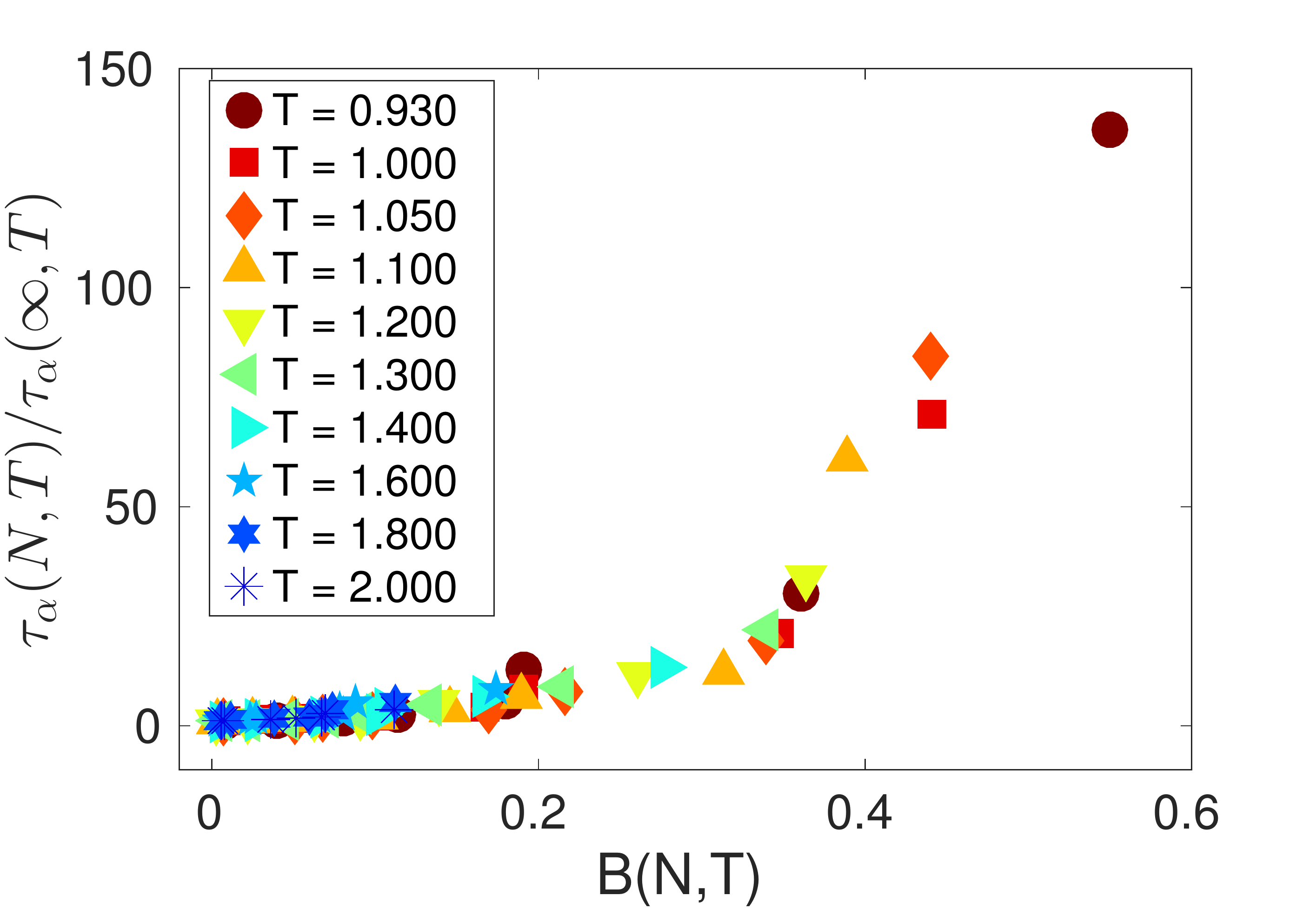}
\vskip -0.1cm               
%{\textbf{b}} 
\includegraphics[scale=0.26]{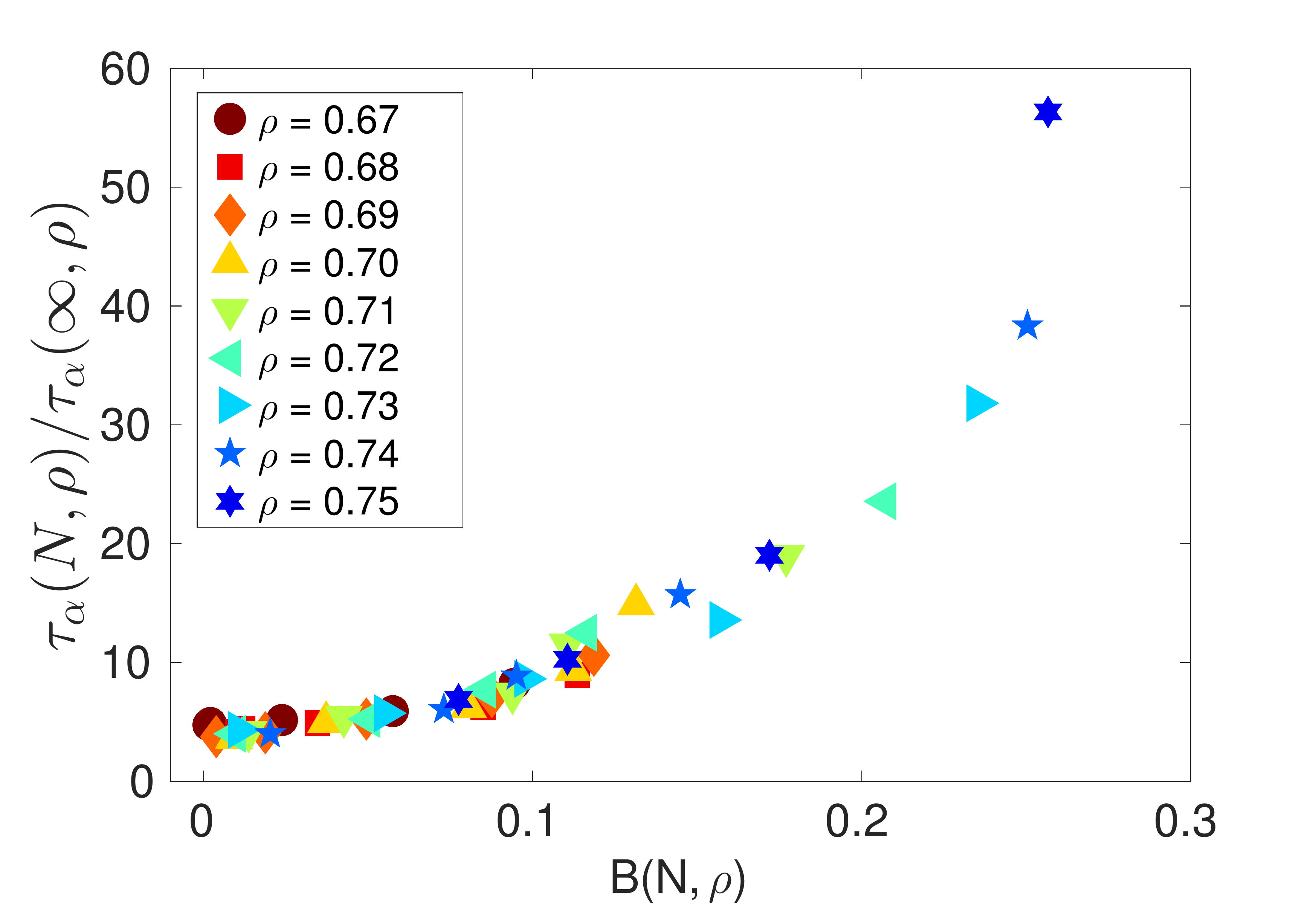}
\vskip -0.1cm 
%\center             
%{\textbf{c}} 
\includegraphics[scale=0.31]{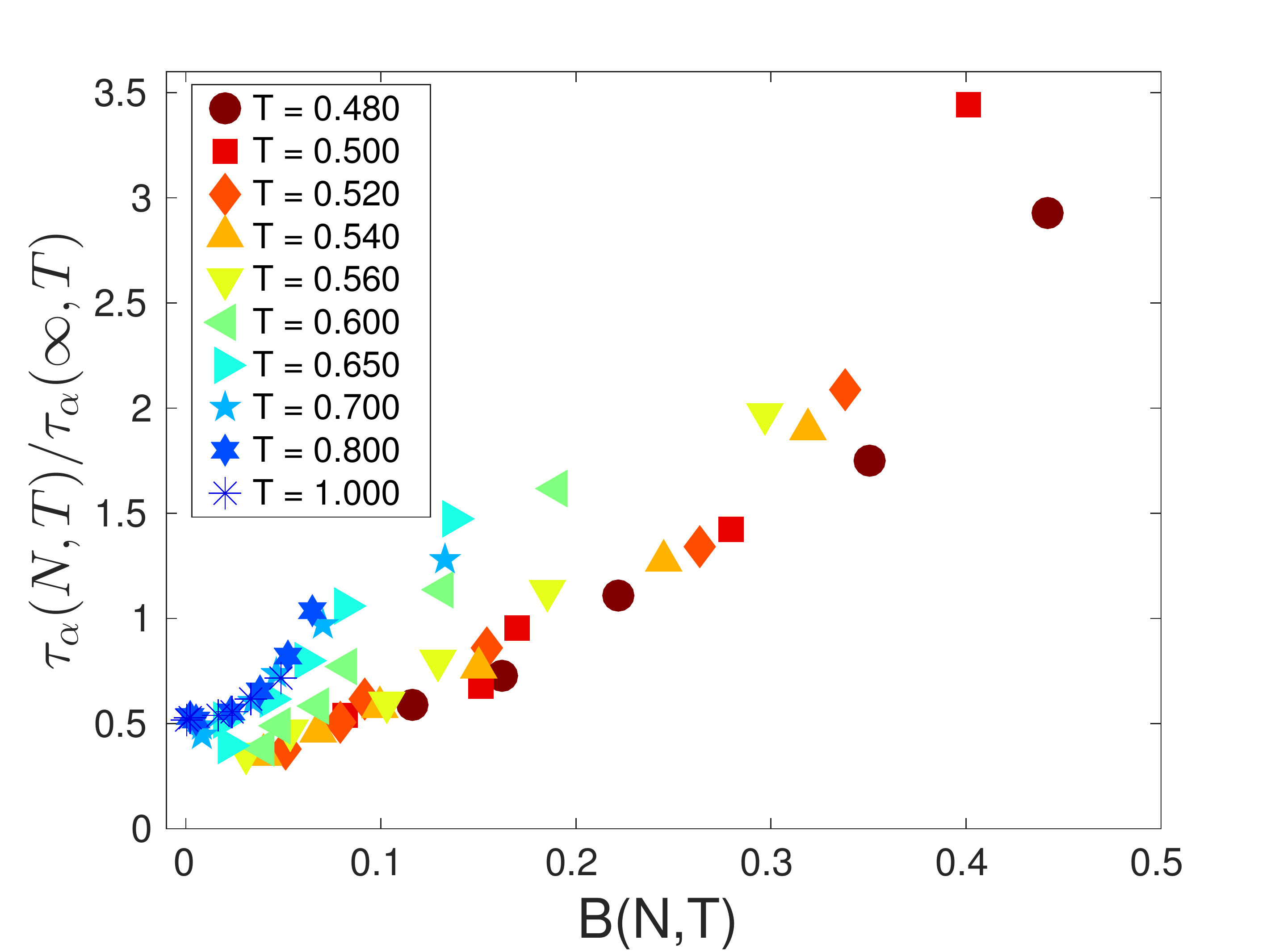}                               
\caption{The $\alpha$ relaxation time $\tau_\alpha(N,T)$ (or $\tau_\alpha(N,\rho)$ 
in the case of the 2dIPL model), scaled by its value for
$N \to \infty$, plotted against the Binder cumulant $B(N,T)$ (or $B(N,\rho)$ in 
the case of the 2dIPL model), for 2dKA system (top) 2dIPL system (middle) and 
2dR10 system (bottom). \label{fig:mastercurve1}}
\end{center}
\end{figure}

We next consider the relationship between the dynamical length scales and 
relaxation times. In Figure.\ref{fig:mastercurve1} we plot the relaxation 
time (scaled by its value in the infinite size limit) against the Binder 
cumulant for the three different models, and for different 
temperatures (or densities for the 2dIPL model) 
and system sizes in each case. If the system studied exhibits 
dynamical finite-size scaling, the data for different system sizes and 
temperatures should fall on a master curve. Further, since we expect the relaxation 
time to be governed by a static length scale (see above), such a data collapse 
would also indicate the equality of static and dynamical length scales, which is 
not generally expected to hold. Figure.~\ref{fig:mastercurve1} shows 
$\tau_{\alpha}(N,T)/\tau_{\alpha}(N\rightarrow \infty, T)$ {\it vs.}  $B(N,T)$ for 2dKA and 2dR10 model and $\tau_{\alpha}(N,\rho)/\tau_{\alpha}(N\rightarrow \infty, \rho)$ {\it vs.}  $B(N,\rho)$ for the 2dIPL model 
systems we study. The top two panels, for 2dKA and 2dIPL, show convincing data 
collapse, indicating that a unique length scale, which we may attribute to 
growing hexatic order, controls the dynamics, as well as the static order. In 
contrast, the bottom panel, for 2dR10, shows no data collapse, and  confirms 
our previous observation in Figure.~\ref{fig:mastercurve1} that the growths of 
dynamic and static length scales are different and decoupled from each other in
this system. 

However, our conclusion about the equality of static, dynamic and structural length scales 
in systems exhibiting
prominent hexatic order has recently been questioned. In Ref.~\cite{PNASUSA2015}, it was argued that 
the PTS method \cite{PhysRevLett.116.098302, PhysRevLett.111.165701} 
fails to capture the growth of the relevant static length scale for systems 
which exhibit growing MRCO.  
The system studied in Ref.\cite{PNASUSA2015} is a 
polydisperse one with $11 \%$ polydispersity. In the PTS method one calculates 
the static overlap as a
function of different cavity radii (R) for a given temperature
or density and then from the decay of the static overlap with
increasing cavity radius one extracts the static length-scale.

%SS: Do better on the data legends ('scaling' is not needed in each legend). Remove the graph title(2dpolydispersity in the bottom panel). Write in full PF in the middle panel. 
%CD changed figure caption

\begin{figure*}[tb]
 \centering 
%\begin{center} 
%\hskip -0.5cm
\includegraphics[scale=0.165]{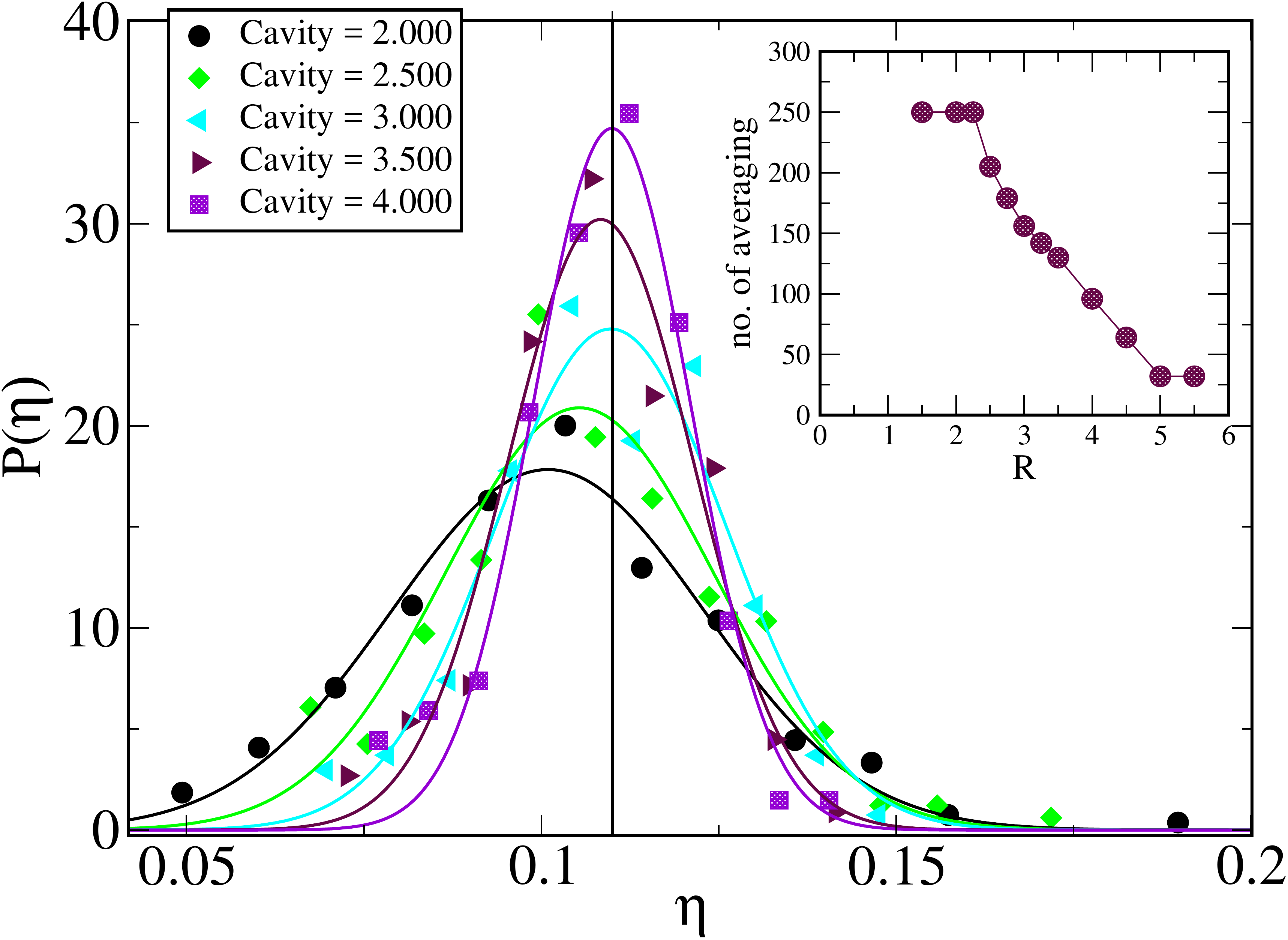} \hskip 0.05cm
\includegraphics[scale=0.18]{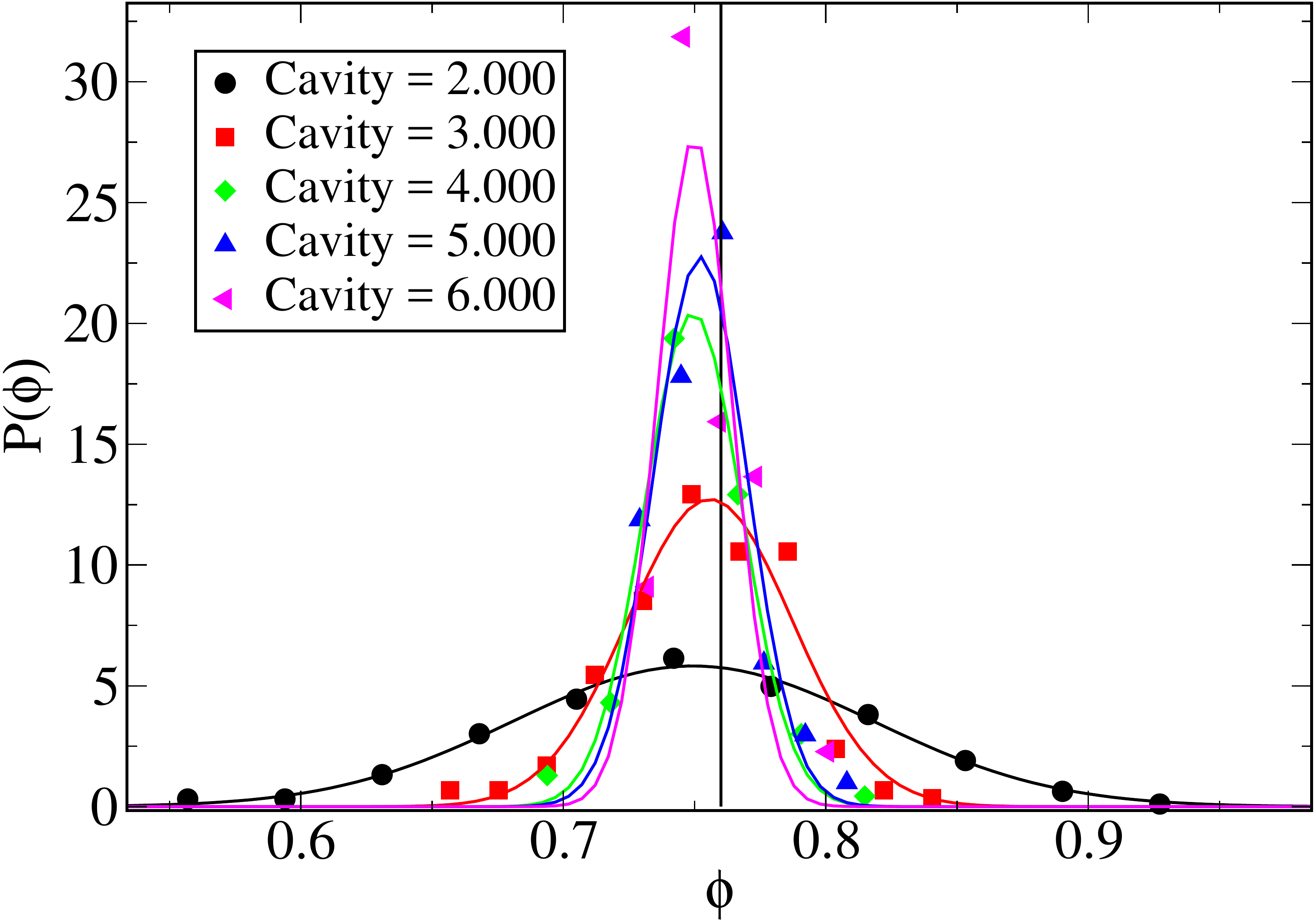}
%\hskip 0.05cm
\includegraphics[scale=0.24]{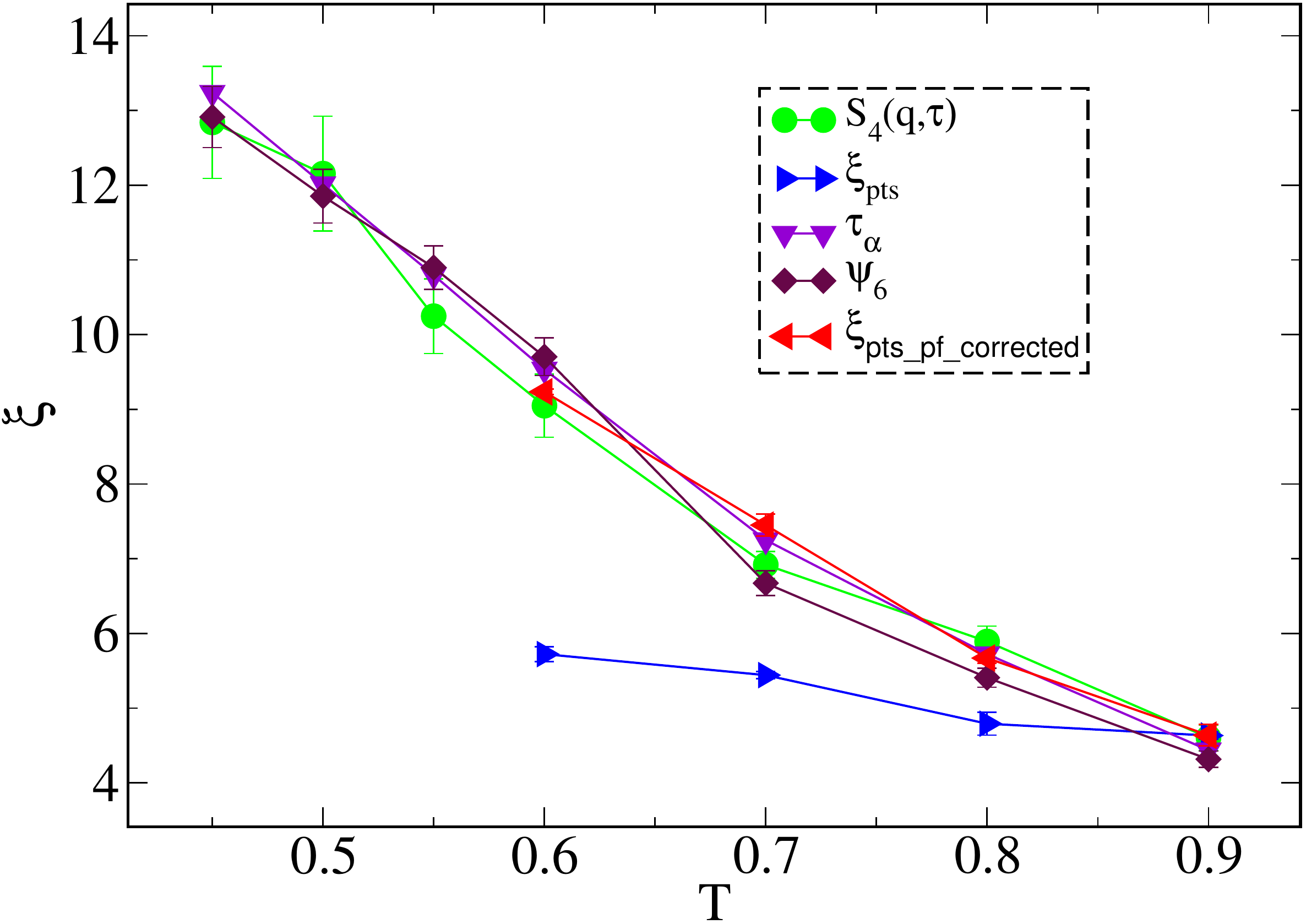}                               
\caption{Left Panel: Probability distribution of polydispersity for 
cavities of different radii. The black vertical line denotes the mean 
polydispersity of 11\%. Inset shows the number of samples required to 
get the average polydispersity within 2\% of the bulk value. Middle Panel:
Probability distribution of packing fraction. The black vertical line 
denotes the bulk packing fraction $0.76$. Labels for different 
data sets are the radii of the cavities. Right Panel: Comparison of 
different length scales for the 2dPoly system showing that the corrected 
PTS length agrees with other static and dynamical length scales computed.}
\label{fig:polydisperse}
%\end{center}
\end{figure*}

For randomly chosen small cavities, the packing fraction and polydispersity 
will be distributed around the mean values, with the variance being larger for 
smaller cavities. It may be expected, however, that for a sufficiently large 
number of realisations, such that the average packing fraction and polydispersity 
lie within a reasonably small tolerance, the estimated properties (in this 
instance, the static correlations to obtain the PTS length scale) will 
converge to the correct values. We thus estimate the number of realisations 
required for the average polydispersity and packing fraction to lie within 
$2\%$ of the bulk values, different values of the PTS length are obtained. In the inset of the left panel of Figure.~\ref{fig:polydisperse} 
we show such an estimate for a system of $N =10000$ polydisperse particles in 
two dimension with $11\%$ polydispersity (henceforth we will refer to this model 
as {\bf 2dPoly}). 
One can clearly see that for small cavities, one needs averaging over as large 
as $250$ different cavities in order to reach the bulk polydispersity. The distributions of the 
polydispersity and packing fraction obtained for different cavity sizes are shown in the left and
middle panels of Figure.~\ref{fig:polydisperse}.

We then obtained the PTS length scale by averaging over the estimated number of 
cavities for each radius. The results obtained this way
%with the estimated number of 
%averaging for individual cavities 
are shown in the right panel of 
Figure.\ref{fig:polydisperse} (blue triangles). Similarly to the results in 
Ref.\cite{PNASUSA2015}, we find that the PTS length scale so estimated falls 
well below the hexatic length scale and other static and dynamical 
length scales that track the hexatic length scale. This result is not expected to change if we average over a larger number of cavities for each $R$.

If, on the other hand, we adopt, following \cite{Nature2008,PhysRevLett.108.225506}, 
the procedure of evaluating the PTS length scale by selecting only those cavities 
for any radius for which the packing fraction and polydispersity within the 
cavity fall within a cutoff of $2\%$ from the bulk values. The PTS length 
scales obtained from this procedure (magenta triangle) match remarkably 
well the hexatic order length, the dynamical heterogeneity length scale 
and the static length scale estimated from finite size scaling of the relaxation 
time. We believe this procedure to be more appropriate because thermodynamic parameters such as temperature, density and  polydispersity of the particles inside the cavity are assumed to be fixed in the theoretical arguments \cite{Nature2008, PhysRevLett.108.225506} that relate the PTS length scale to the mosaic scale of RFOT. We thus conclude that the difference between the PTS length scale and the
hexatic order length scale observed in Ref. \cite{PNASUSA2015} arises from 
the method of calculation, rather than from any fundamental limitations 
of the PTS method in determining static length scales. The PTS method appears 
to be a robust, order agnostic approach, and while the issues surrounding what 
constitutes a proper way of calculating $\xi_{PTS}$ need to be better understood, there 
may be no need to modify it, as recently proposed \cite{PhysRevE.94.032605}.  

In conclusion, we have shown that the growth of different measures of static 
and dynamic length scales in supercooled liquids with medium range crystalline 
order (MRCO) is fundamentally different from the generic behaviour of glass 
forming liquids where local crystalline order is not prominent. For liquids 
with MRCO, the growth of dynamic and static length scales tracks very closely 
the MRCO length scale. The growth of relaxation times in these systems are 
apparently controlled by the growth of the MRCO length scale. On the other 
hand, for generic glass formers where the growth of MRCO as temperature or 
density is varied is not prominent, the dynamic length scales grow more 
rapidly than static length scales and the increase in relaxation 
times is controlled by the static length scale. Finally we have conclusively 
shown that the point-to-set correlation method is indeed order agnostic and 
captures the relevant length scale for all model systems including those 
with prominent MRCO.

\bibliography{mrco_ss}

\begin{thebibliography}{10}

\bibitem{11BB}
L.~Berthier and G.~Biroli, ``Theoretical perspective on the glass transition
  and amorphous materials,'' {\em Rev. Mod. Phys.}, vol.~83, pp.~587--645, Jun
  2011.

\bibitem{arcmp}
S.~Karmakar, C.~Dasgupta, and S.~Sastry, ``Growing length scales and their
  relation to timescales in glass-forming liquids,'' {\em Annu. Rev. Condens.
  Matter Phys.}, vol.~5, pp.~255--284, Mar 2014.

\bibitem{ROPPSMA}
S.~Karmakar, C.~Dasgupta, and S.~Sastry, ``Length scales in glass-forming
  liquids and related systems: a review,'' {\em Reports on Progress in
  Physics}, vol.~79, p.~016601, Dec 2015.

\bibitem{annurev.physchem.51.1.99}
M.~D. Ediger, ``Spatially heterogeneous dynamics in supercooled liquids.,''
  {\em Annual Review of Physical Chemistry}, vol.~51, pp.~99--128, Oct 2000.

\bibitem{05Berthier}
L.~Berthier, G.~Biroli, J.-P.~C. Bouchaud, L.~Masri, D.~E.~L. ˇote, D.~Ladieu,
  F.~Pierno, and M.~Pierno, ``Direct experimental evidence of a growing length
  scale accompanying the glass transition,'' {\em Science}, vol.~310,
  pp.~1797--1800, Jun 2005.

\bibitem{RFOT2}
P.~G. Wolynes and V.~Lubchenko, {\em Structural Glasses and Supercooled
  Liquids: Theory, Experiment, and Applications}.
\newblock John Wiley \& Sons, 2012.

\bibitem{paddyRoyal1}
C.~P. Royall and S.~R. Williams, ``The role of local structure in dynamical
  arrest,'' {\em PhysicsReports}, vol.~560, pp.~1--75, 2015.

\bibitem{paddyRoyal2}
C.~P. Royall, S.~R. Williams, T.~Ohtsuka, and H.~Tanaka, ``Direct observation
  of a local structural mechanism for dynamic arrest,'' {\em Nat Mater},
  vol.~7, pp.~556--561, Jul 2008.

\bibitem{PhysRevLett.99.215701}
T.~Kawasaki, T.~Araki, and H.~Tanaka, ``Correlation between dynamic
  heterogeneity and medium-range order in two-dimensional glass-forming
  liquids,'' {\em Phys. Rev. Lett.}, vol.~99, p.~215701, Nov 2007.

\bibitem{Nat2010}
H.~Tanaka, T.~Kawasaki, H.~Shintani, and K.~Watanabe, ``Critical-like behaviour
  of glass-forming liquids,'' {\em Nat. Mater.}, vol.~99, pp.~324--331, Apr
  2010.

\bibitem{J.Phys.Condens.Matter23-194121}
T.~Kawasaki and H.~Tanaka, ``Structural signature of slow dynamics and dynamic
  heterogeneity in two-dimensional colloidal liquids: glassy structural
  order,'' {\em Journal of Physics: Condensed Matter}, vol.~23, Apr 2011.

\bibitem{PNASUSA2015}
J.~Russo and H.~Tanaka, ``Assessing the role of static length scales behind
  glassy dynamics in polydisperse hard disks,'' {\em Proc. Natl. Acad. Sci.
  USA}, vol.~112, pp.~6920--6924, June 2015.

\bibitem{PhysRevLett.79.2827}
W.~Kob, C.~Donati, S.~J. Plimpton, P.~H. Poole, and S.~C. Glotzer, ``Dynamical
  heterogeneities in a supercooled lennard-jones liquid,'' {\em Phys. Rev.
  Lett.}, vol.~79, pp.~2827--2830, Oct 1997.

\bibitem{PhysRevE.58.3515}
R.~Yamamoto and A.~Onuki, ``Dynamics of highly supercooled liquids:
  Heterogeneity, rheology, and diffusion,'' {\em Phys. Rev. E}, vol.~58,
  pp.~3515--3529, Sep 1998.

\bibitem{science.1166665}
L.~O. Hedges1, R.~L. Jack1, J.~P. Garrahan, and D.~Chandler, ``Dynamic
  order-disorder in atomistic models of structural glass formers,'' {\em
  science}, vol.~323, pp.~1309--1313, Mar 2009.

\bibitem{JCP119-14}
N.~Lacˇevic, W.~F. Star, B.~T. Schrøder, and C.~S. Glotzer, ``Spatially
  heterogeneous dynamics investigated via a time-dependent four-point density
  correlation function,'' {\em The Journal of Chemical Physics}, vol.~119,
  pp.~7372--7387, Jan 2003.

\bibitem{PhysRevE.75.041503}
T.~Hamanaka and A.~Onuki, ``Heterogeneous dynamics in polycrystal and glass in
  a binary mixture with changing size dispersity and composition,'' {\em Phys.
  Rev. E}, vol.~75, p.~041503, Apr 2007.

\bibitem{JCM14}
K.~Binder, S.~Sengupta, and P.~Nielaba, ``The liquid-solid transition of hard
  discs: first-order transition or kosterlitz-thouless-halperin-nelson-young
  scenario?,'' {\em Journal of Physics: Condensed Matter}, vol.~14,
  pp.~2323--2333, Feb 2002.

\bibitem{Nature2008}
G.~Biroli, J.-P. Bouchaud, A.~Cavagna, T.~S. Grigera, and P.~Verrocchio,
  ``Thermodynamic signature of growing amorphous order in glass-forming
  liquids,'' {\em Nat Phys}, vol.~99, pp.~771--775, Oct 2008.

\bibitem{PhysRevLett.108.225506}
G.~M. Hocky, T.~E. Markland, and D.~R. Reichman, ``Growing point-to-set length
  scale correlates with growing relaxation times in model supercooled
  liquids,'' {\em Phys. Rev. Lett.}, vol.~108, p.~225506, Jun 2012.

\bibitem{RevModPhys.83.587}
L.~Berthier and G.~Biroli, ``Theoretical perspective on the glass transition
  and amorphous materials,'' {\em Rev. Mod. Phys.}, vol.~83, pp.~587--645, Jun
  2011.

\bibitem{PhysRevLett.97.195701}
G.~Biroli, J.-P. Bouchaud, K.~Miyazaki, and D.~R. Reichman, ``Inhomogeneous
  mode-coupling theory and growing dynamic length in supercooled liquids,''
  {\em Phys. Rev. Lett.}, vol.~97, p.~195701, Nov 2006.

\bibitem{ARPC58}
V.~Lubchenko and G.~P. Wolyne, ``Theory of structural glasses and supercooled
  liquids,'' {\em Annual Review of Physical Chemistry}, vol.~58, pp.~235--266,
  Oct 2006.

\bibitem{AIP2003}
F.~Ritort and P.~Sollich, ``Glassy dynamics of kinetically constrained
  models,'' {\em Advances in Physics}, vol.~52, p.~219–342, 2003.

\bibitem{J.Phys.Soc.Jpn}
R.~Yamamoto and A.~Onuki, ``Kinetic heterogeneities in a highly supercooled
  liquid.,'' {\em J. Phys. Soc. Jpn}, vol.~66, pp.~2545--2548, Sep 1997.

\bibitem{PhysRevE.52.1694}
M.~M. Hurley and P.~Harrowell, ``Kinetic structure of a two-dimensional
  liquid,'' {\em Phys. Rev. E}, vol.~52, pp.~1694--1698, Aug 1995.

\bibitem{PhysRevLett.80.2338}
C.~Donati, J.~F. Douglas, W.~Kob, S.~J. Plimpton, P.~H. Poole, and S.~C.
  Glotzer, ``Stringlike cooperative motion in a supercooled liquid,'' {\em
  Phys. Rev. Lett.}, vol.~80, pp.~2338--2341, Mar 1998.

\bibitem{PhysRevE.51.4626}
W.~Kob and H.~C. Andersen, ``Testing mode-coupling theory for a supercooled
  binary lennard-jones mixture i: The van hove correlation function,'' {\em
  Phys. Rev. E}, vol.~51, pp.~4626--4641, May 1995.

\bibitem{PhysRevLett.105.157801}
U.~R. Pedersen, T.~B. Schroder, and J.~C. Dyre, ``Repulsive reference potential
  reproducing the dynamics of a liquid with attractions,'' {\em Phys. Rev.
  Lett.}, vol.~105, p.~157801, Oct 2010.

\bibitem{PhysRevE.82.031301}
S.~Karmakar, E.~Lerner, I.~Procaccia, and J.~Zylberg, ``Statistical physics of
  elastoplastic steady states in amorphous solids: Finite temperatures and
  strain rates,'' {\em Phys. Rev. E}, vol.~82, p.~031301, Sep 2010.

\bibitem{PhysicaA}
S.~Karmakar, E.~Lerner, and I.~Procaccia, ``Direct estimate of the static
  length-scale accompanying the glass transition,'' {\em Physica A: Statistical
  Mechanics and its Applications}, vol.~391, pp.~1001--1008, Jun 2012.

\bibitem{JOPCMEPP}
E.~Tamborini1, C.~P. Royall, and P.~Cicuta, ``Correlation between crystalline
  order and vitrification in colloidal monolayers,'' {\em Journal of Physics:
  Condensed Matter}, vol.~27, p.~194124, Nov 2010.

\bibitem{PhysRevLett.105.015701}
S.~Karmakar, C.~Dasgupta, and S.~Sastry, ``Analysis of dynamic heterogeneity in
  a glass former from the spatial correlations of mobility,'' {\em Phys. Rev.
  Lett.}, vol.~105, p.~015701, Jul 2010.

\bibitem{PhysRevLett.105.019801}
S.~Karmakar, C.~Dasgupta, and S.~Sastry, ``Comment on ``scaling analysis of
  dynamic heterogeneity in a supercooled lennard-jones liquid'','' {\em Phys.
  Rev. Lett.}, vol.~105, p.~019801, Jul 2010.

\bibitem{PhysRevLett.105.217801}
E.~Flenner and G.~Szamel, ``Dynamic heterogeneity in a glass forming fluid:
  Susceptibility, structure factor, and correlation length,'' {\em Phys. Rev.
  Lett.}, vol.~105, p.~217801, Nov 2010.

\bibitem{PNASUSA2009}
S.~Karmakar, C.~Dasgupta, and S.~Sastry, ``Growing length and time scales in
  glass-forming liquids,'' {\em Proc. Natl. Acad. Sci. USA}, vol.~106,
  pp.~3675--3679, Jan 2009.

\bibitem{Zeitschrift}
K.~Binder, ``Finite size scaling analysis of ising model block distribution
  functions,'' {\em Zeitschrift f{ü}r Physik B Condensed Matter}, vol.~43,
  pp.~119--140, 1981.

\bibitem{JChemPhys.2013Mar28}
F.~W. Starr, J.~F. Douglas, and S.~Sastry, ``The relationship of dynamical
  heterogeneity to the adam-gibbs and random first-order transition theories of
  glass formation,'' {\em The Journal of Chemical Physics}, vol.~138, Mar 2013.

\bibitem{JCP121-7347}
J.-P. Bouchaud and G.~Biroli, ``On the
  adam-gibbs-kirkpatrick-thirumalai-wolynes scenario for the viscosity increase
  in glasses,'' {\em The Journal of Chemical Physics}, vol.~121,
  pp.~7347--7354, July 2004.

\bibitem{PhysRevLett.116.098302}
B.~Zhang and X.~Cheng, ``Structures and dynamics of glass-forming colloidal
  liquids under spherical confinement,'' {\em Phys. Rev. Lett.}, vol.~116,
  p.~098302, Mar 2016.

\bibitem{PhysRevLett.111.165701}
G.~Biroli, S.~Karmakar, and I.~Procaccia, ``Comparison of static length scales
  characterizing the glass transition,'' {\em Phys. Rev. Lett.}, vol.~111,
  p.~165701, Oct 2013.

\bibitem{PhysRevE.94.032605}
S.~Yaida, L.~Berthier, P.~Charbonneau, and G.~Tarjus, ``Point-to-set lengths,
  local structure, and glassiness,'' {\em Phys. Rev. E}, vol.~94, p.~032605,
  Sep 2016.

\end{thebibliography}


\begin{thebibliography}{20}%
\makeatletter
\providecommand \@ifxundefined [1]{%
 \@ifx{#1\undefined}
}%
\providecommand \@ifnum [1]{%
 \ifnum #1\expandafter \@firstoftwo
 \else \expandafter \@secondoftwo
 \fi
}%
\providecommand \@ifx [1]{%
 \ifx #1\expandafter \@firstoftwo
 \else \expandafter \@secondoftwo
 \fi
}%
\providecommand \natexlab [1]{#1}%
\providecommand \enquote  [1]{``#1''}%
\providecommand \bibnamefont  [1]{#1}%
\providecommand \bibfnamefont [1]{#1}%
\providecommand \citenamefont [1]{#1}%
\providecommand \href@noop [0]{\@secondoftwo}%
\providecommand \href [0]{\begingroup \@sanitize@url \@href}%
\providecommand \@href[1]{\@@startlink{#1}\@@href}%
\providecommand \@@href[1]{\endgroup#1\@@endlink}%
\providecommand \@sanitize@url [0]{\catcode `\\12\catcode `\$12\catcode
  `\&12\catcode `\#12\catcode `\^12\catcode `\_12\catcode `\%12\relax}%
\providecommand \@@startlink[1]{}%
\providecommand \@@endlink[0]{}%
\providecommand \url  [0]{\begingroup\@sanitize@url \@url }%
\providecommand \@url [1]{\endgroup\@href {#1}{\urlprefix }}%
\providecommand \urlprefix  [0]{URL }%
\providecommand \Eprint [0]{\href }%
\providecommand \doibase [0]{http://dx.doi.org/}%
\providecommand \selectlanguage [0]{\@gobble}%
\providecommand \bibinfo  [0]{\@secondoftwo}%
\providecommand \bibfield  [0]{\@secondoftwo}%
\providecommand \translation [1]{[#1]}%
\providecommand \BibitemOpen [0]{}%
\providecommand \bibitemStop [0]{}%
\providecommand \bibitemNoStop [0]{.\EOS\space}%
\providecommand \EOS [0]{\spacefactor3000\relax}%
\providecommand \BibitemShut  [1]{\csname bibitem#1\endcsname}%
\let\auto@bib@innerbib\@empty
%</preamble>
\bibitem [{\citenamefont {Kob}\ and\ \citenamefont {Andersen}(1995)}]{KA}%
  \BibitemOpen
  \bibfield  {author} {\bibinfo {author} {\bibfnamefont {W.}~\bibnamefont
  {Kob}}\ and\ \bibinfo {author} {\bibfnamefont {H.~C.}\ \bibnamefont
  {Andersen}},\ }\href {\doibase 10.1103/PhysRevE.52.4134} {\bibfield
  {journal} {\bibinfo  {journal} {Phys. Rev. E}\ }\textbf {\bibinfo {volume}
  {52}},\ \bibinfo {pages} {4134} (\bibinfo {year} {1995})}\BibitemShut
  {NoStop}%
\bibitem [{\citenamefont {Pedersen}\ \emph {et~al.}(2010)\citenamefont
  {Pedersen}, \citenamefont {Schr\o{}der},\ and\ \citenamefont
  {Dyre}}]{10PSDPRL}%
  \BibitemOpen
  \bibfield  {author} {\bibinfo {author} {\bibfnamefont {U.~R.}\ \bibnamefont
  {Pedersen}}, \bibinfo {author} {\bibfnamefont {T.~B.}\ \bibnamefont
  {Schr\o{}der}}, \ and\ \bibinfo {author} {\bibfnamefont {J.~C.}\ \bibnamefont
  {Dyre}},\ }\href {\doibase 10.1103/PhysRevLett.105.157801} {\bibfield
  {journal} {\bibinfo  {journal} {Phys. Rev. Lett.}\ }\textbf {\bibinfo
  {volume} {105}},\ \bibinfo {pages} {157801} (\bibinfo {year}
  {2010})}\BibitemShut {NoStop}%
\bibitem [{\citenamefont {Karmakar}\ \emph {et~al.}(2012)\citenamefont
  {Karmakar}, \citenamefont {Lerner},\ and\ \citenamefont {Procaccia}}]{2dR10}%
  \BibitemOpen
  \bibfield  {author} {\bibinfo {author} {\bibfnamefont {S.}~\bibnamefont
  {Karmakar}}, \bibinfo {author} {\bibfnamefont {E.}~\bibnamefont {Lerner}}, \
  and\ \bibinfo {author} {\bibfnamefont {I.}~\bibnamefont {Procaccia}},\ }\href
  {\doibase 10.1016/j.physa.2011.11.020} {\bibfield  {journal} {\bibinfo
  {journal} {Physica A: Statistical Mechanics and its Applications}\ }\textbf
  {\bibinfo {volume} {391}},\ \bibinfo {pages} {1001} (\bibinfo {year}
  {2012})}\BibitemShut {NoStop}%
\bibitem [{\citenamefont {Weeks}\ \emph {et~al.}(1971)\citenamefont {Weeks},
  \citenamefont {Chandler},\ and\ \citenamefont {Andersen}}]{WCA}%
  \BibitemOpen
  \bibfield  {author} {\bibinfo {author} {\bibfnamefont {D.~J.}\ \bibnamefont
  {Weeks}}, \bibinfo {author} {\bibfnamefont {D.}~\bibnamefont {Chandler}}, \
  and\ \bibinfo {author} {\bibfnamefont {C.~H.}\ \bibnamefont {Andersen}},\
  }\href {\doibase 10.1063/1.1674820} {\bibfield  {journal} {\bibinfo
  {journal} {The Journal of Chemical Physics}\ }\textbf {\bibinfo {volume}
  {54}},\ \bibinfo {pages} {5237} (\bibinfo {year} {1971})}\BibitemShut
  {NoStop}%
\bibitem [{\citenamefont {Hamanaka}\ and\ \citenamefont {Onuki}(2006)}]{hop1}%
  \BibitemOpen
  \bibfield  {author} {\bibinfo {author} {\bibfnamefont {T.}~\bibnamefont
  {Hamanaka}}\ and\ \bibinfo {author} {\bibfnamefont {A.}~\bibnamefont
  {Onuki}},\ }\href {\doibase 10.1103/PhysRevE.74.011506} {\bibfield  {journal}
  {\bibinfo  {journal} {Phys. Rev. E}\ }\textbf {\bibinfo {volume} {74}},\
  \bibinfo {pages} {011506} (\bibinfo {year} {2006})}\BibitemShut {NoStop}%
\bibitem [{\citenamefont {Binder}\ \emph {et~al.}(2002)\citenamefont {Binder},
  \citenamefont {Sengupta},\ and\ \citenamefont {Nielaba}}]{hop2}%
  \BibitemOpen
  \bibfield  {author} {\bibinfo {author} {\bibfnamefont {K.}~\bibnamefont
  {Binder}}, \bibinfo {author} {\bibfnamefont {S.}~\bibnamefont {Sengupta}}, \
  and\ \bibinfo {author} {\bibfnamefont {P.}~\bibnamefont {Nielaba}},\ }\href
  {http://iopscience.iop.org/0953-8984/14/9/321} {\bibfield  {journal}
  {\bibinfo  {journal} {Journal of Physics: Condensed Matter}\ }\textbf
  {\bibinfo {volume} {14}},\ \bibinfo {pages} {2323} (\bibinfo {year}
  {2002})}\BibitemShut {NoStop}%
\bibitem [{\citenamefont {Kawasaki}\ \emph {et~al.}(2007)\citenamefont
  {Kawasaki}, \citenamefont {Araki},\ and\ \citenamefont {Tanaka}}]{g6}%
  \BibitemOpen
  \bibfield  {author} {\bibinfo {author} {\bibfnamefont {T.}~\bibnamefont
  {Kawasaki}}, \bibinfo {author} {\bibfnamefont {T.}~\bibnamefont {Araki}}, \
  and\ \bibinfo {author} {\bibfnamefont {H.}~\bibnamefont {Tanaka}},\ }\href
  {\doibase 10.1103/PhysRevLett.99.215701} {\bibfield  {journal} {\bibinfo
  {journal} {Phys. Rev. Lett.}\ }\textbf {\bibinfo {volume} {99}},\ \bibinfo
  {pages} {215701} (\bibinfo {year} {2007})}\BibitemShut {NoStop}%
\bibitem [{\citenamefont {Kawasaki}\ and\ \citenamefont {Tanaka}(2011)}]{g61}%
  \BibitemOpen
  \bibfield  {author} {\bibinfo {author} {\bibfnamefont {T.}~\bibnamefont
  {Kawasaki}}\ and\ \bibinfo {author} {\bibfnamefont {H.}~\bibnamefont
  {Tanaka}},\ }\href {\doibase 10.1088/0953-8984/23/19/194121} {\bibfield
  {journal} {\bibinfo  {journal} {Journal of Physics: Condensed Matter}\
  }\textbf {\bibinfo {volume} {23}} (\bibinfo {year} {2011}),\
  10.1088/0953-8984/23/19/194121}\BibitemShut {NoStop}%
\bibitem [{\citenamefont {Karmakar}\ \emph {et~al.}(2014)\citenamefont
  {Karmakar}, \citenamefont {Dasgupta},\ and\ \citenamefont
  {Sastry}}]{SMANNUAL}%
  \BibitemOpen
  \bibfield  {author} {\bibinfo {author} {\bibfnamefont {S.}~\bibnamefont
  {Karmakar}}, \bibinfo {author} {\bibfnamefont {C.}~\bibnamefont {Dasgupta}},
  \ and\ \bibinfo {author} {\bibfnamefont {S.}~\bibnamefont {Sastry}},\ }\href
  {\doibase 10.1146/annurev-conmatphys-031113-133848} {\bibfield  {journal}
  {\bibinfo  {journal} {Annual Review of Condensed Matter Physics}\ }\textbf
  {\bibinfo {volume} {5}},\ \bibinfo {pages} {255} (\bibinfo {year}
  {2014})}\BibitemShut {NoStop}%
\bibitem [{\citenamefont {Karmakar}\ \emph {et~al.}(2009)\citenamefont
  {Karmakar}, \citenamefont {Dasgupta},\ and\ \citenamefont {Sastry}}]{SMPNAS}%
  \BibitemOpen
  \bibfield  {author} {\bibinfo {author} {\bibfnamefont {S.}~\bibnamefont
  {Karmakar}}, \bibinfo {author} {\bibfnamefont {C.}~\bibnamefont {Dasgupta}},
  \ and\ \bibinfo {author} {\bibfnamefont {S.}~\bibnamefont {Sastry}},\ }\href
  {\doibase 10.1073/pnas.0811082106} {\bibfield  {journal} {\bibinfo  {journal}
  {Proc. Natl. Acad. Sci. USA}\ }\textbf {\bibinfo {volume} {106}},\ \bibinfo
  {pages} {3675} (\bibinfo {year} {2009})}\BibitemShut {NoStop}%
\bibitem [{\citenamefont {Dasgupta}\ \emph {et~al.}(1991)\citenamefont
  {Dasgupta}, \citenamefont {Indrani}, \citenamefont {Ramaswamy},\ and\
  \citenamefont {Phani}}]{chandan92}%
  \BibitemOpen
  \bibfield  {author} {\bibinfo {author} {\bibfnamefont {C.}~\bibnamefont
  {Dasgupta}}, \bibinfo {author} {\bibfnamefont {V.~A.}\ \bibnamefont
  {Indrani}}, \bibinfo {author} {\bibfnamefont {S.}~\bibnamefont {Ramaswamy}},
  \ and\ \bibinfo {author} {\bibfnamefont {K.~M.}\ \bibnamefont {Phani}},\
  }\href@noop {} {\bibfield  {journal} {\bibinfo  {journal} {Europhys. Lett.}\
  }\textbf {\bibinfo {volume} {15}},\ \bibinfo {pages} {307} (\bibinfo {year}
  {1991})}\BibitemShut {NoStop}%
\bibitem [{\citenamefont {Binder}(1981)}]{BINDER}%
  \BibitemOpen
  \bibfield  {author} {\bibinfo {author} {\bibfnamefont {K.}~\bibnamefont
  {Binder}},\ }\href {\doibase 10.1007/BF01293604} {\bibfield  {journal}
  {\bibinfo  {journal} {Zeitschrift f{ü}r Physik B Condensed Matter}\ }\textbf
  {\bibinfo {volume} {43}},\ \bibinfo {pages} {119} (\bibinfo {year}
  {1981})}\BibitemShut {NoStop}%
\bibitem [{\citenamefont {Lacˇevic}\ \emph {et~al.}(2003)\citenamefont
  {Lacˇevic}, \citenamefont {Star}, \citenamefont {Schrøder},\ and\
  \citenamefont {Glotzer}}]{S4Lacevic}%
  \BibitemOpen
  \bibfield  {author} {\bibinfo {author} {\bibfnamefont {N.}~\bibnamefont
  {Lacˇevic}}, \bibinfo {author} {\bibfnamefont {W.~F.}\ \bibnamefont {Star}},
  \bibinfo {author} {\bibfnamefont {B.~T.}\ \bibnamefont {Schrøder}}, \ and\
  \bibinfo {author} {\bibfnamefont {C.~S.}\ \bibnamefont {Glotzer}},\ }\href
  {\doibase http://dx.doi.org/10.1063/1.1796231} {\bibfield  {journal}
  {\bibinfo  {journal} {The Journal of Chemical Physics}\ }\textbf {\bibinfo
  {volume} {119}},\ \bibinfo {pages} {7372} (\bibinfo {year}
  {2003})}\BibitemShut {NoStop}%
\bibitem [{\citenamefont {Shiba}\ \emph {et~al.}(2016)\citenamefont {Shiba},
  \citenamefont {Yamada}, \citenamefont {Kawasaki},\ and\ \citenamefont
  {Kim}}]{shiba}%
  \BibitemOpen
  \bibfield  {author} {\bibinfo {author} {\bibfnamefont {H.}~\bibnamefont
  {Shiba}}, \bibinfo {author} {\bibfnamefont {Y.}~\bibnamefont {Yamada}},
  \bibinfo {author} {\bibfnamefont {T.}~\bibnamefont {Kawasaki}}, \ and\
  \bibinfo {author} {\bibfnamefont {K.}~\bibnamefont {Kim}},\ }\href {\doibase
  10.1103/PhysRevLett.117.245701} {\bibfield  {journal} {\bibinfo  {journal}
  {Phys. Rev. Lett.}\ }\textbf {\bibinfo {volume} {117}},\ \bibinfo {pages}
  {245701} (\bibinfo {year} {2016})}\BibitemShut {NoStop}%
\bibitem [{\citenamefont {Yamamoto}\ and\ \citenamefont {Onuki}(1997)}]{YO}%
  \BibitemOpen
  \bibfield  {author} {\bibinfo {author} {\bibfnamefont {R.}~\bibnamefont
  {Yamamoto}}\ and\ \bibinfo {author} {\bibfnamefont {A.}~\bibnamefont
  {Onuki}},\ }\href@noop {} {\bibfield  {journal} {\bibinfo  {journal} {J.
  Phys. Soc. Jpn.}\ }\textbf {\bibinfo {volume} {66}},\ \bibinfo {pages} {2545}
  (\bibinfo {year} {1997})}\BibitemShut {NoStop}%
\bibitem [{\citenamefont {Karmakar}\ \emph {et~al.}(2015)\citenamefont
  {Karmakar}, \citenamefont {Dasgupta},\ and\ \citenamefont {Sastry}}]{SMFSS}%
  \BibitemOpen
  \bibfield  {author} {\bibinfo {author} {\bibfnamefont {S.}~\bibnamefont
  {Karmakar}}, \bibinfo {author} {\bibfnamefont {C.}~\bibnamefont {Dasgupta}},
  \ and\ \bibinfo {author} {\bibfnamefont {S.}~\bibnamefont {Sastry}},\ }\href
  {\doibase 10.1088/0034-4885/79/1/016601} {\bibfield  {journal} {\bibinfo
  {journal} {Reports on Progress in Physics}\ }\textbf {\bibinfo {volume}
  {79}},\ \bibinfo {pages} {016601} (\bibinfo {year} {2015})}\BibitemShut
  {NoStop}%
\bibitem [{\citenamefont {Biroli}\ \emph {et~al.}(2008)\citenamefont {Biroli},
  \citenamefont {Bouchaud}, \citenamefont {Cavagna}, \citenamefont {Grigera},\
  and\ \citenamefont {Verrocchio}}]{PTS1}%
  \BibitemOpen
  \bibfield  {author} {\bibinfo {author} {\bibfnamefont {G.}~\bibnamefont
  {Biroli}}, \bibinfo {author} {\bibfnamefont {J.-P.}\ \bibnamefont
  {Bouchaud}}, \bibinfo {author} {\bibfnamefont {A.}~\bibnamefont {Cavagna}},
  \bibinfo {author} {\bibfnamefont {T.~S.}\ \bibnamefont {Grigera}}, \ and\
  \bibinfo {author} {\bibfnamefont {P.}~\bibnamefont {Verrocchio}},\ }\href
  {\doibase http://dx.doi.org/10.1038/nphys1050} {\bibfield  {journal}
  {\bibinfo  {journal} {Nat Phys}\ }\textbf {\bibinfo {volume} {99}},\ \bibinfo
  {pages} {771} (\bibinfo {year} {2008})}\BibitemShut {NoStop}%
\bibitem [{\citenamefont {Hocky}\ \emph {et~al.}(2012)\citenamefont {Hocky},
  \citenamefont {Markland},\ and\ \citenamefont {Reichman}}]{PTS2}%
  \BibitemOpen
  \bibfield  {author} {\bibinfo {author} {\bibfnamefont {G.~M.}\ \bibnamefont
  {Hocky}}, \bibinfo {author} {\bibfnamefont {T.~E.}\ \bibnamefont {Markland}},
  \ and\ \bibinfo {author} {\bibfnamefont {D.~R.}\ \bibnamefont {Reichman}},\
  }\href {\doibase 10.1103/PhysRevLett.108.225506} {\bibfield  {journal}
  {\bibinfo  {journal} {Phys. Rev. Lett.}\ }\textbf {\bibinfo {volume} {108}},\
  \bibinfo {pages} {225506} (\bibinfo {year} {2012})}\BibitemShut {NoStop}%
\bibitem [{\citenamefont {Grigera}\ and\ \citenamefont {Parisi}(2001)}]{PSA}%
  \BibitemOpen
  \bibfield  {author} {\bibinfo {author} {\bibfnamefont {T.~S.}\ \bibnamefont
  {Grigera}}\ and\ \bibinfo {author} {\bibfnamefont {G.}~\bibnamefont
  {Parisi}},\ }\href {\doibase 10.1103/PhysRevE.63.045102} {\bibfield
  {journal} {\bibinfo  {journal} {Phys. Rev. E}\ }\textbf {\bibinfo {volume}
  {63}},\ \bibinfo {pages} {045102} (\bibinfo {year} {2001})}\BibitemShut
  {NoStop}%
\bibitem [{\citenamefont {Russo}\ and\ \citenamefont {Tanaka}(2015)}]{POL}%
  \BibitemOpen
  \bibfield  {author} {\bibinfo {author} {\bibfnamefont {J.}~\bibnamefont
  {Russo}}\ and\ \bibinfo {author} {\bibfnamefont {H.}~\bibnamefont {Tanaka}},\
  }\href {\doibase 10.1073/pnas.1501911112} {\bibfield  {journal} {\bibinfo
  {journal} {Proc. Natl. Acad. Sci. USA}\ }\textbf {\bibinfo {volume} {112}},\
  \bibinfo {pages} {6920} (\bibinfo {year} {2015})}\BibitemShut {NoStop}%
\end{thebibliography}%
\bibliographystyle{ieeetr}
%\begin{thebibliography}{10}
\end{document}

\title{Glass Transition in Supercooled Liquids with Medium Range 
Crystalline Order -- Supplementary Information}
\author{Indrajit Tah$^{1}$}
\author{Shiladitya Sengupta$^{2}$}
\author{Srikanth Sastry$^{3}$}
\author{Chandan Dasgupta$^{4}$}
\author{Smarajit Karmakar$^{1}$}
%\email{smarajit@tifrh.res.in}
\affiliation{$^1$ Centre for Interdisciplinary Sciences,
Tata Institute of Fundamental Research, 
21 Brundavan Colony, Narisingi, Hyderabad, 500075, India,\\
$^2$ Department of Chemical Physics, Weizmann Institute of Science, Israel,\\
$^3$ Jawaharlal Nehru Centre for Advanced 
Scientific Research, Bangalore 560064, India,\\
$^4$ Centre for Condensed Matter Theory, 
Department of Physics, Indian Institute of Science, Bangalore, 560012, 
India}

\maketitle

\section{Models and Simulation Details}
\label{modelsAndSim}
We have studied four different model glass forming liquids in two 
dimensions. The model details are given below:
\vskip +0.3cm
\noindent{\textbf{2dKA:}} The model glass former we have studied is 
the Kob-Anderson $80:20$ \cite{KA} Lenard-Jones Binary mixture. 
This model was first introduced by Kob and Anderson to simulate 
$Ni_{80}P_{20}$. This model has been studied extensively  
and found to be a very good glass former in three dimensions. In two 
dimensions this model shows strong growth of medium range crystalline 
order (MRCO).

\noindent The interaction potential is given by
\[ V_{\alpha\beta}(r) = 4.0\epsilon_{\alpha\beta}[(\frac{\sigma_{\alpha\beta}}{r})^{12} - (\frac{\sigma_{\alpha\beta}}{r})^6]\]
where $\alpha,\beta \in \{ A,B \}$ and $\epsilon_{AA} = 1.0$, $\epsilon_{AB} = 1.5$, $\epsilon_{BB} = 0.5$; 
$\sigma_{AA} = 1.0$, $\sigma_{AB} = 0.80$,   $\sigma_{BB} = 0.88$. The interaction potential was cut off at 
$2.50\sigma_{\alpha\beta}$ and the number density is $\rho = 1.20$. Length, energy and time scale are measured in 
units of $\sigma_{AA}, \epsilon_{AA}$ and $\sqrt{\frac{\sigma_{AA}^2}{\epsilon_{AA}}}$. For Argon these units 
corresponds to a length of $3.4 \AA$, an energy of $k_{B}(120K)$ and time of  $3\times 10^{-13} s$. We have done 
simulation in the temperature range $T \in \{ 0.930, 2.200 \}$.

\noindent{\textbf{2dIPL:}} This model glass forming liquid is the two dimensional version
of the Inverse Power Law model studied in \cite{10PSDPRL}. Here the interaction potential is given by 
\[ V_{\alpha\beta}(r) = 1.945\epsilon_{\alpha\beta}[(\frac{\sigma_{\alpha\beta}}{r})^{15.48}]\]
All the parameters and the interaction cut-off are the same as those for the 2dKA model. The 
temperature of the system is fixed at $T = 0.025$. We have performed simulations in 
the density range $\rho \in \{ 0.65,0.75 \}$.

\noindent{\textbf{2dR10:}} This is a \textbf{50:50} binary mixture \cite{2dR10} 
with the pair wise interaction potential
\[ V_{\alpha\beta}(r) = \epsilon_{\alpha\beta}[(\frac{\sigma_{\alpha\beta}}{r})^{10}]\]
Here $\epsilon_{\alpha\beta} = 1.0$, $\sigma_{AA} = 1.0$, $\sigma_{AB} = 1.22$,   
$\sigma_{BB} = 1.40$. The interaction potential is cut-off at 
$1.38\sigma_{\alpha\beta}$. The number density $\rho = 0.85$ and the 
temperature range is $T \in \{ 0.480, 1.000 \}$.

\noindent{\textbf{2dPoly:}} This is a polydisperse mixture of soft disks with the 
diameter $\sigma$ of the disks chosen from a Gaussian distribution. The 
polydispersity $(\Delta)$ of the model is defined as
\[ \Delta = \frac{\sqrt{\langle \sigma^2 \rangle - \langle \sigma \rangle^2}}{\langle \sigma \rangle}\]
For our case we fix $\Delta = 11 \%$. The particles in this polydisperse 
model systems interact with the Weeks-Chandle-Anderson potential \cite{WCA}
\[ V_{ij}(r) = 4.0\epsilon_{ij}\{(\frac{\sigma_{ij}}{r})^{12} - (\frac{\sigma_{ij}}{r})^6 + \frac{1}{4}\}\]
for $r < 2^{1/6}\sigma_{ij}$, otherwise $V_{ij}(r) = 0$, where $\sigma_{ij} = \frac{(\sigma_{i} +\sigma_{j})}{2}$.

We study the behavior of the polydisperse system at different 
temperatures in the range $T \in \{ 0.450, 0.900 \}$. While calculating the static length 
scale using finite size scaling of $\tau_{\alpha}$ (see Sec~\ref{pol2d} B for
further details), we fix the packing fraction to be the same for all system 
sizes. The packing fraction $\phi$ is defined as $\phi = \frac{1}{L^2} \sum_{i=1}^{N}\pi(\frac{\sigma_i}{2})^2$ where 
$L$ is the box size. We fix the packing fraction $\phi = 0.76$, the corresponding number density 
being $\approx 0.95$. 

\section{Calculation of $\xi_6$}
\label{psi6Calc}
\begin{figure*}[htbp]
%\hspace*{0 cm}                                                           
\includegraphics[scale=0.222]{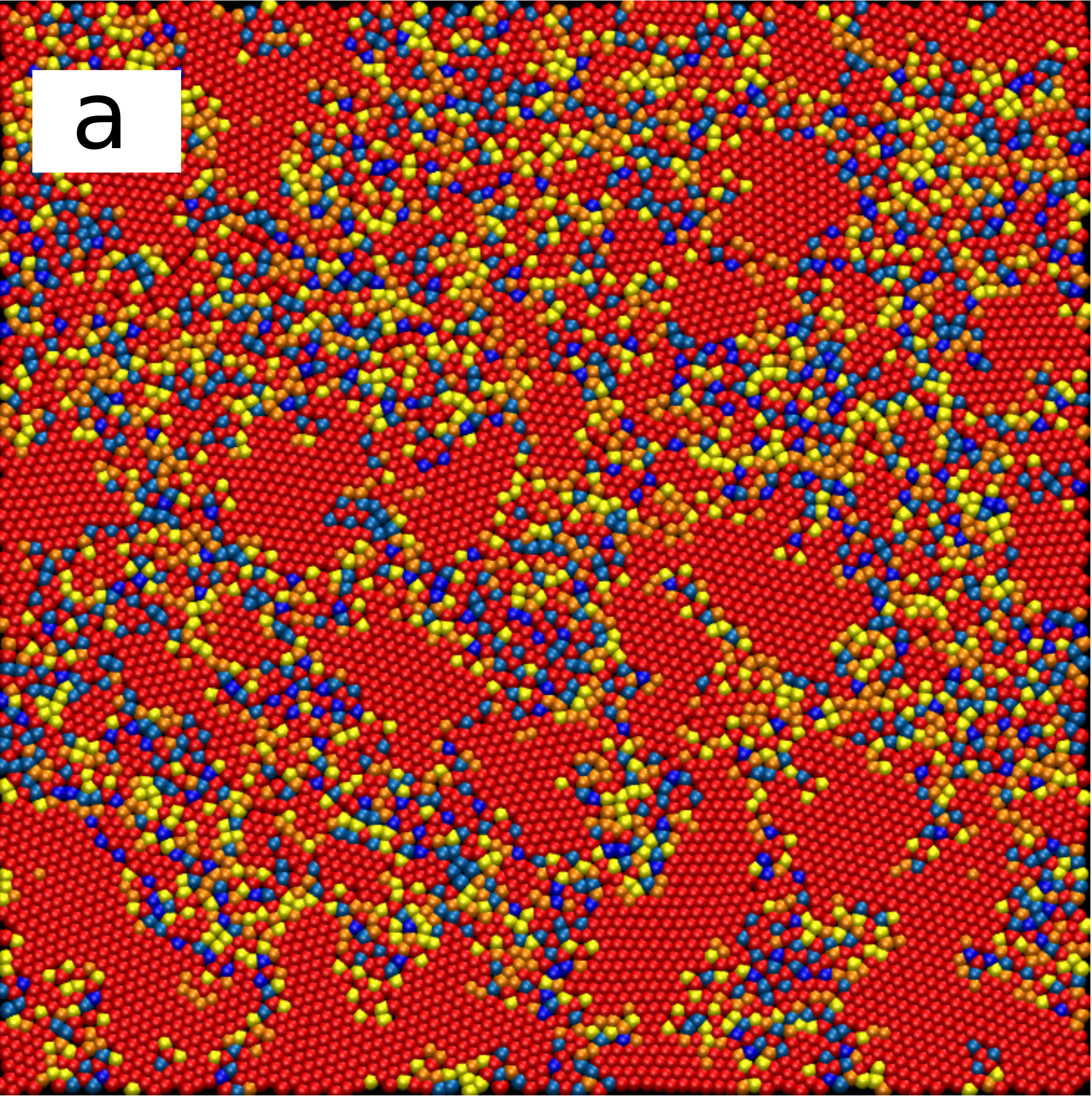}                
\includegraphics[scale=0.22]{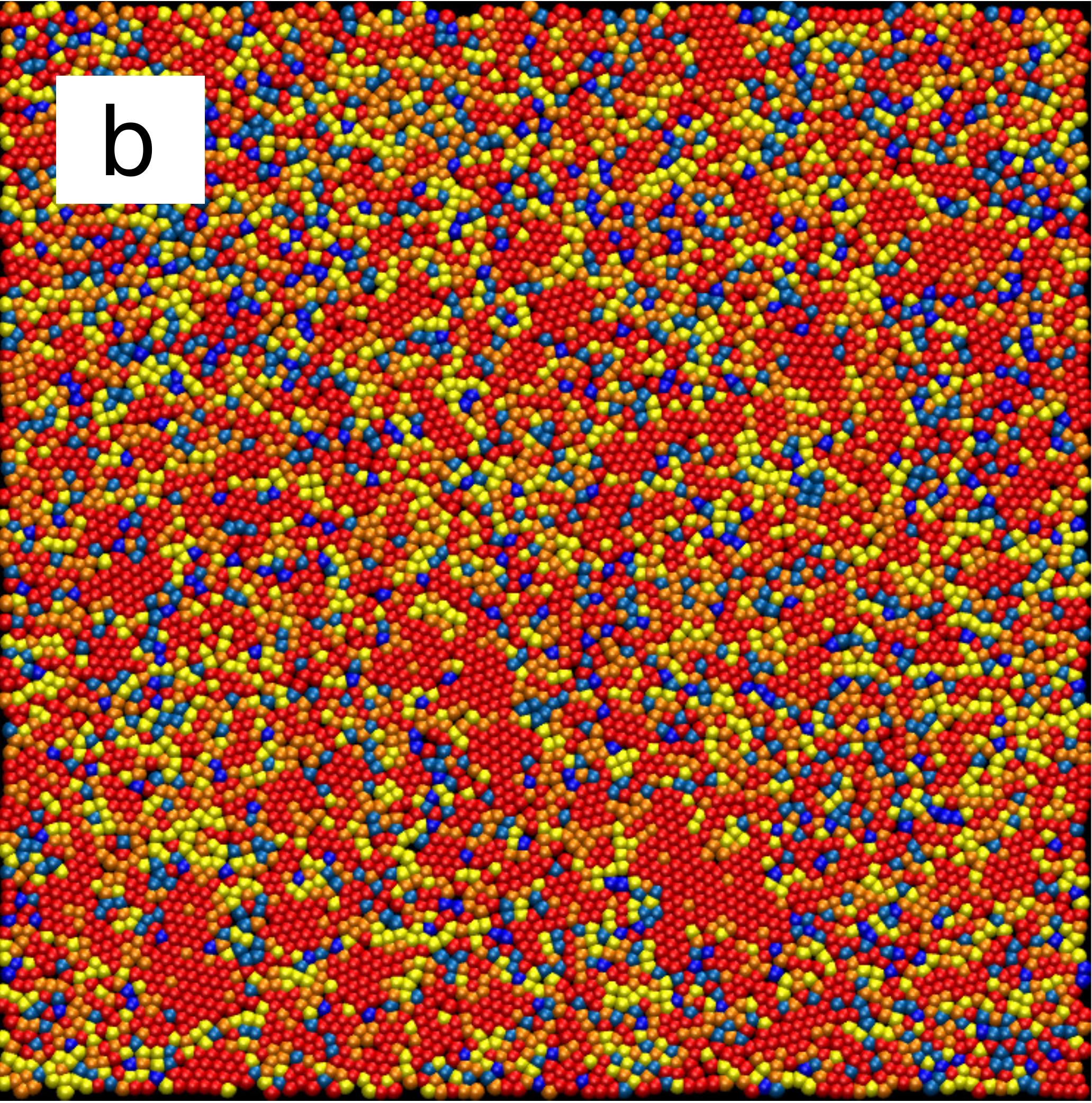} 
\includegraphics[scale=0.28]{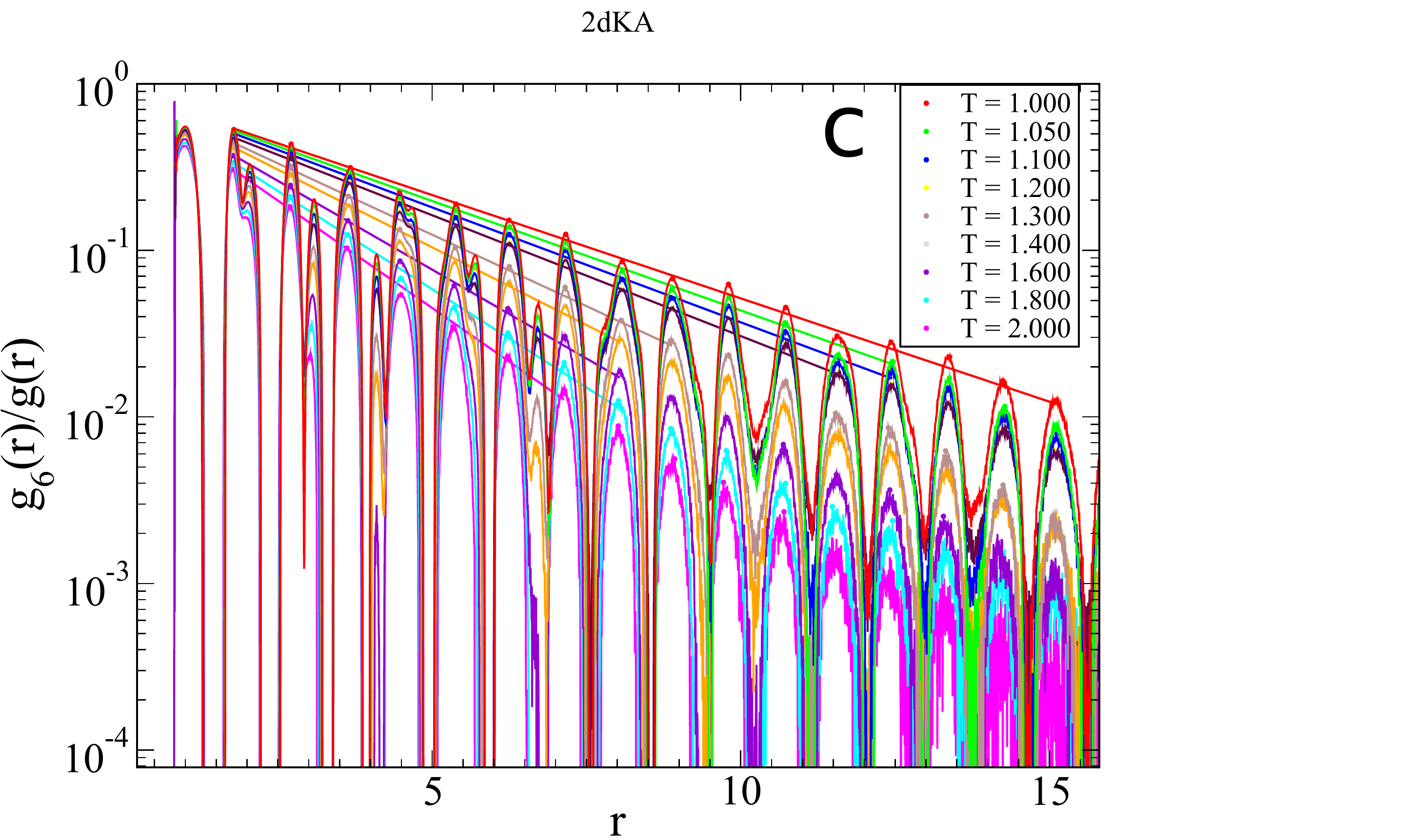}
\caption{Snapshots of configurations for the 2dKA model (a) at a low temperature, T = 1.000 and (b) at a high 
temperature, T = 2.000. The particles are colored according the hexatic order parameter. Red corresponds to higher values of 
$|\psi_6^i|$. The size of the correlation region increases with decreasing temperature. (c) Decay of the hexatic order 
correlation function $g_6(r)/g(r)$ for 2dKA systems. The solid straight lines are obtained by fitting the peaks of the 
correlation function to the 2D version of the Ornstein-Zernike (OZ) function.}%CDQ the fits are to an exponential function or to the 2D version of the Ornstein-Zernike (OZ) function?
\label{2dKAhop}
%\vskip +0.8cm
\includegraphics[scale=0.203]{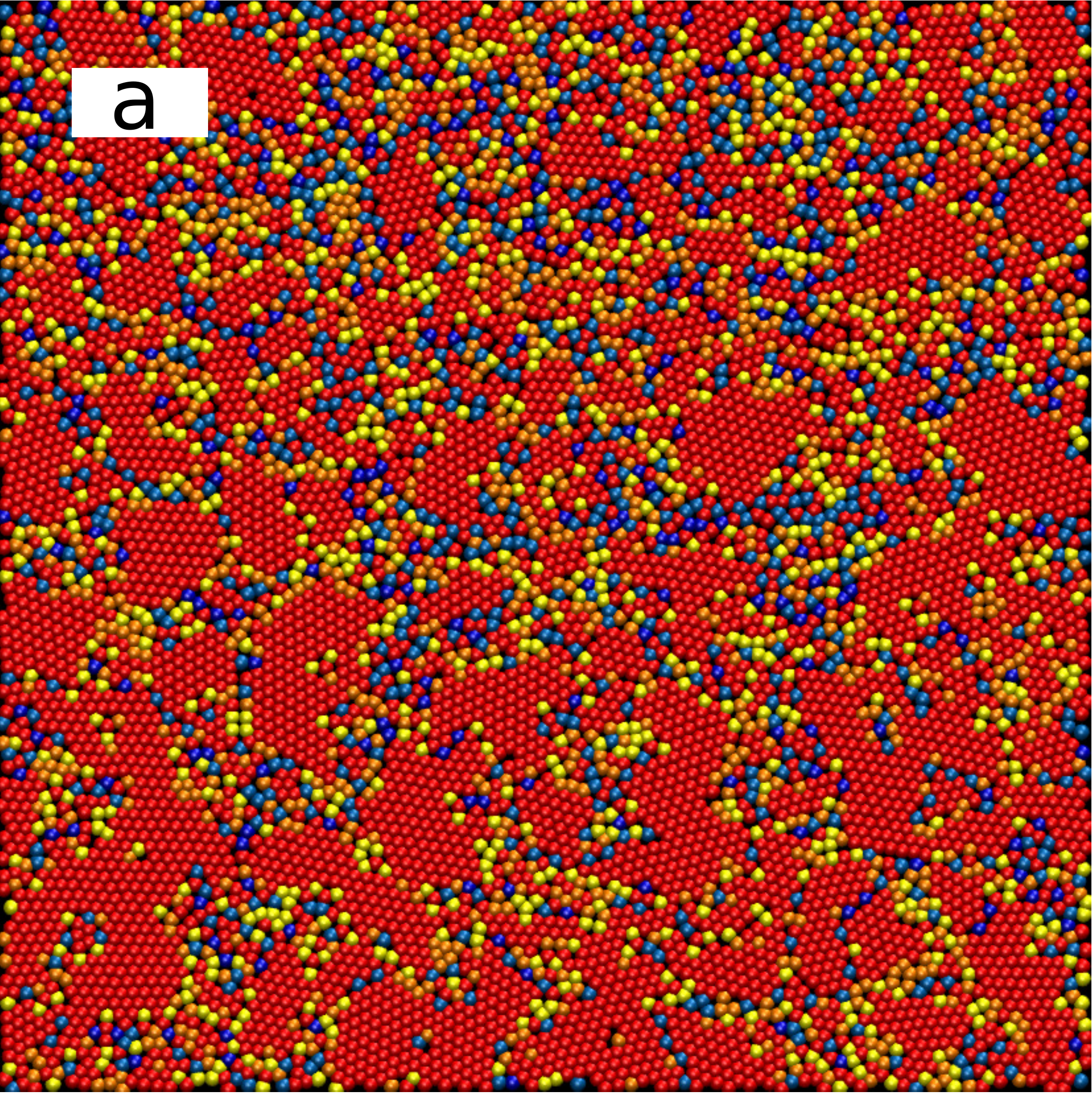}                
\includegraphics[scale=0.23]{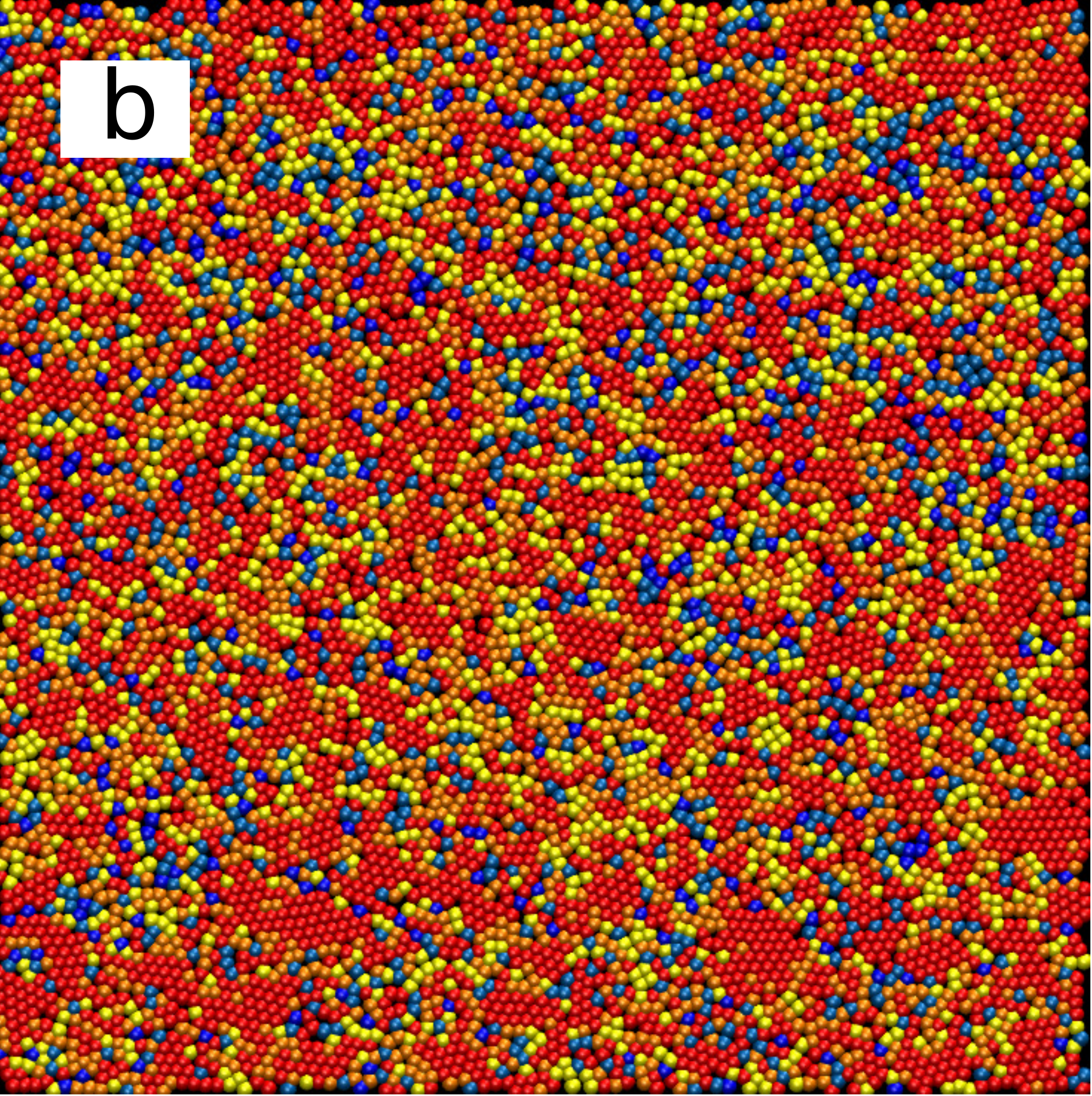}
\includegraphics[scale=0.26]{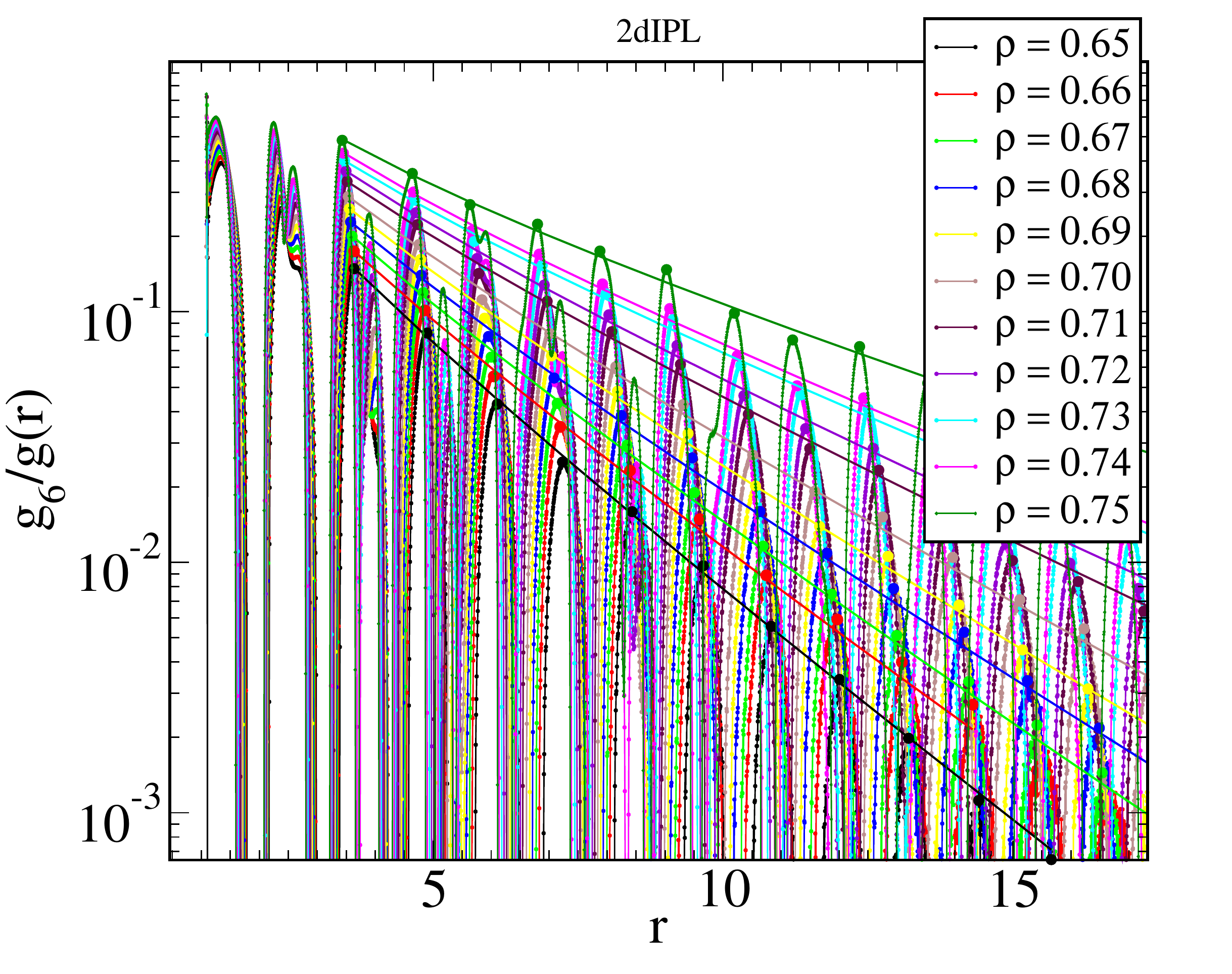}
\caption{Snapshots of configurations for the 2dIPL model (a) at high density, $\rho = 0.74$ and (b) at low density, $\rho = 
0.67$. The particles are colored according the hexatic order parameter. Red corresponds to higher values of $|\psi_6^i|$. 
The size of the correlation region increases with increasing density. (c) Decay of the hexatic order correlation 
function $g_6(r)/g(r)$ for 2dIPL systems. The solid straight lines are obtained by fitting the peaks of the correlation 
function to the 2D version of the Ornstein-Zernike (OZ) function.} %CDQ the fits are to an exponential function or to the 2D version of the Ornstein-Zernike (OZ) function?
\label{2dIPLhop}
%\vskip +0.8cm                                                           
\includegraphics[scale=0.26]{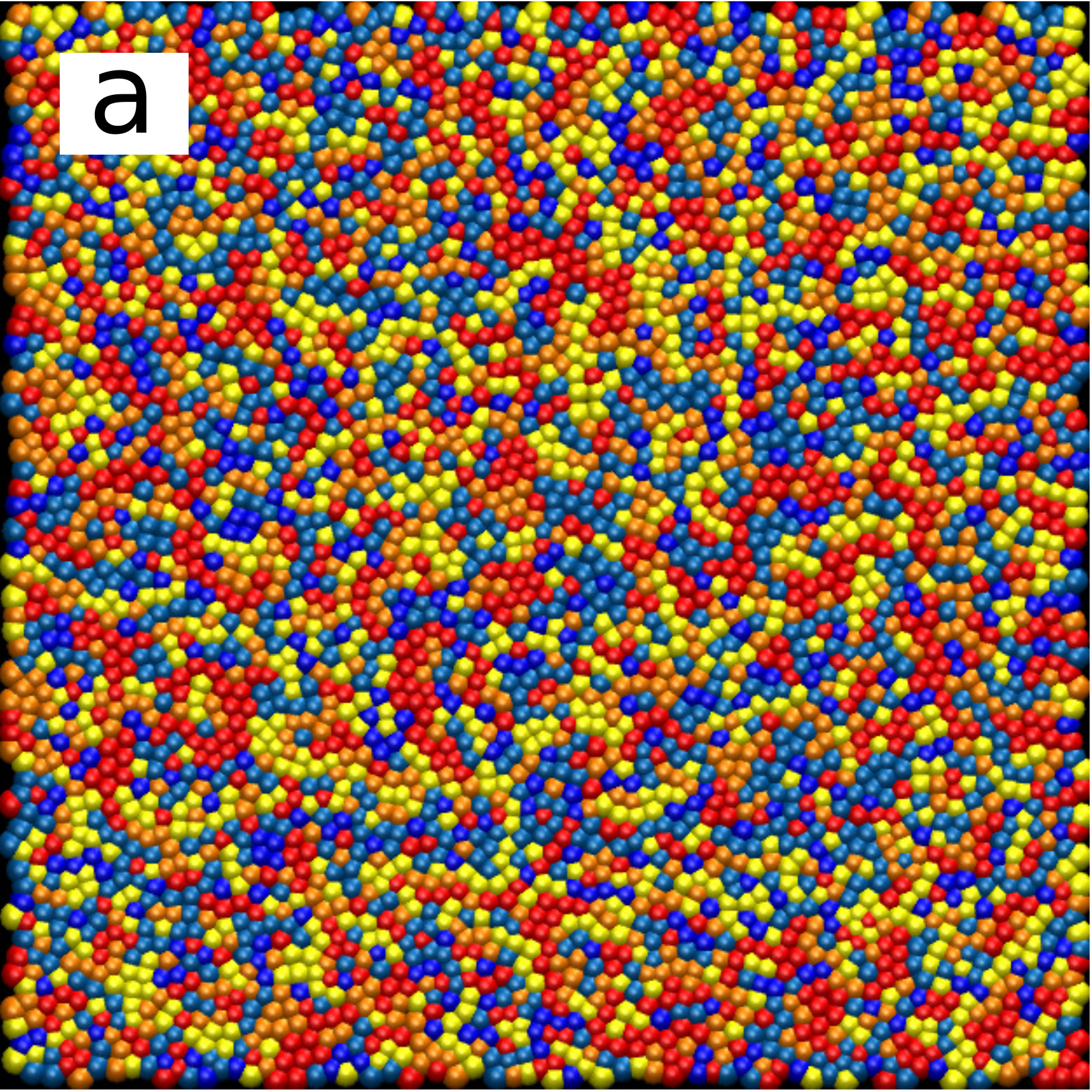}                
\includegraphics[scale=0.262]{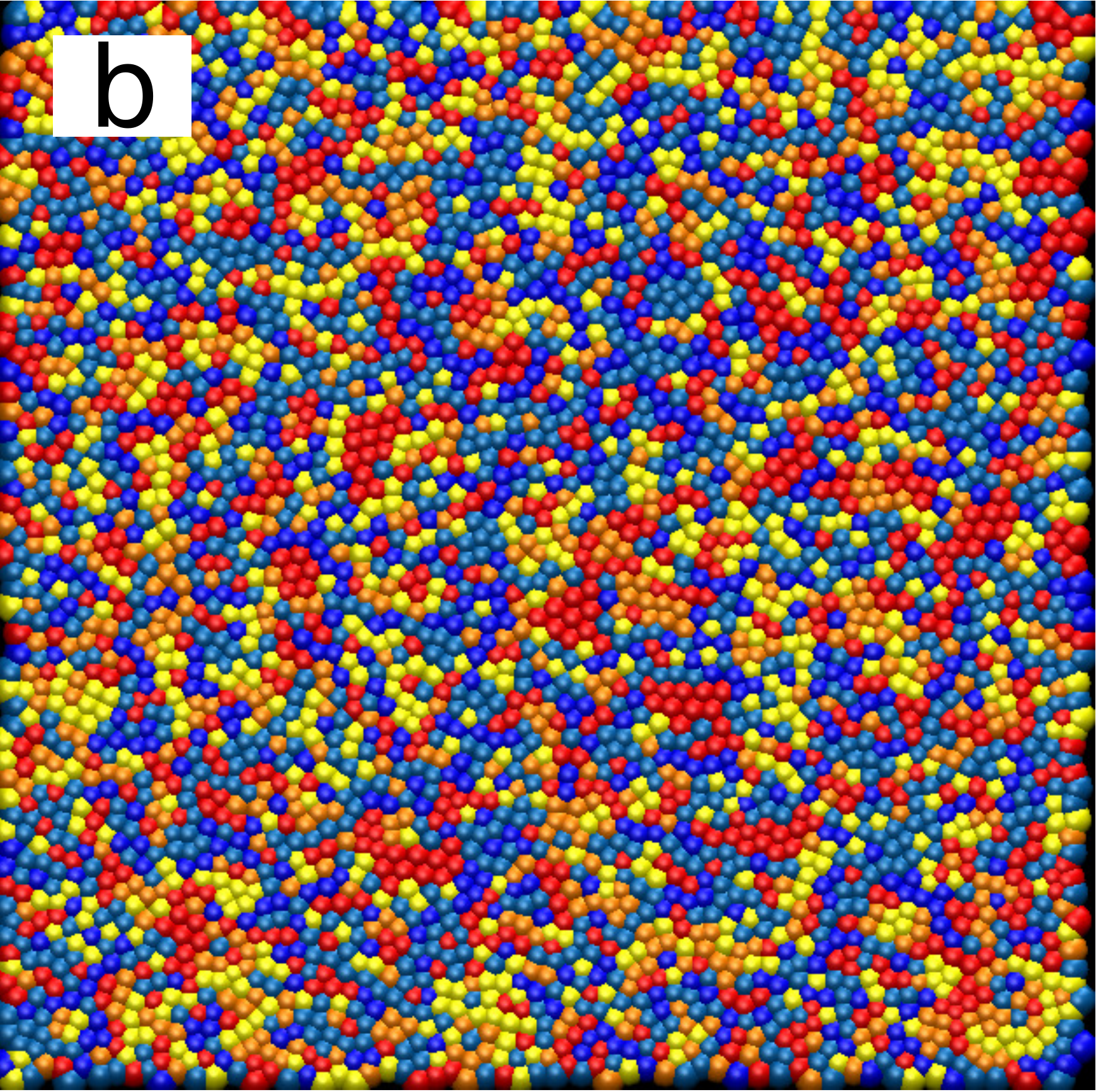} 
\includegraphics[scale=0.17]{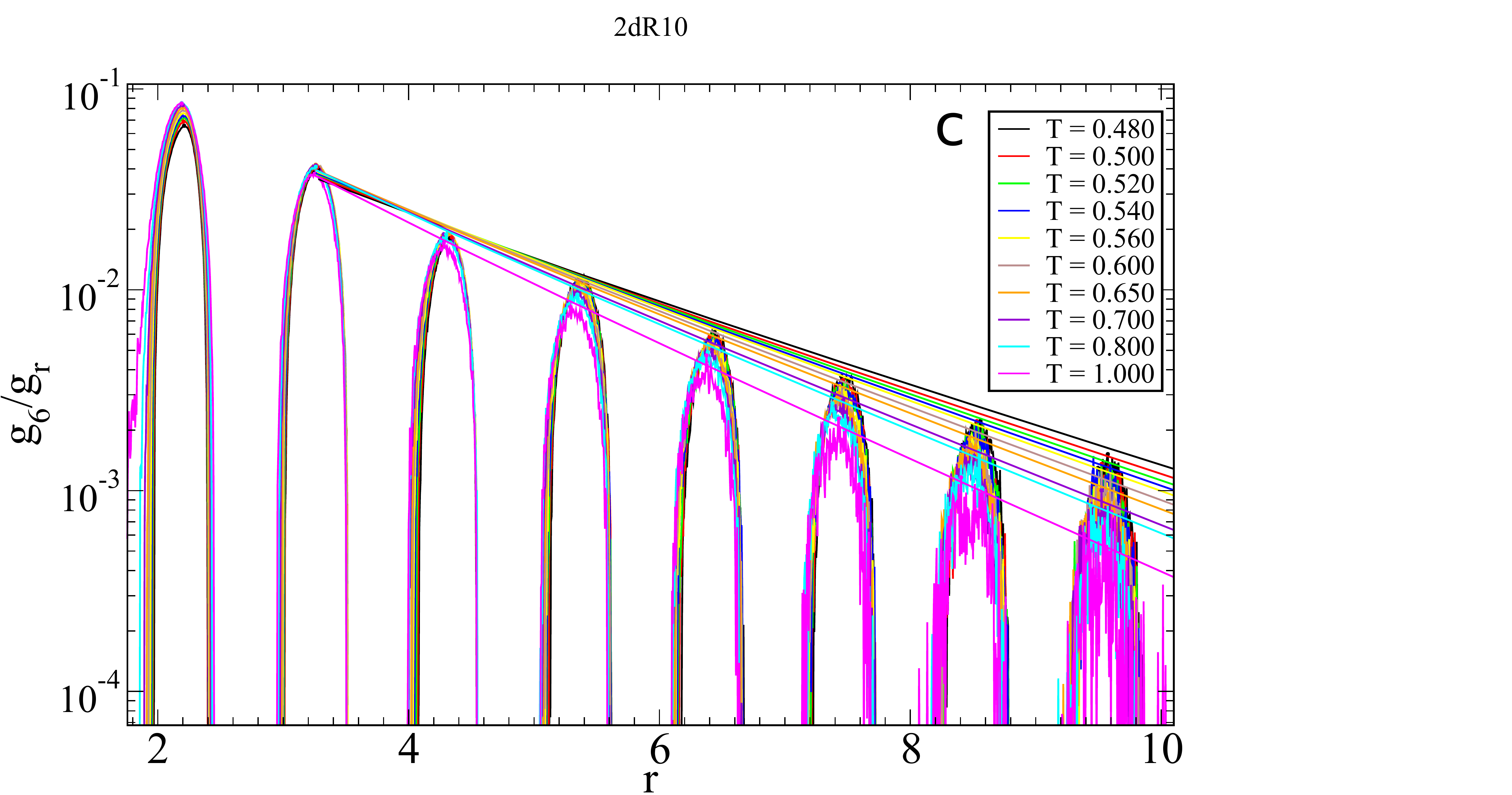}
\caption{Snapshots of configurations for the 2dR10 model (a) at a low temperature, T = 0.500 and (b) at a low 
temperature, T = 1.000. The particles are colored according the hexatic order parameter. Red corresponds to higher values of 
$|\psi_6^i|$. Here, there is no growth of local crystalline order with decreasing temperature. (c) Decay of the hexatic order 
correlation function $g_6(r)/g(r)$ for 2dR10 systems. The solid straight lines are obtained by fitting the peaks of the 
correlation function to the 2D version of the Ornstein-Zernike (OZ) function.}%CDQ the fits are to an exponential function or to the 2D version of the Ornstein-Zernike (OZ) function?
\label{2dR10hop}
\end{figure*}
In this section we describe the calculation of the hexatic order correlation 
length $\xi_6$. We calculate the bond orientational order for each particle \cite{hop1,hop2} as 
\[ \psi_6^j = \frac{1}{n_j} \sum_{k=1}^{n_j} \exp(i6\theta_{jk}).\]	
Here $\psi_6^j$ is the hexatic order parameter for the $j^{th}$ particle, 
$n_j$ is the number of nearest neighbours of particle $j$, and $\theta_{jk}$ 
is the angle made by the position vectors of two particles $j$ and $k$ 
with a reference axis. Here we have taken the cutoff nearest neighbour distance 
to be the distance of the first minimum of the pair distribution function 
$g(r)$.

The snapshots in Fig.~\ref{2dKAhop}, Fig.~\ref{2dIPLhop} and Fig.~\ref{2dR10hop} 
show the growth of hexatic order from high temperature to low temperature 
(2dKA and 2dR10 models) and low density to high density (2dIPL model). For 
the 2dR10 model it is clearly seen from Fig.\ref{2dR10hop} that there is no 
prominent growth of hexatic order at low temperatures and for the other two 
models (2dKA and 2dIPL) there is prominent growth of hexatic order 
at low temperatures and high densities. In these figures we use the same colour 
scheme as described in the main text. 
 
By defination $|\psi_6^j| = 1$ for a perfect triangular lattice and 
$|\psi_6^j |< 1$ implies that the structure deviates from the perfect triangular lattice  
configuration. We calculate the spatial correlation of local hexatic order, $g_6(r)$ \cite{g6,g61} as:
\[ g_6(r) = \frac{L^2}{2\pi r \Delta r N(N-1)} \sum_{j\neq k} \delta (r-\left|r_{jk}\right|)\psi_6^j \psi_6^{k*}\]
The hexatic correlation length $\xi_6$ is extracted by fitting the peak 
values of the normalized hexatic correlation function, $g_6(r)/g(r)$. We 
have also used the two dimensional version of the Ornstein-Zernike (OZ) function to fit the peaks of the correlation function. In the right panels
of Figs.\ref{2dKAhop}, \ref{2dIPLhop} and \ref{2dR10hop} we show the hexatic 
order correlation function for the three model systems.
%\begin{figure}[htbp]
%\hspace*{0 cm}
%\centering                                                               
%  \includegraphics[scale=0.18]{ALLFIGURE/2dKA/g62dKA-eps-converted-to.pdf}%
%\hspace{1.5mm}%
%\end{figure}  
%\begin{figure}[htbp]                  
%   \includegraphics[scale=0.18]{ALLFIGURE/2dR10/g62dR10-eps-converted-to.pdf}% 
%\centering 
%\hspace{1.5mm}              
% \includegraphics[scale=0.18]{ALLFIGURE/2dIPL/g62dIPL-eps-converted-to.pdf}
% \caption{Decay of the hexatic order correlation function $g_6(r)/g(r)$ for 2dKA, 2dR10 and 2dIPL systems. solid straight lines are obtained by fitting the peaks of the correlation function.}
%  \label{hexatic}
%\end{figure}

\section{Calculation of Dynamical Length scales} \label{xiDynamic}

\noindent{\bf Binder Cummulant: }
We calculate the dynamical correlation length $\xi_d$ in various ways 
to assess the robustness of our results. We extract the dynamical length scale 
$\xi_d$ \cite{SMANNUAL,SMPNAS} from finite size scaling of the peak height
of the four-point dynamic susceptibility $\chi_4(t)$. Four-point dynamic 
correlations (defined below) can be thought of as the correlation between two 
relaxation processes at two spatial points separated by some distance 
and the dynamical susceptibility $\chi_4(t)$ is the integrated effect of these 
correlations over the whole volume. The relaxation processes are 
characterized by the two point correlation function $Q(t)$, which gives 
the amount of overlap between two configurations which are separated 
by time t, 
\[ Q(t) = \sum_{i=1}^{N}w(\left| r_i(0) - r_i(t)\right|)\]
where $w(r)=1$ when $r\leq a$ and $0$ otherwise. The parameter $a$ is
chosen to eliminate from consideration the decorrelation that might arise due to the  
vibrational motion of particles inside the cages formed by other,
neighboring, particles. The particular choice of $a$ is not very
important, and in our studies we choose $a=0.3$ which corresponds to
the plateau value of the mean square displacement.
\begin{figure}[!h]
\includegraphics[scale=0.28]{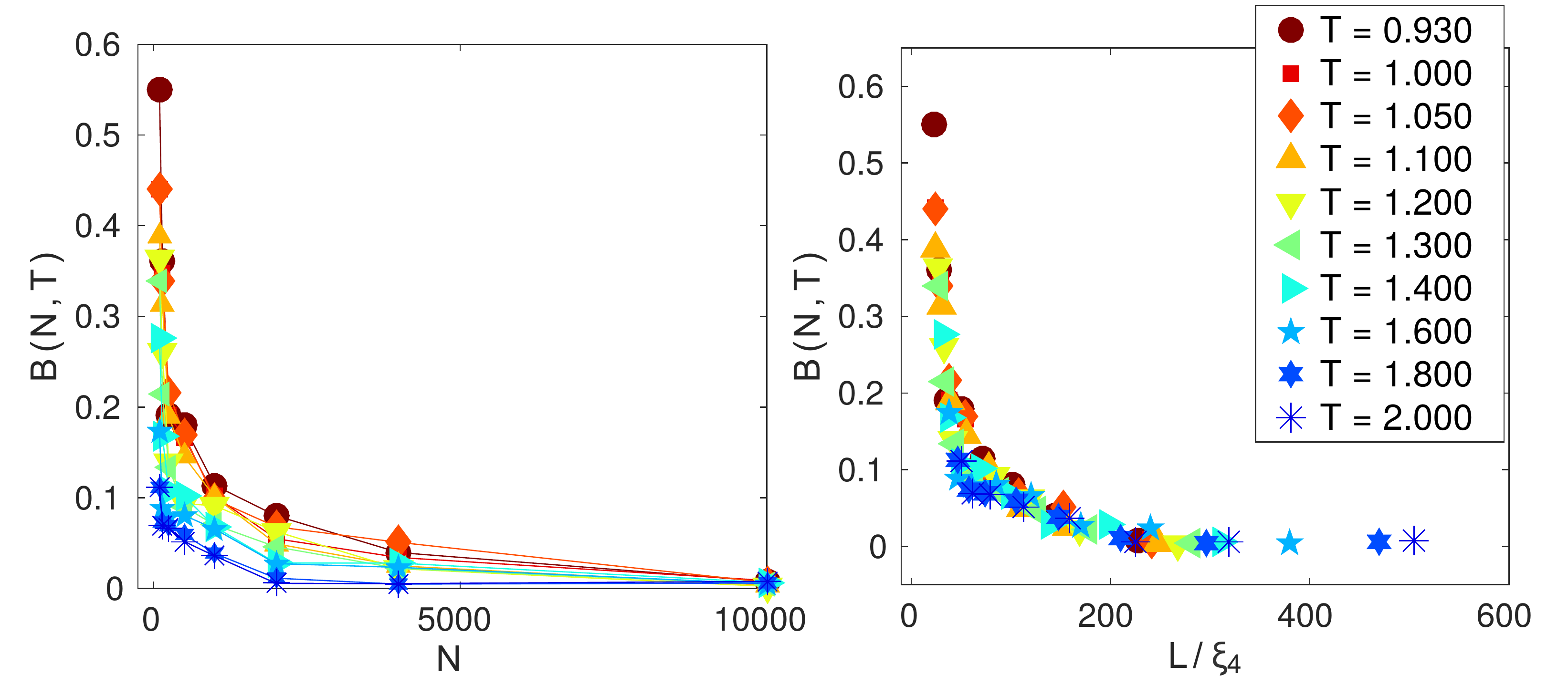}
\includegraphics[scale=0.29]{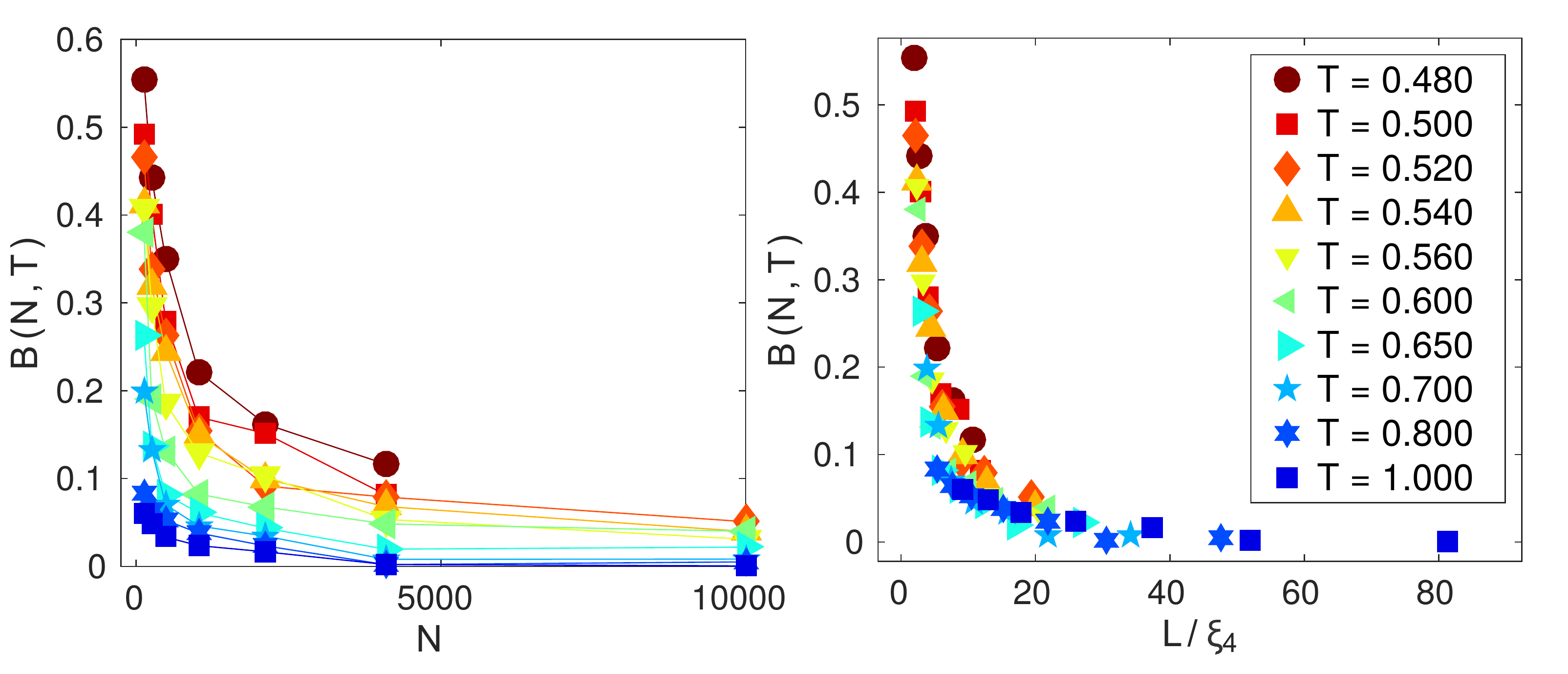}               
\includegraphics[scale=0.28]{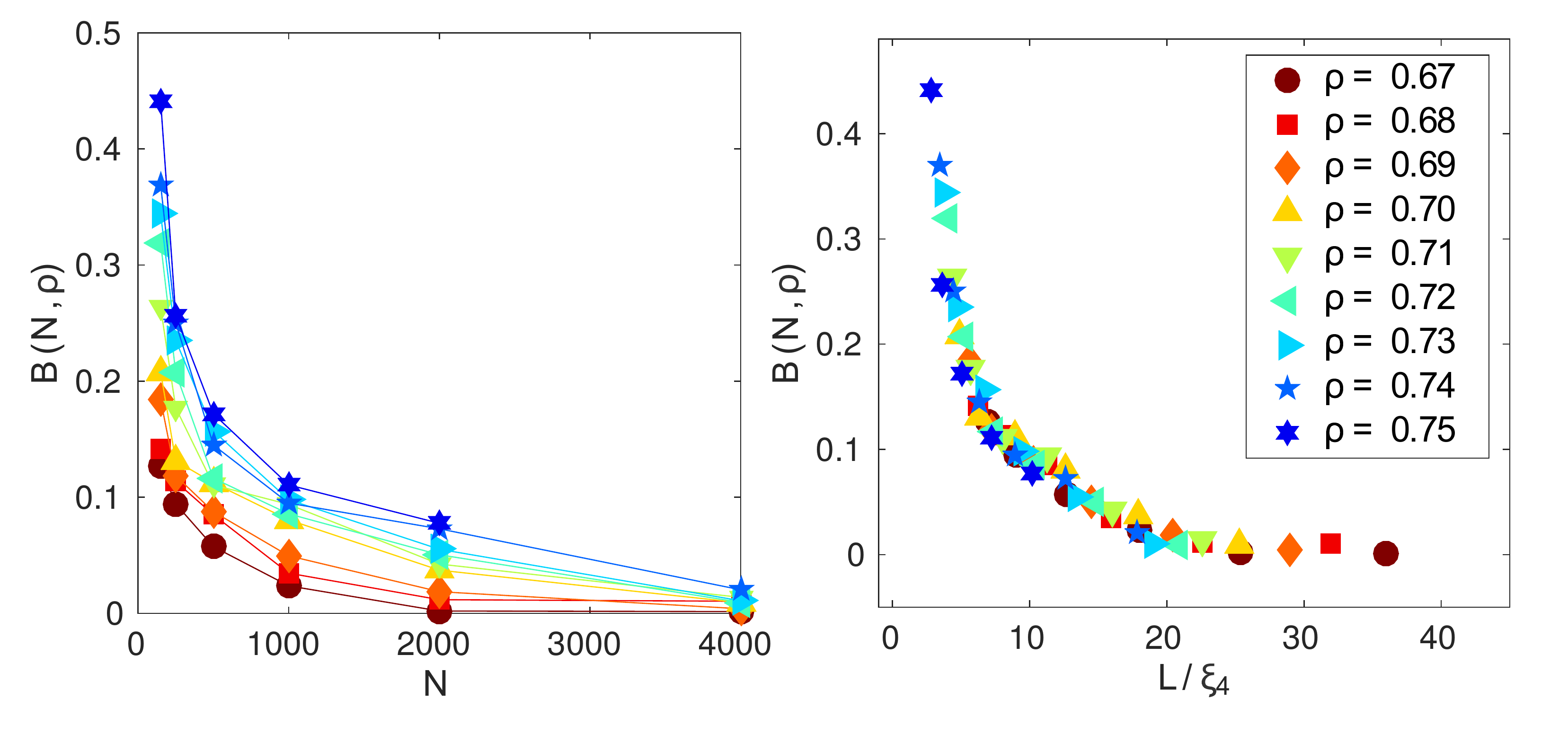}
\caption{Left panel shows the Binder cumulant B(N,T) and $B(N,\rho)$ as a function of system 
size for the three systems. The top panel is for 2dKA, the middle one is for 2dR10 and the bottom one 
is for 2dIPL model systems. The right panel shows the data collapse for B(N,T) and $B(N,\rho)$ 
plotted as a function of $L/\xi$. The correlation length $\xi$ is an unknown quantity 
which is determined by requiring data collapse for the values of the Binder cumulant at different temperature or density
and system size.}
\label{bc}
\end{figure}

The four point susceptibility, $\chi_4(t)$, \cite{chandan92} is 
defined as
\[ \chi_4(t) = \frac{1}{N}[\langle Q^2(t)\rangle] - \langle Q(t) \rangle^2] \]
%We consider $\chi_4^P(N,T)$, the peak value of $\chi_4 P(t,N,T)$.
Finite size scaling (FSS) is done for the system size 
dependence of the peak value of $\chi_4(t)$, $\chi_4^P$ for each 
temperature using the following scaling ansatz
\begin{equation}
\chi_4^P(N,T) = \chi_4^P(\infty, T)\mathcal{F}\left(\frac{N}{\xi_d^d}\right),
\end{equation}
where $\chi_4^P(\infty,T)$ is the asymptotic value of $\chi_4^P$
in the infinite system size limit. In this scaling analysis both
$\chi_4^P(\infty, T)$ and the length scale $\xi_d$ are treated 
as parameters to be determined by demanding data collapse. 
$\mathcal{F}(x)$ is the unknown scaling function.  

A better method to estimate this dynamical length scale $\xi_d$ is 
from the Binder cumulant \cite{BINDER}  which is defined in terms of the fourth and 
second moments of the distribution of $Q(\tau)$ 
where $\tau$ is the time associate with peak height of $\chi_4$. 
$\tau$ is proportional to the structural relaxation time 
$\tau_{\alpha}$ with a proportionality constant very close to $1$. 
The binder cumulant is defined as 
\[ B(N,T) = 1 - \frac{\langle [Q(\tau) - \langle Q(\tau) \rangle]^4\rangle}{3\langle[Q(\tau) - \langle  Q(\tau)\rangle]^2 \rangle ^2}. \]
By its definition, $B(N,T) = 0$ for a Gaussian distribution of $P(Q(\tau))$ 
and it is a scaling function of only $N/\xi_d^d$  without any 
pre-factor:
\begin{equation}
B(L,T) = \mathcal{G}\left( \frac{L}{\xi_d}\right),
\end{equation}
where $L = (N/\rho)^{1/d}$ is the linear size of the system, 
$\xi_d$ is the 
correlation length and $\mathcal{G}(x)$ is the scaling function. 
In Fig.~\ref{bc} (left panels) we show the system size 
dependence of the Binder Cumulant at different temperatures and 
densities for the 2dKA, 2dIPL and 2dR10 models. Right panels of the same
figures show the data collapse achieved by estimate values of  the dynamical length 
scales $\xi_d$. The observed good data collapse confirms the 
reliability of the extracted length scales. 

\vskip +0.3cm
\noindent{\bf Four-point Structure Factor: }
\begin{figure}[!h]
\includegraphics[scale=0.198]{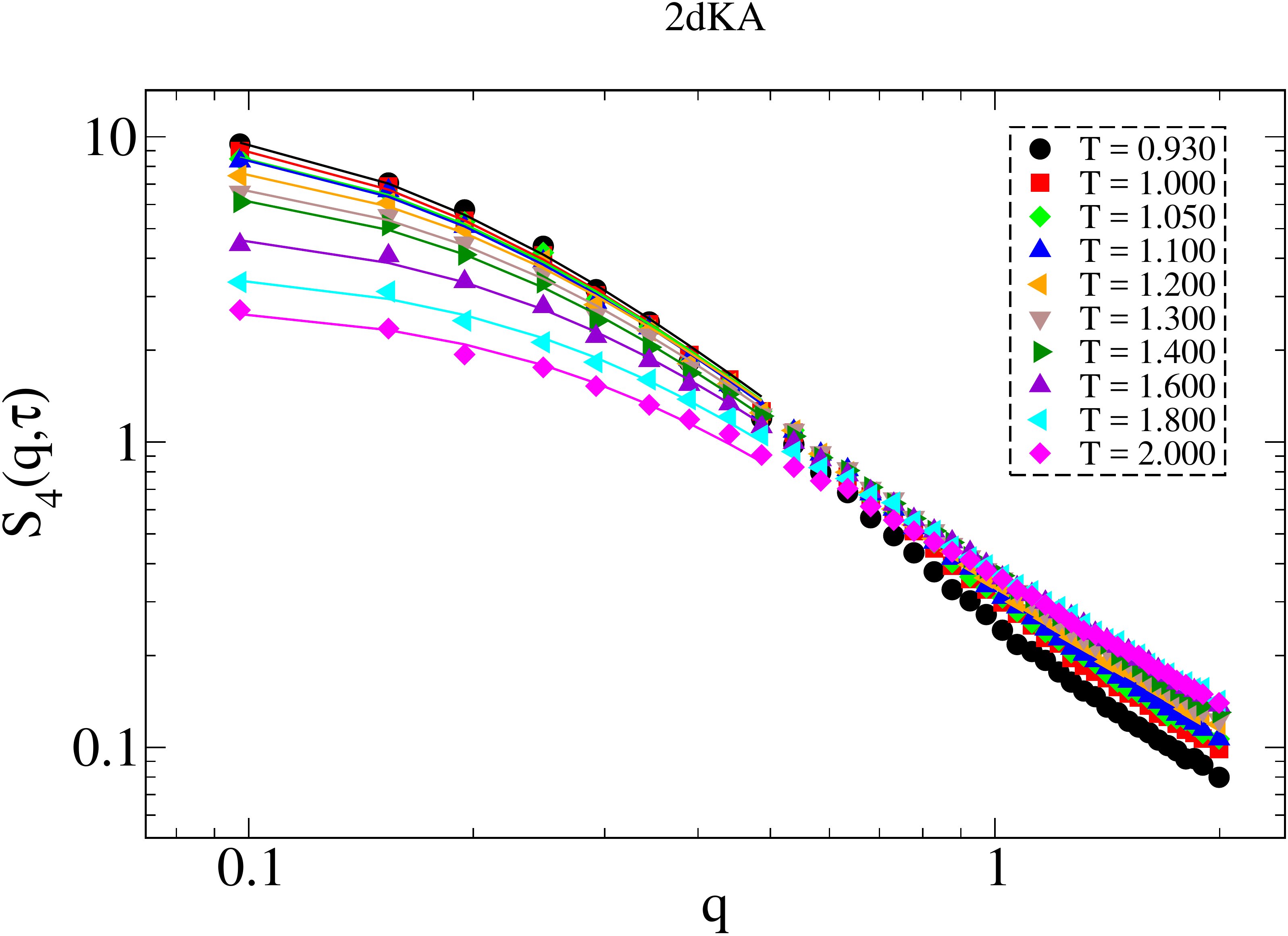}         
\includegraphics[scale=0.214]{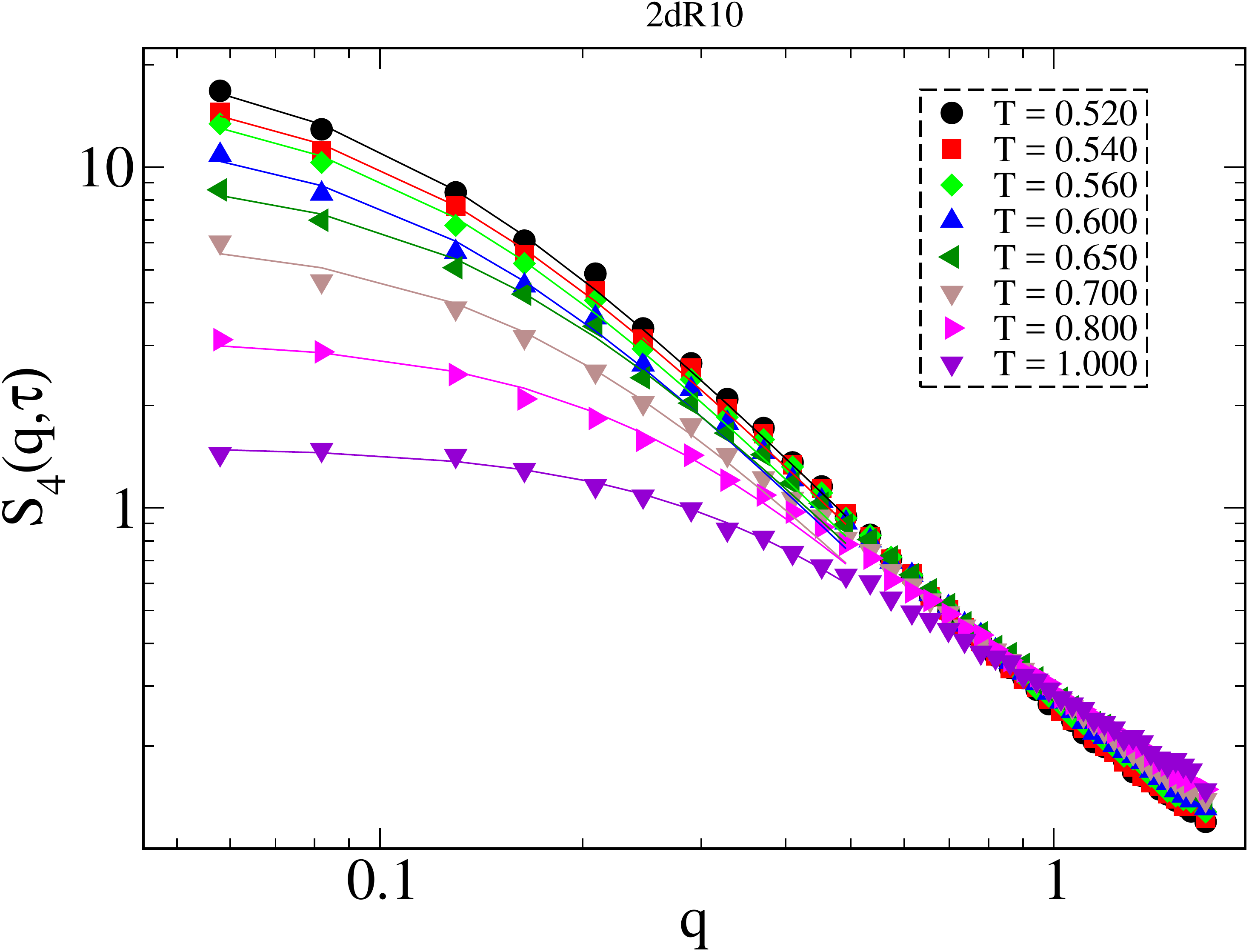}              
\includegraphics[scale=0.218]{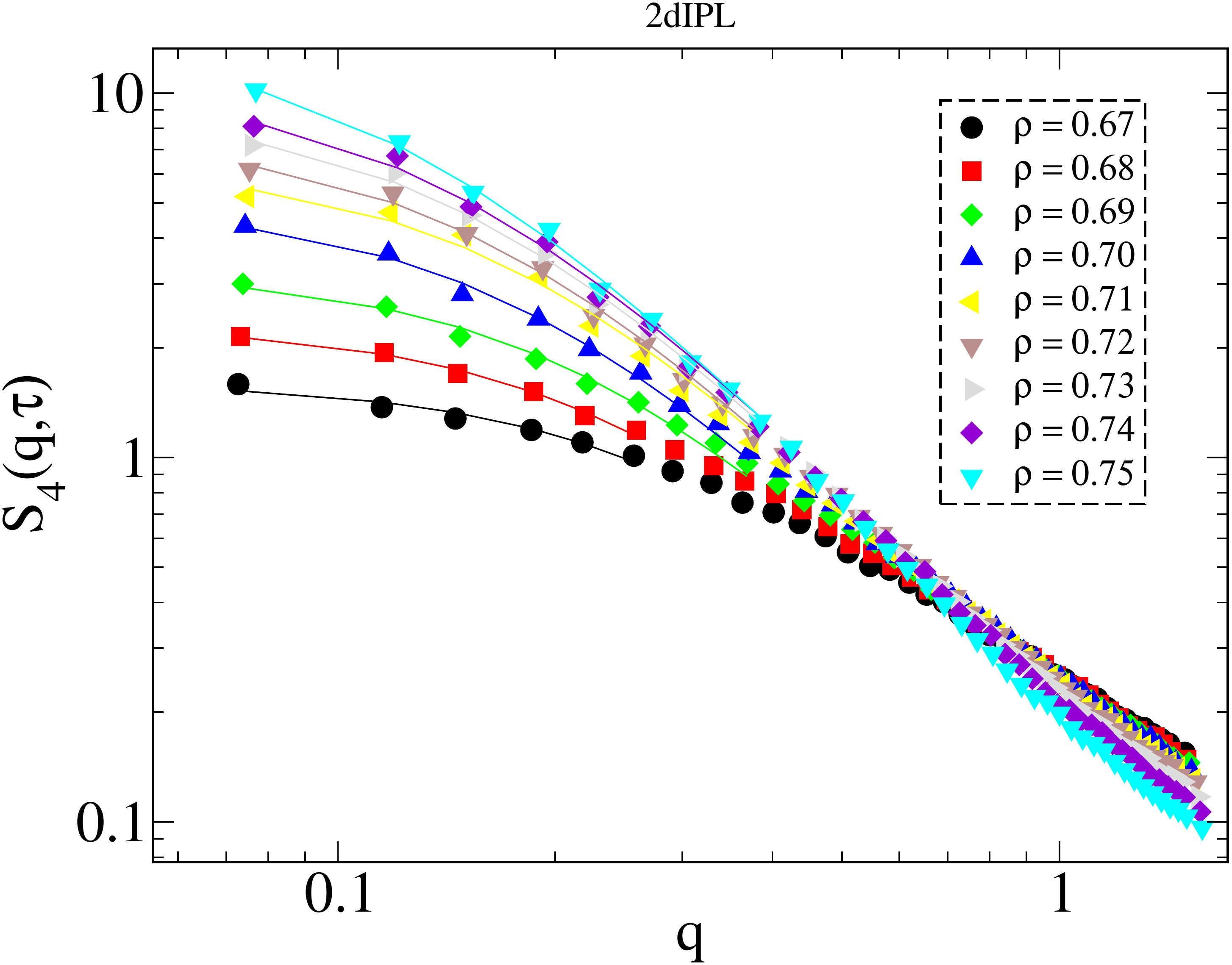}
\caption{$S_4(q,\tau)$ for 2dKA (for different temperature), 2dR10 (for different temperature) and 2dIPL (for different density). The solid lines are fits to the Ornstein-Zernike function.}
\label{S4qt}
\end{figure}
The four point structure factor \cite{S4Lacevic} is defined as
\[ S_4(q,\tau) = N\left[\langle \tilde{Q}(q,\tau) \tilde{Q}(-q,\tau)\rangle - |\langle \tilde{Q}(q,\tau)\rangle|^2\right]. \]
Here, the spatial Fourier decomposition of the local mobility is defined as 
\[\tilde{Q}(q, t) = \frac{1}{N}\sum_{i=1}^{N} \exp(i {\bf q}.{\bf r}_i(o)) w\left(|\vec{r}_i(t) - \vec{r}_i(0)|\right).\]

We plot the $S_4(q,\tau)$ as a function $q$ in Fig.
~\ref{S4qt}. We use the Ornstein-Zernike (OZ) form for the $q$
dependence of $S_4(q,\tau)$:
\begin{equation}
S_4(q,\tau) = \frac{S_4(q=0,\tau)}{1+[q\xi_4]^2}
\end{equation}
where $S_4(q=0,\tau)$ and $\xi_4$ are treated as fitting parameters. 
We find a good fit in the range $q \in [0.097, 0.486]$ for the 2dKA model system, 
$q \in[0.076, 0.384]$ for the 2dIPL system and $q \in[0.058, 0.492]$ for the  
2dR10 system. The system size used for this analysis is $N = 10000$.
One finds the fits to the OZ equation to be very good and the obtained
length-scales are in good agreement with the ones obtained using FSS of
$\chi_4^P$ and the Binder cummulant.

\noindent{\bf Four-point Structure Factor from Bond-breakage Correlation:}
The dynamical length scale $\xi_d$ is also calculated (for the 2dKA model)
from the susceptibility of a bond breakage correlation function following 
the procedure of Ref.\cite{YO,shiba}. It was pointed out in \cite{shiba}
that in two dimensions, due to long wave length density fluctuations 
(consequence of Mermin-Wagner Theorem), $\chi_4$ and the corresponding structure
factor $S_4(q,\tau)$ show a strong contribution from phonon like modes. This
masks the contribution of the growth of the dynamic length scale and extracting 
the dynamic length scale from $S_4(q,\tau_4)$ becomes erroneous. So to check
the reliability of the extracted dynamic heterogeneity length scale from
different methods, we have also calculated the same using the four-point
structure factor from the bond-breakage correlation function.

The bond-breakage correlation function is defined as follows.
Initially we consider a pair of atom $i$ and $j$ to be bonded if 
\begin{equation}
r_{ij}(t_0) = |{\bf r}_i(t_0) - {\bf r}_j(t_0)| \leq1.0 \sigma_{\alpha\beta}.
\end{equation} 
where $i$ and $j$ belong to the species $\alpha$ and $\beta$ with 
$\sigma_{\alpha\beta}=\frac{1}{2}(\sigma_{\alpha}+\sigma_{\beta})$. 
After some elapsed time we calculate the surviving bonds which 
satisfy the criterion 
$r_{ij}(t_0+t)\leq 1.35\sigma_{\alpha\beta}$. 
The bond breakage correlation is defined as the number of surviving bonds 
normalized by the total number of bond at the initial time. From this correlation function, 
we extract the bond breakage time, $t_B$, as the time at which the 
correlation function becomes $1/e$ of its initial value.

\begin{figure}[!h]
\includegraphics[scale=0.256]{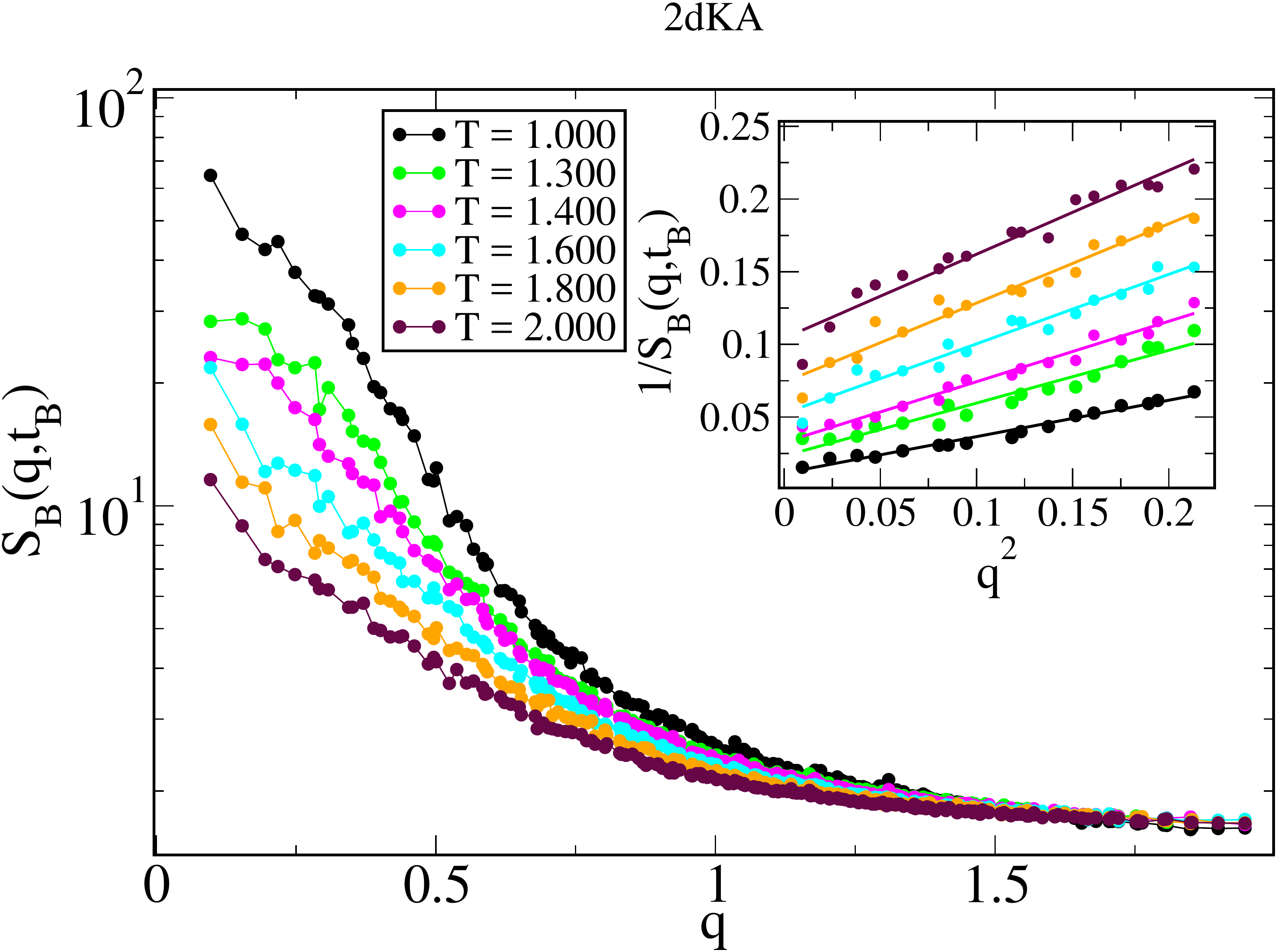}               
\caption{$S_B(q)$, the bond-beakage structure factor, for the 2dKA model. The inset shows $1/S_B(q,t_B)$ vs $q^2$  from which 
$\xi_B$ is determined.}
\label{S4qtbb}
\end{figure}

We next calculate the structure factor $S_B(q,t_B)$ which is defined as 
\begin{equation}
S_B(q,t_B) = \frac{1}{N_b}\langle|\sum \exp(i{\bf q}.{\bf R}_{ij})|^2\rangle_a
\end{equation}
where the summation runs over broken pairs, $N_b$ is the number of broken bonds and 
${\bf R}_{ij}=\frac{1}{2}({\bf r}_i(t_0)+{\bf r}_j(t_0))$ is the center position of the broken 
pair at the initial time $t_0$. $\langle..
\rangle_a$ is the angular average over the direction of the wave vector ${\bf q}=\frac{2\pi}
{L}(n_x,n_y)$ where $n_x,n_y = \pm1,\pm2,\pm3,..$. Then we plot the $S_B(q,t_B)$ as 
a function q in Fig.~\ref{S4qtbb}. We use the same  Ornstein-Zernike (OZ) form to 
fit the function.
\begin{equation}
S_B(q,t_B) = \frac{S_B(q=0,t_B)}{1+[q\xi_B]^2}
\end{equation}
$S_B(q,t_B)$ and the $\xi_B$ 
obtained from the fits exhibit no significant size dependence.
The obtained length scale $\xi_B$ is also in good agreement with the dynamical 
length scales obtained from the Binder cumulant. 
\section{Static Length Scale from Finite Size Scaling of 
$\tau_{\alpha}$ }
\label{tauAlphaFSS}

\begin{figure}[!h]
\includegraphics[scale=0.215]{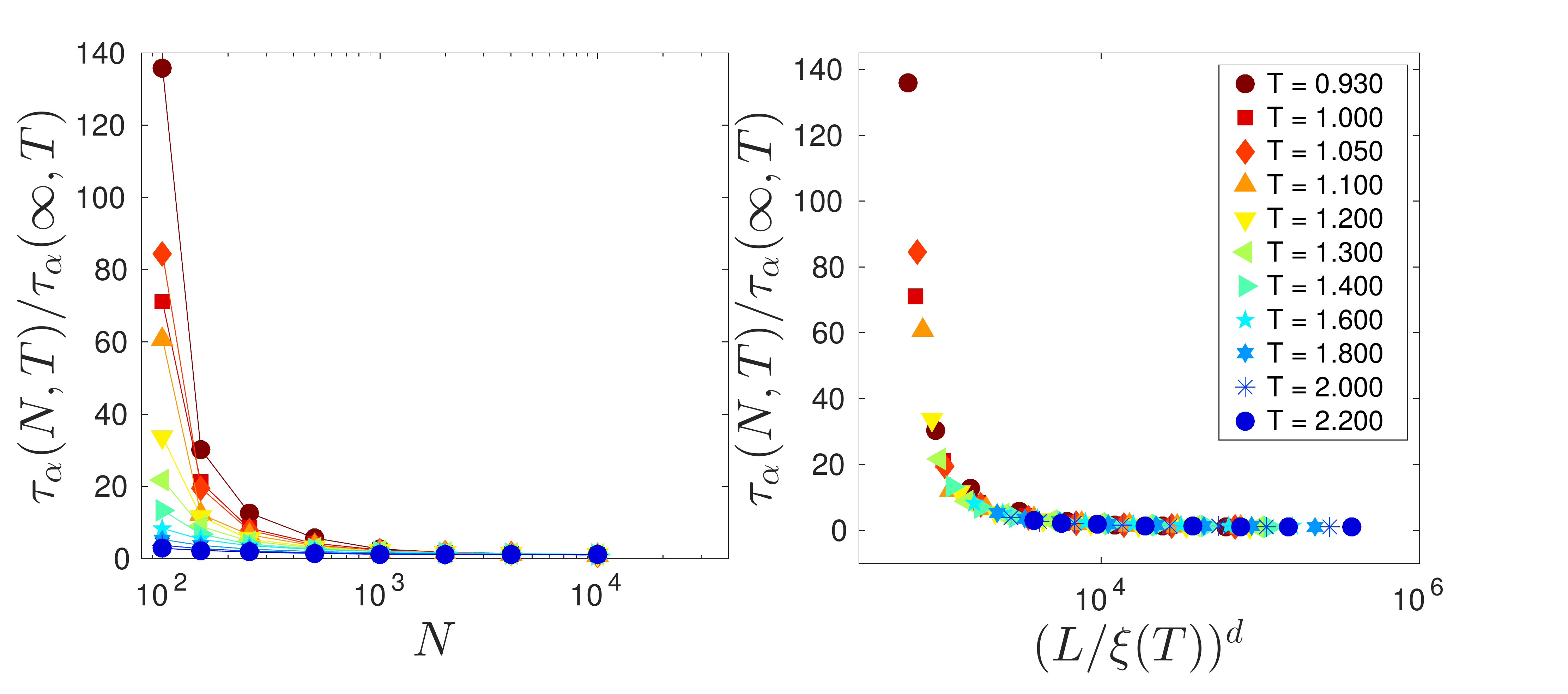}
\vspace{-0.3cm}                
\includegraphics[scale=0.21]{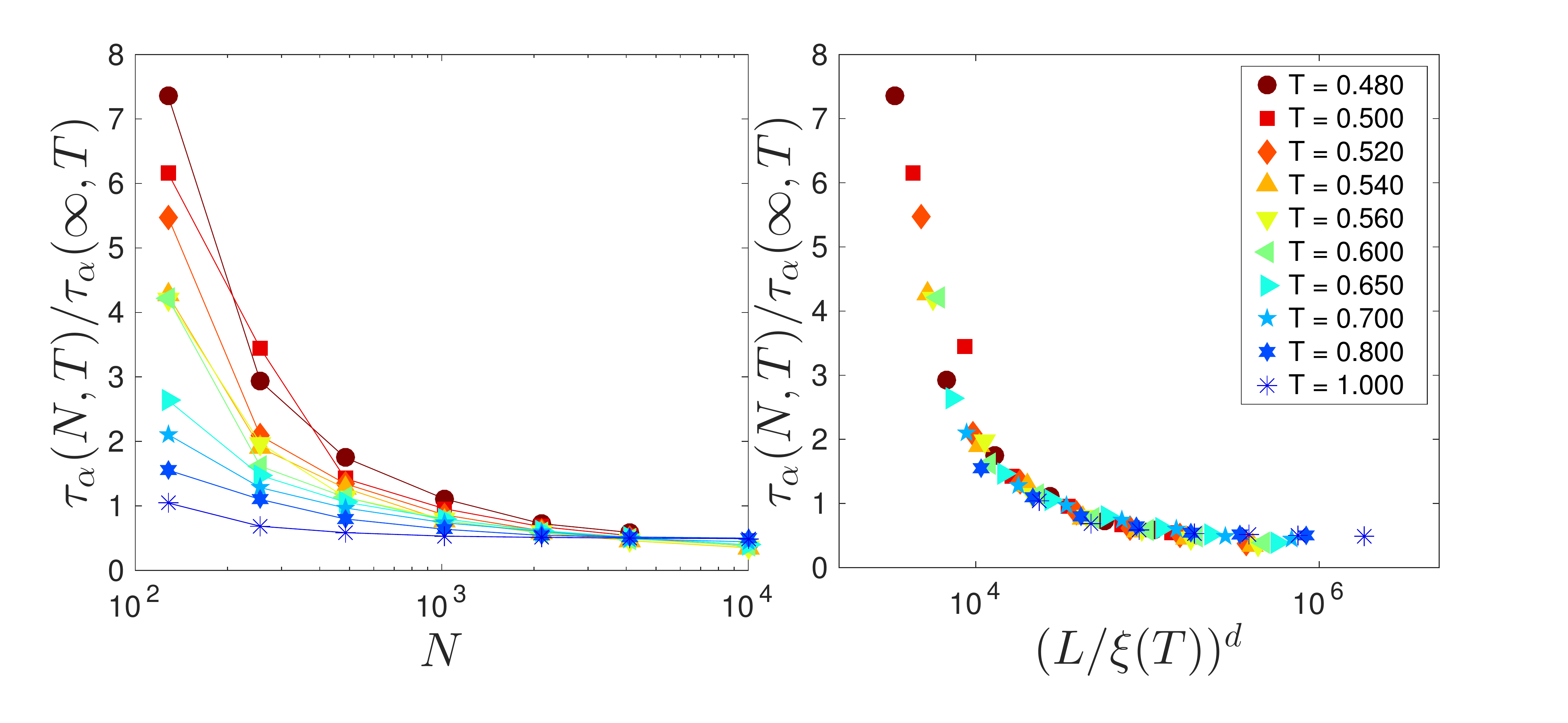}
\vspace{-0.3cm}                 
\includegraphics[scale=0.215]{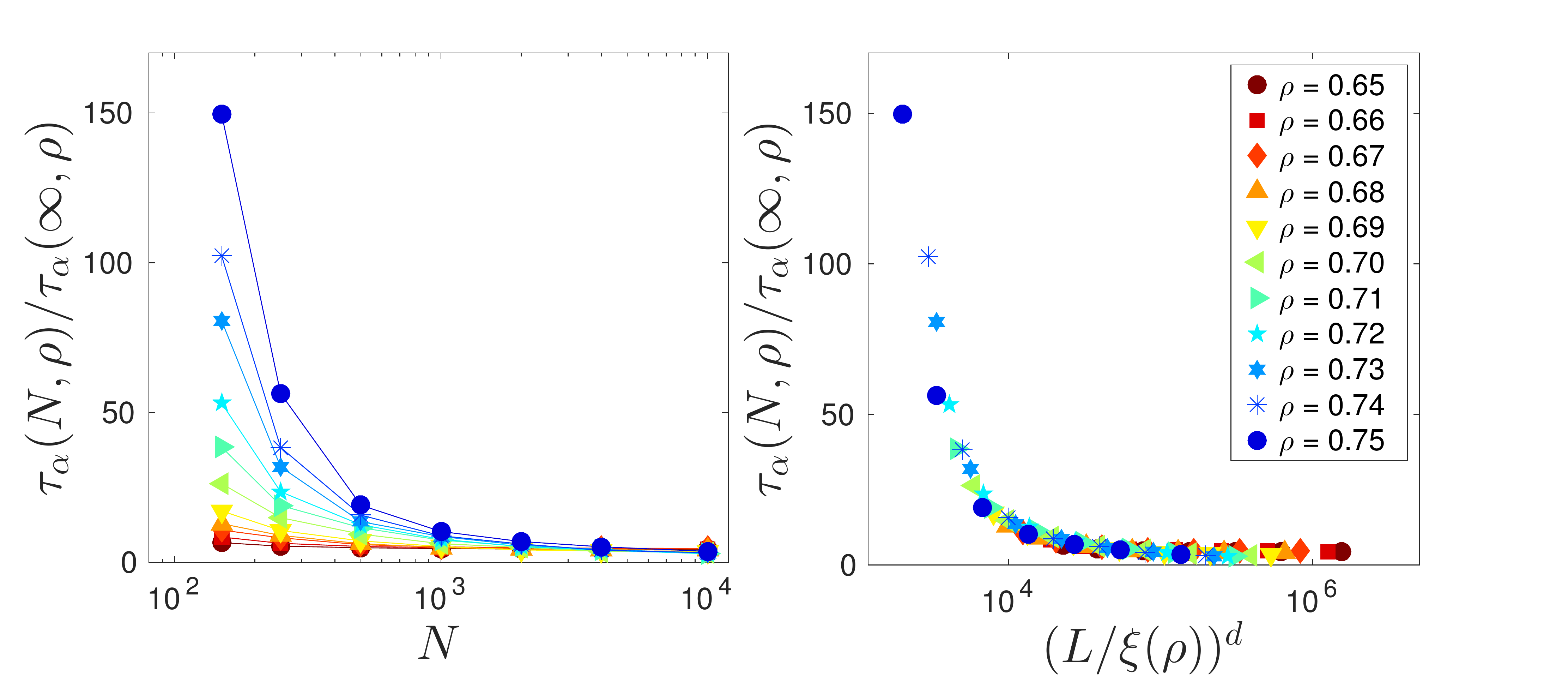}
\caption{Top panel shows the finite system size dependence of the $\alpha$ 
relaxation time for the 2dKA model and data collapse to obtain 
the static length scale. The middle one shows the same 
analysis for the 2dR10 model system and the bottom one for the 2dIPL 
system.}
 \label{tauscale}
\end{figure}

We calculate the static length scale from finite size scaling of 
the $\alpha$ relaxation time, $\tau_{\alpha}$ \cite{SMFSS}. The relaxation 
time $\tau_{\alpha}$ is defined from $Q(\tau_{\alpha}) = 1/e$. 
The $\alpha$ relaxation time shows strong system size dependence, 
especially at low temperatures. It decreases monotonically with 
increasing system size and reaches an asymptotic value at large
system sizes. The system sizes dependence of the $\alpha$ relaxation time 
becomes more prominent at low temperatures or at high densities. We use the 
following scaling form to obtain the static length scale:
\begin{equation}
\tau_\alpha(N,T) = \tau_\alpha(\infty,T)
\mathcal{G} \left(\frac{N}{\xi_s^d(T)}\right),
\end{equation} 
where the form of $\mathcal{G}(x)$ is unknown and one does not need 
the details of this scaling function to obtain the static length scale
$\xi_s$. In the left panels of Fig.~\ref{tauscale}, we plot
$\tau_{\alpha}(N,T)$ scaled by $\tau_{\alpha}(\infty,T)$ for the 2dKA 
and 2dR10 models and $\tau_{\alpha}(N,\rho)$ scaled by $\tau(\infty,
\rho)$ for the 2dIPL model as a function of system size, $N$. To estimate 
the value of $\tau_{\alpha}(\infty,T)$ or $\tau_{\alpha}(\infty,\rho)$ 
we have used the following functional form $f(x) = a+ b/x$. 
The right panels show the data collapse for the three model systems, 
which is observed to be good. Reasonably good collapse of the 
data gives confidence in the reliability of the extracted static length scale 
$\xi_s$.

\section{Static Length Scale from Finite Size Scaling of the Minimum 
Eigenvalue of the Hessian Matrix }
\label{minEigenvalue}
\begin{figure}[!h]                                                           
\includegraphics[scale=0.36]{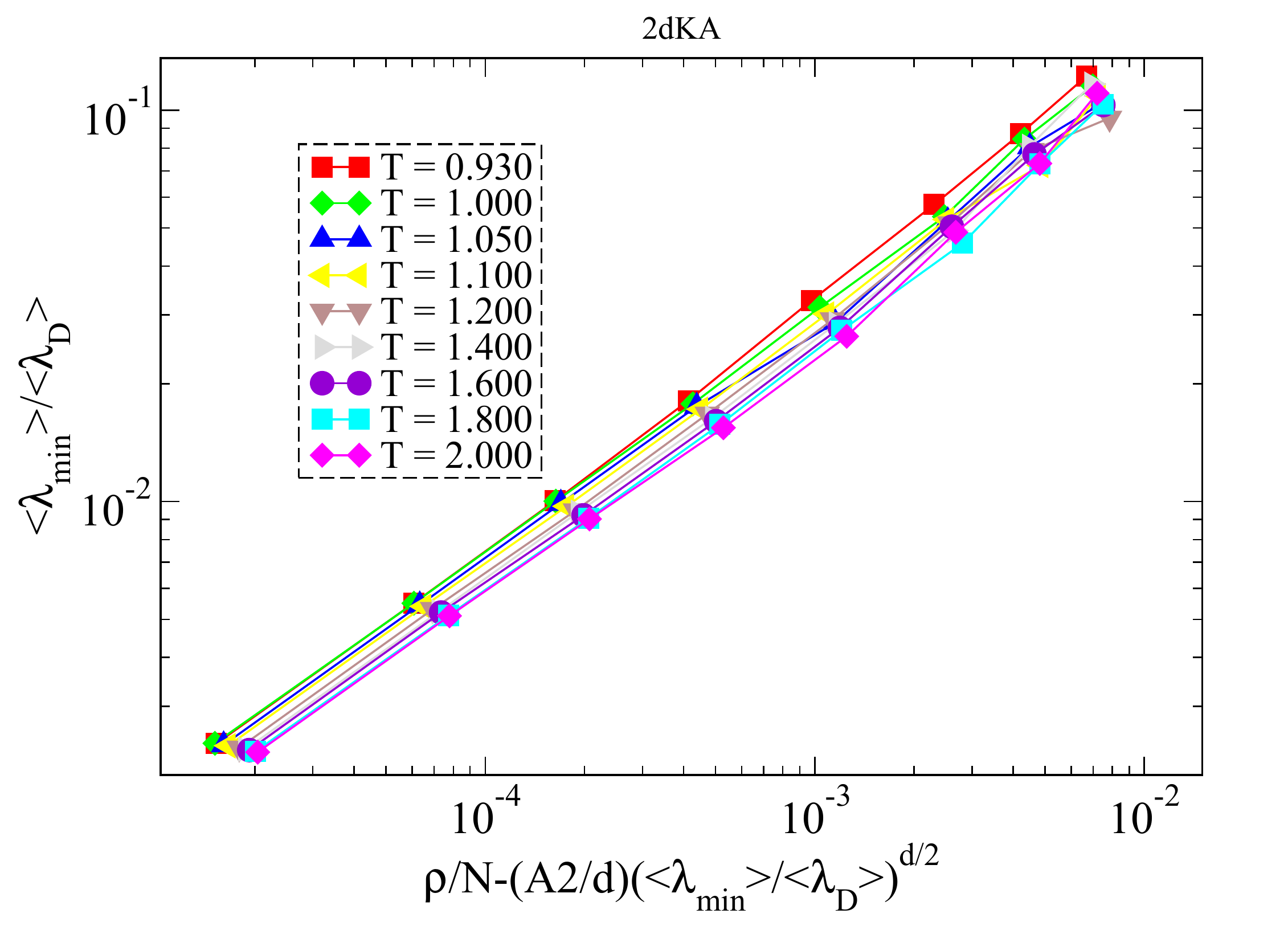}                
\includegraphics[scale=0.41]{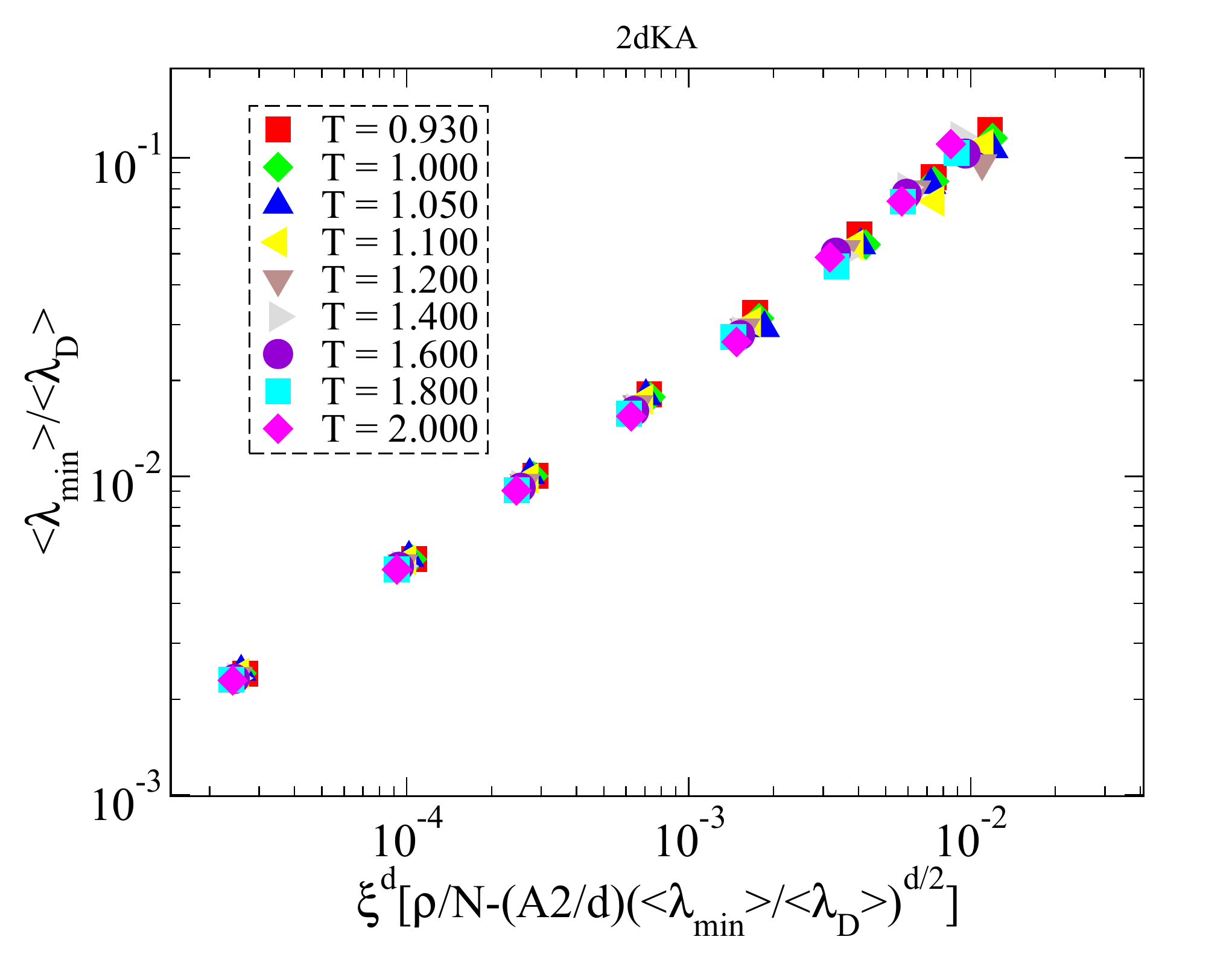}% 
\caption{ Scaling of the minimal eigenvalue calculated for the 2dKA model.}
\label{2dKAeigen}
\end{figure}
Yet another method used to obtain the static length scale is the 
finite size scaling of the minimum eigenvalue, $\lambda_{min}$ of 
the Hessian matrix \cite{2dR10} evaluated at the minima of the
potential energy (Inherent Structure or IS) explored by
the system at a given temperature or density. It is argued in 
\cite{2dR10}, that $\lambda_{min}$ will have a cross-over in the
system size dependence with increasing system sizes and the cross
over length scale is the static length scale. This method is 
useful for low temperature or at high density.  At high 
temperature or low density, due to an-harmonic effects, the scaling
arguments do not hold and the obtained length scale becomes increasingly
less reliable. 
To obtain the static length scale we have used the following scaling 
form as suggested in \cite{2dR10}: 
\begin{eqnarray*}
\frac{\langle\lambda_{min}(T)\rangle}{\langle\lambda_D(T)\rangle}
=\mathcal{F}\left[
\xi_s^d(T)\left(
\frac{1}{V}-\frac{A2}{d}\left(
\frac{\langle\lambda_{min}(T)\rangle}{\langle\lambda_D(T)\rangle}
\right)^{d/2}
\right)
\right],
\end{eqnarray*}
%CDQ: The x-label of figure 8 has \rho/N instead of 1/N
where $\mathcal{F}$ is an unknown scaling function and $\lambda_D \approx 
\mu\rho^{(2/(d-1))}$ where $\rho$ is the Debye frequency and $\mu$ is the 
shear modulus. We obtain the length scale $\xi_s$ by demanding that 
all data should collapse onto a master curve for different system 
sizes and temperature (or density) by correctly choosing $\xi_s(T)$.
In the top panel of Fig.~\ref{2dKAeigen}, we plot the data 
according to the above scaling ansatz. In the bottom panel, we 
show the collapsed data in a master curve to extract the  
correlation length scale. Similar data for the 2dR10 model are taken from
Ref.~\cite{2dR10}.

\section{Point to set length scale: }
To measure the length scale $\xi_{pts}$, we follow the methods 
elaborated in Refs.\cite{PTS1,PTS2}. 
\label{ptsMethod}
\begin{figure}[!h]
\includegraphics[scale=0.29]{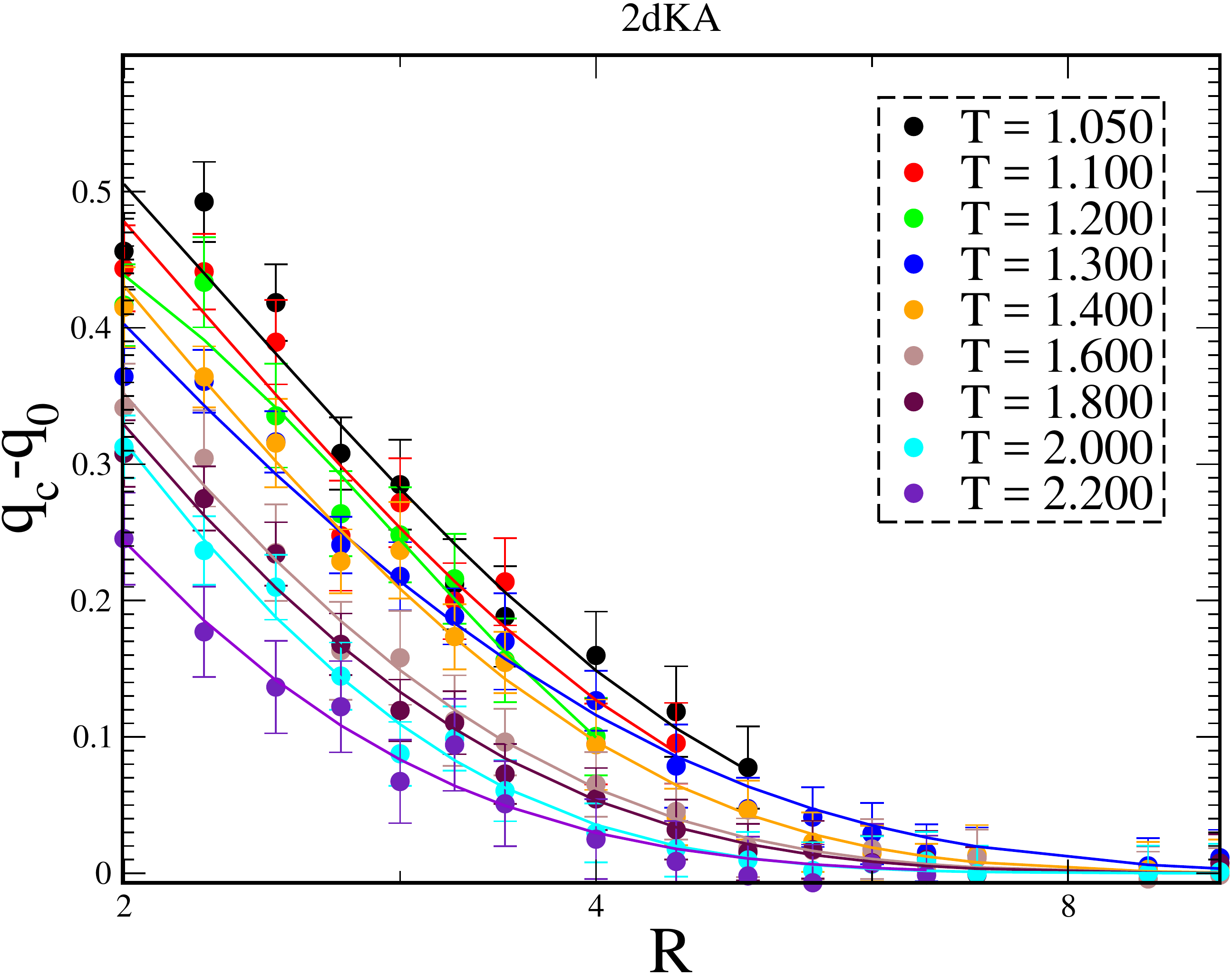}\\                
\includegraphics[scale=0.27]{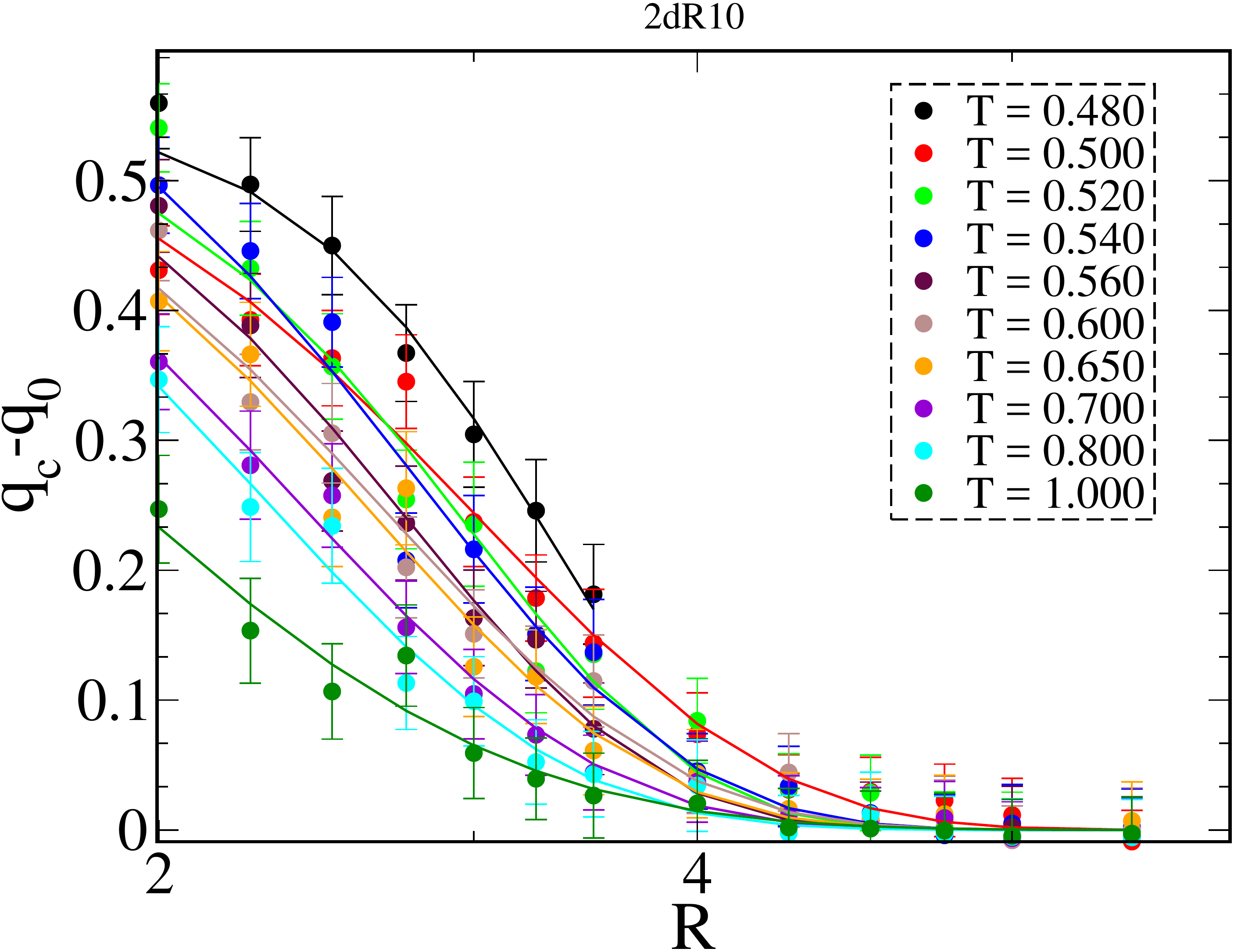}\\                
\includegraphics[scale=0.255]{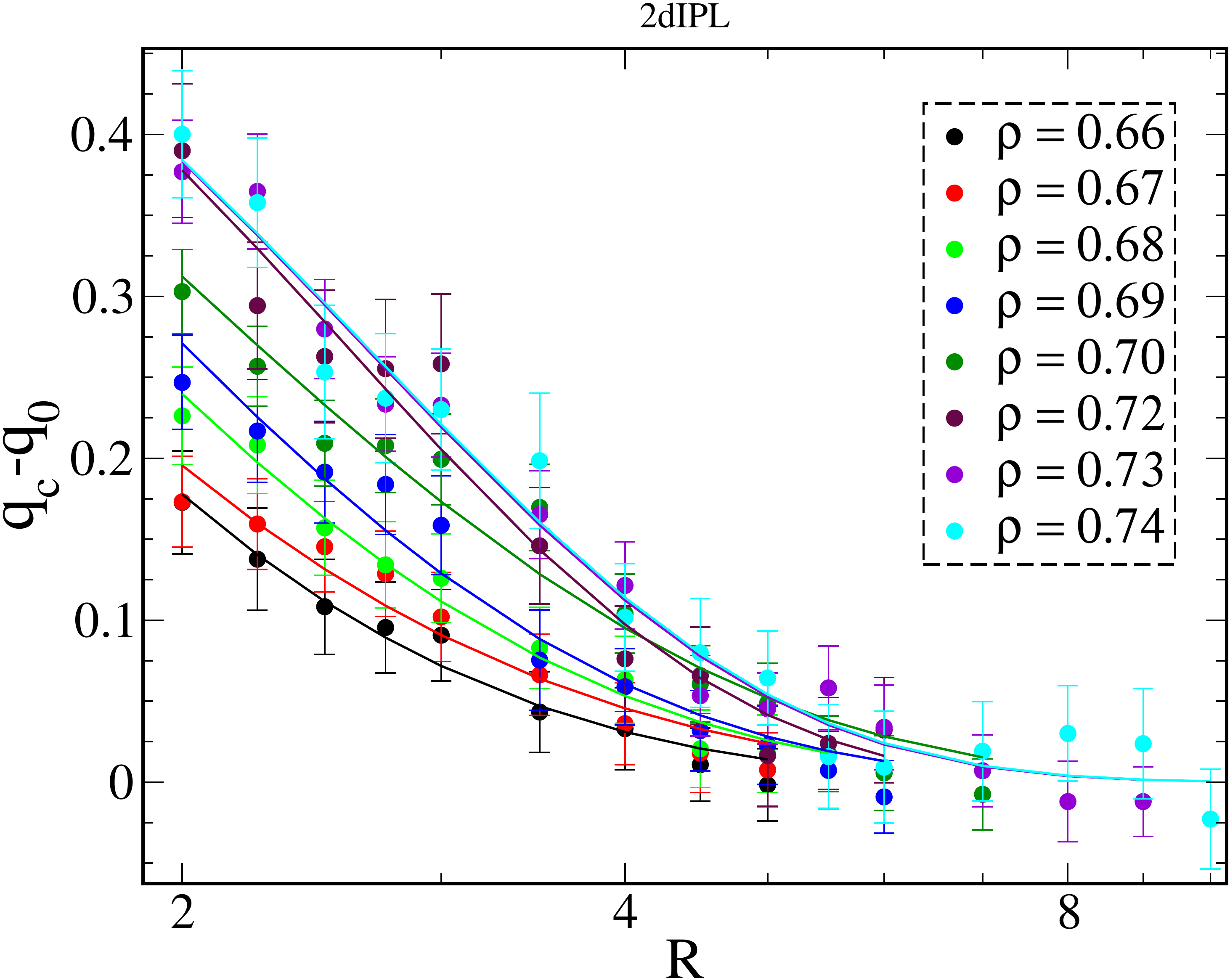}   
\caption{Overlap as a function of cavity size R for different 
temperature and density for the three model systems. Lines are 
compressed exponential fits of the data with the form discussed 
in text.}
\label{ptsfigure}
\end{figure} 
First, we equilibrate a system of $N$ particles and then cavities are generated by freezing the 
particles outside a sphere of radius $R$. We study how the 
thermodynamics of the particles inside the sphere of radius $R$ is affected by 
the amorphous boundary condition generated by the pinned particles. 
While choosing the cavities we demand that the density or packing 
fraction of the set of particles inside the cavity should be equal 
to the bulk density or packing fraction within a tolerance of $2 \%$. 

%SS: I don't understand what this means. 

%For  binary mixtures 
%the density of the particles inside the cavities are close to the 
%bulk value due to large number of ensemble averaging.

We then partition the central region of the cavity into $M$ cubic 
boxes of side $l$ to calculate the static overlap defined below. 
The size of the boxes is such that probability
of finding more than one particle in a single box is negligibly small.
The static overlap is defined as 
\begin{eqnarray*}
q_c(R) = \lim_{t \to \infty} \frac{1}{Ml^2\rho} \sum\limits_{i=1}^{M} \langle n_i(0) n_i(t)\rangle.
\end{eqnarray*}
Here $\langle .. \rangle$ implies both thermal average and ensemble 
average. The maximum overlap so defined of two configurations  is $1$ and that of two 
uncorrelated configurations  is $q_0 = \rho l^2$. To extract the 
PTS length scale we choose $M = 6 \times 6$ boxes of side $l = 0.36$.
We use the particle swap annealing (PSA) method \cite{PSA,PTS2} 
with molecular dynamics simulation to extract the overlap correlation 
function for different cavity sizes. The length scale $\xi_{pts}$ is 
obtained by fitting the overlap function with a compressed exponential 
form as given below
\begin{eqnarray*}
 \tilde{q}(R) = q_c(R) - q_0 = A\exp\left[-\left(\frac{R-a}{\xi_{pts}}\right)^\eta\right],
\end{eqnarray*}
where $A$ is a fixed number for all temperatures following \cite{PTS2}.
In our case, we find that $A=0.55$ fits the data  well for the 2dKA, 
2dR10 and 2dIPL model systems.

\section{Polydisperse system: 2dPoly }
In this section some of the details of our results for the polydisperse system (2dPoly)
are given.
\label{pol2d}  
\subsection{Hexatic and Dynamic correlation length scale:}
\begin{figure}[!h]
\includegraphics[scale=0.33]{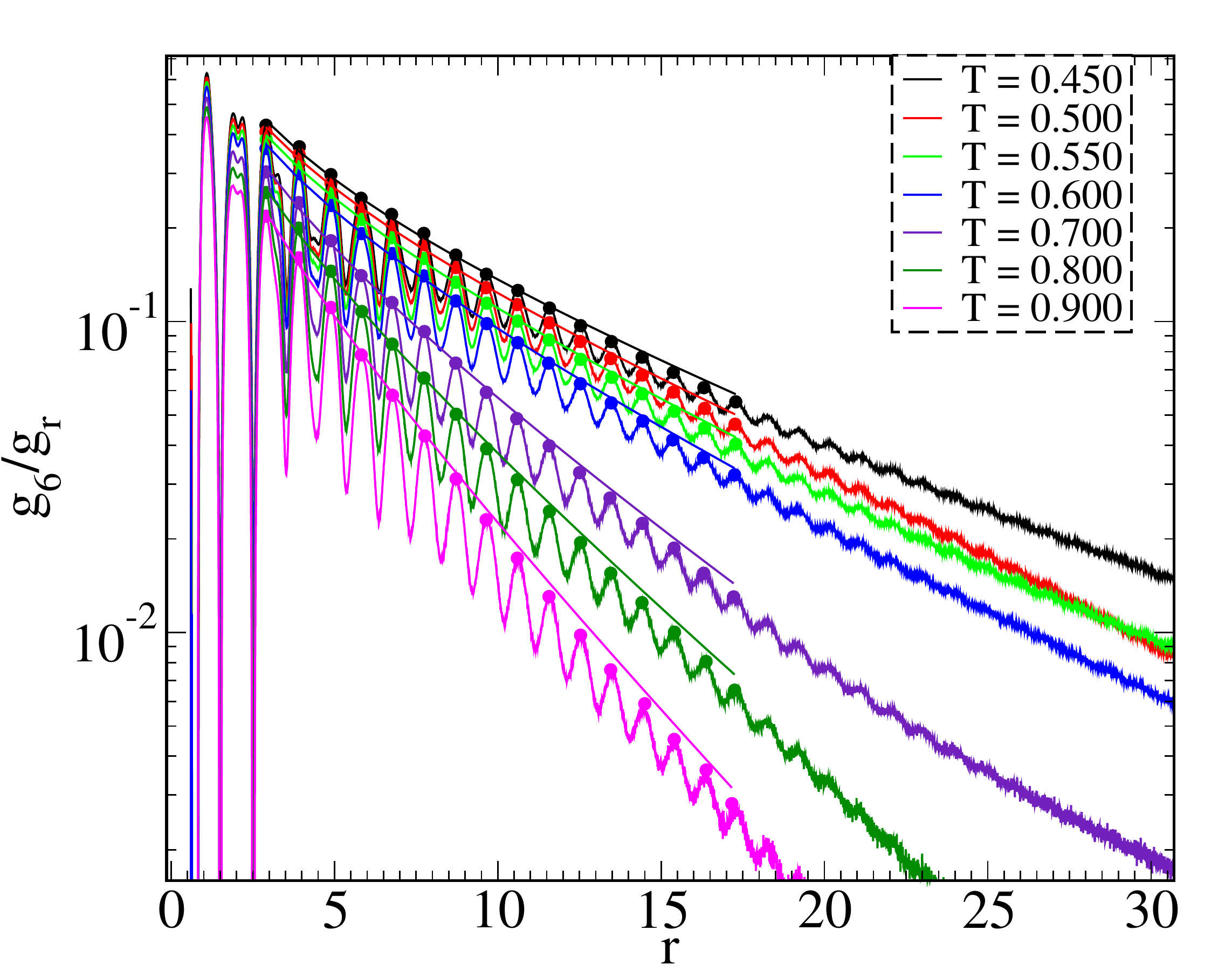}
\includegraphics[scale=0.34]{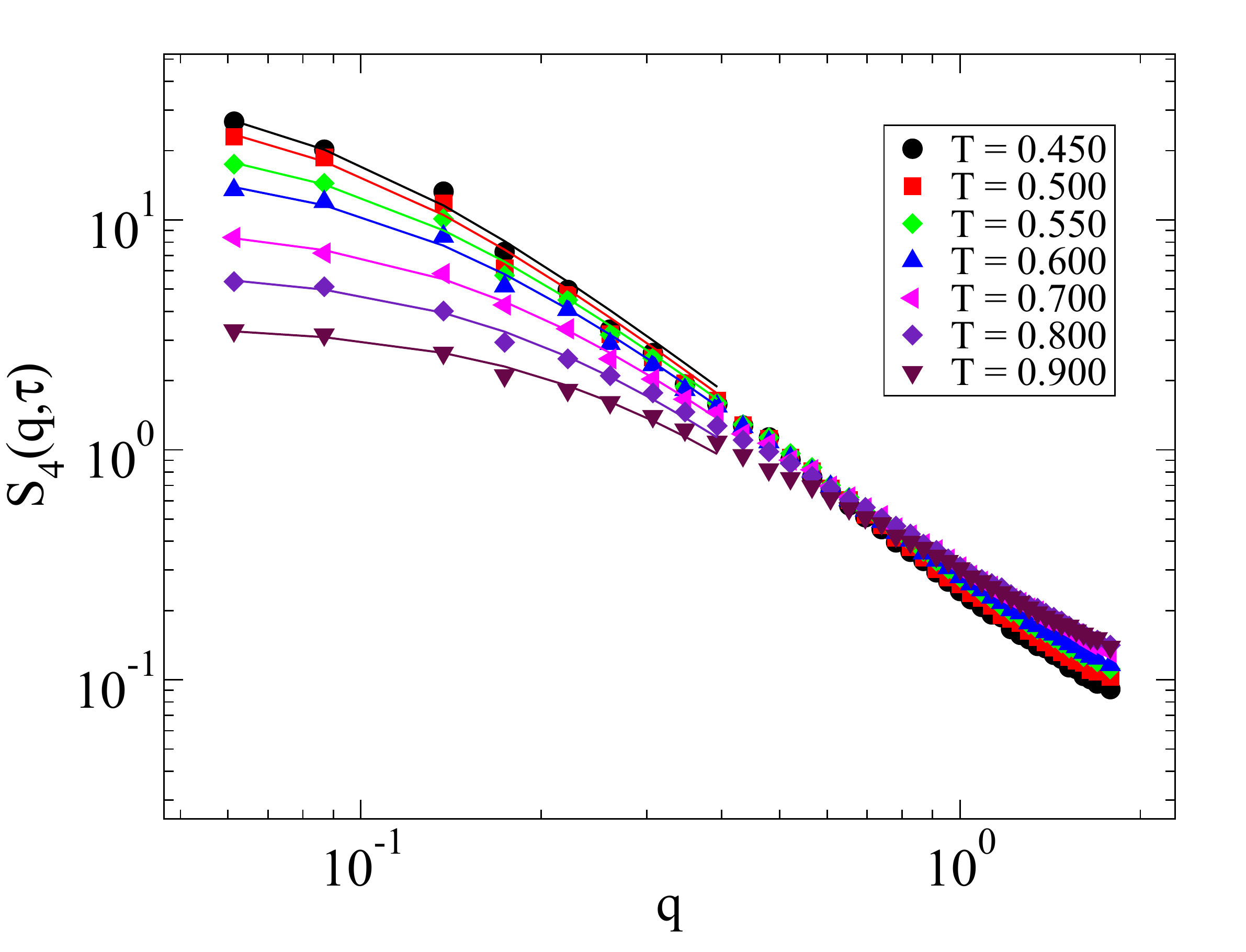}
 \caption{ Top panel: Decay of the hexatic order correlation function $g_6(r)/g(r)$ for the 2d polydisperse system. Bottom 
 panel: $S_4(q,\tau)$ for the 2d polydisperse  system for different temperatures. The solid lines are fits with the Ornstein-
 Zernike function.}
  \label{hoppoly}
\end{figure}

For the calculation of the hexatic correlation length $\xi_6$ we follow the 
same procedure as described in Sec.\ref{psi6Calc}. In the top panel of
Fig.~\ref{hoppoly} we show the hexatic correlation function
normalized by $g(r)$ for the 2dPoly model. The lines are  fits
of the peak values of $g_6(r)/g(r)$ to the OZ form, which we employ to extract the 
hexatic correlation length, $\xi_6$. 
In the bottom panel of Fig.\ref{hoppoly} we  plot
$S_4(q,\tau)$ as a function of $q$. We find a good fit to the OZ form in the 
region of $q \in [0.0614, 0.3935]$ for this model. From this we 
obtain the dynamical length scale $\xi_{4}$.
\subsection{Static Length-scales}
\begin{figure}[!h]
\begin{center}
\hskip -0.70cm
\includegraphics[scale=0.28]{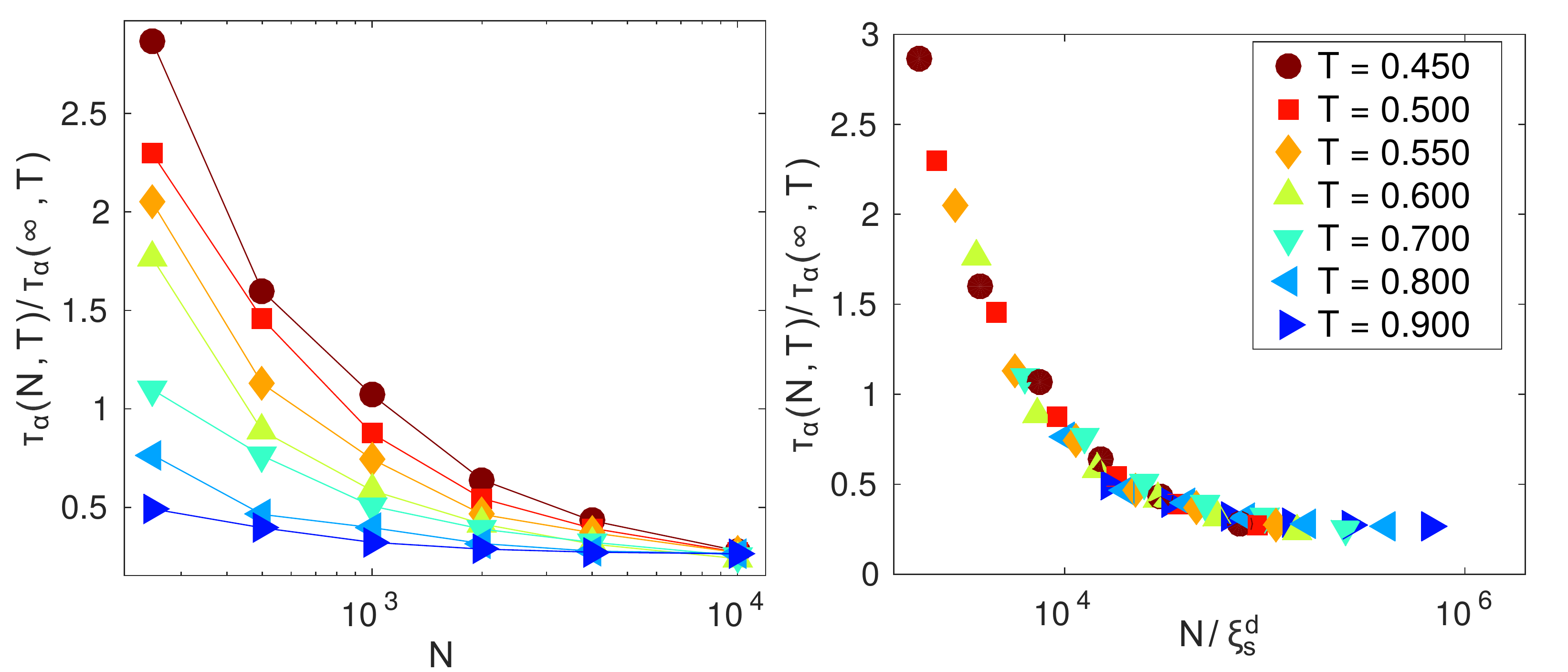}%
 \caption{System size dependence of the $\alpha$ relaxation time for the 2d polydisperse model system (left panel)
 and data collapse to obtain 
 the static length scale (right panel).}
  \label{taupoly}
\end{center}
%\begin{figure}[!h] 
%\vskip -0.3cm                                                 
\includegraphics[scale=0.22]{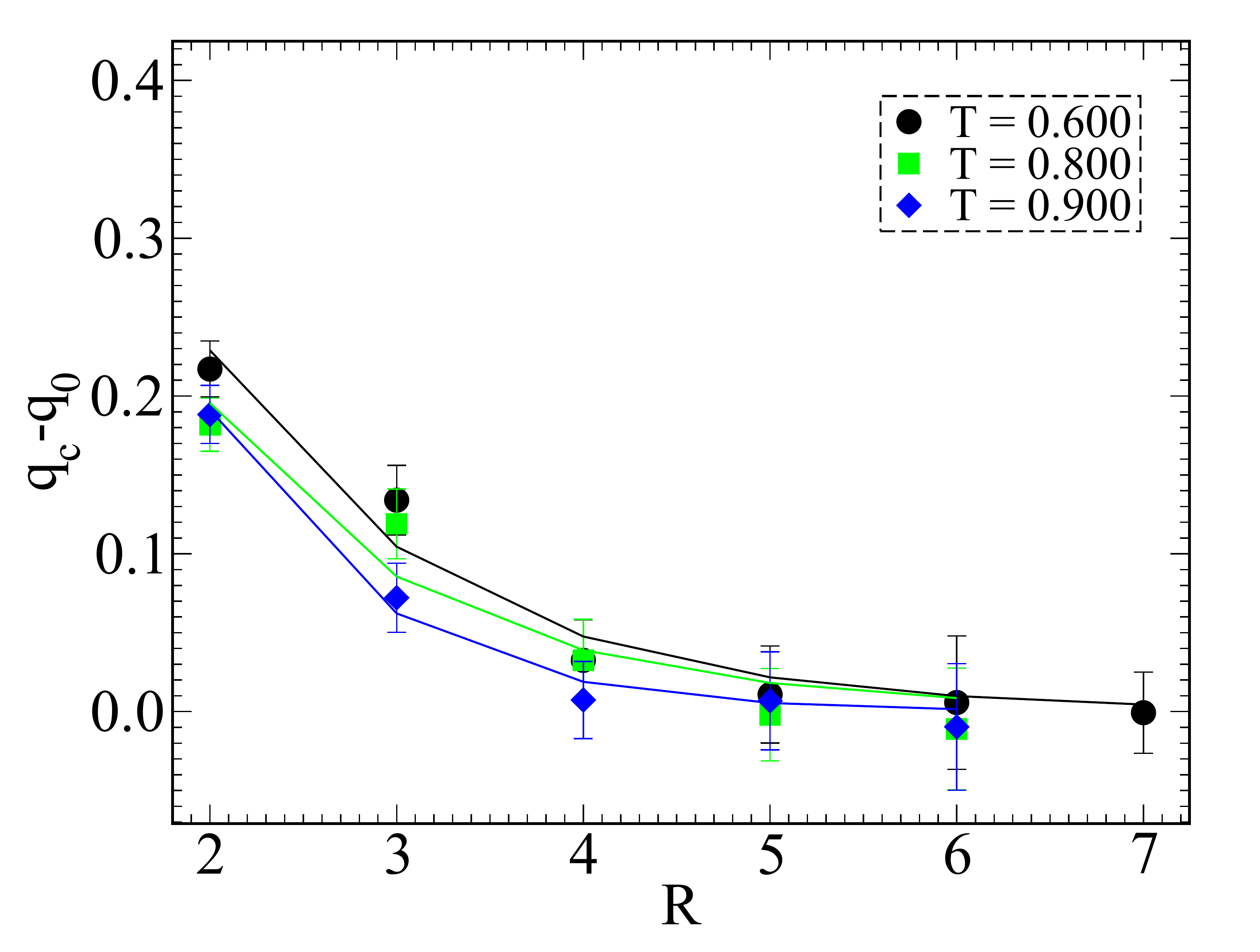}               
\includegraphics[scale=0.24]{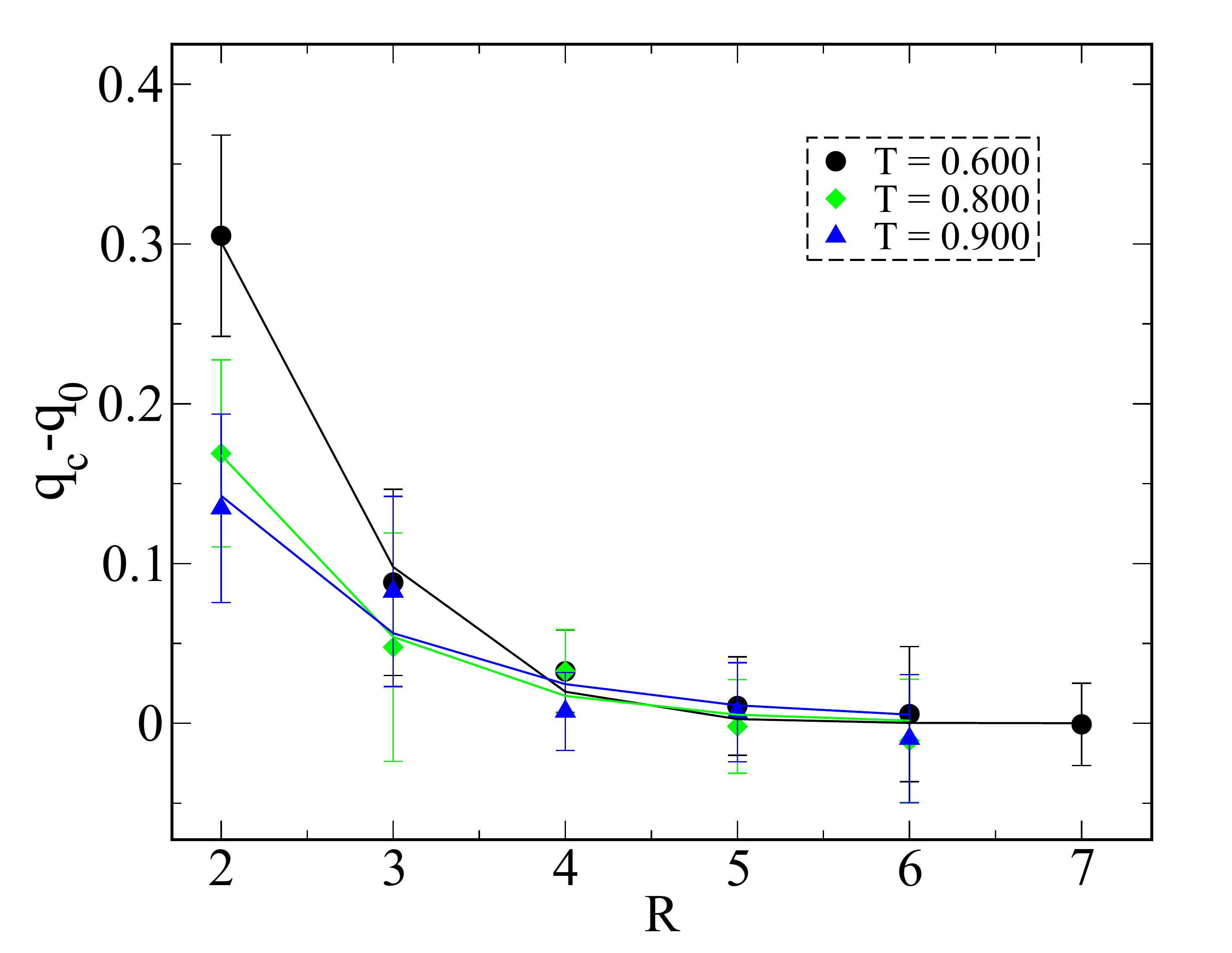} 
 \caption{ Top panel shows the overlap as a function of cavity radius R for different temperatures, averaged over all 250 
 cavities. Bottom panel shows the results obtained by considering cavities whose packing fraction is similar to the bulk packing fraction.}
  \label{polPTS}
%\end{figure}
\end{figure}

\noindent{\bf Finite size scaling of $\tau_{\alpha}$:}
We have calculated the static length scale from finite size scaling of 
the $\alpha$ relaxation time $\tau_{\alpha}$ following the 
procedure of Sec.\ref{tauAlphaFSS}. In Fig.~\ref{taupoly} we show
the finite size scaling of $\tau_{\alpha}$.
\vskip +0.5cm
\noindent{\bf Point-to-set length scale:}
%SS: This section is not written very well, and could use a round of rewriting. 
For calculating the point-to-set length scale we follow the 
procedure described in Sec.\ref{ptsMethod}.

The point-to-set analysis for the polydisperse system is demanding. 
For this model system it is very difficult to obtain the density or 
the packing fraction to be close to the bulk value due to large 
fluctuations of particle sizes in these cavities. This is particularly 
severe for small cavities and small cavities are the ones for which one
expects to have large values of the static overlap in the studied
temperature range. Another important feature is the polydispersity 
inside the cavity, which also has to be constrained to be close to the bulk 
polydispersity. Obtaining small cavities with packing fraction and polydispersity close 
to the bulk values requires a large sampling of statistically independent cavities. 
As described in the main text, not restricting the cavity packing fraction and polydispersity 
to be close to the bulk values results 
in  point-to-set length scales $\xi_{pts}$ that show very little variation with temperature. 

This result is very similar to the results obtained in 
Ref.\cite{POL}. If we choose only those cavities for which the packing 
fraction and 
polydispersity are close to the bulk values,  the obtained PTS length scale $\xi_{pts}$ exhibits stronger temperature 
dependence which matches the temperature dependence of other static length scales
obtained by the methods mentioned above. We have averaged over 
$250$ realizations (for small cavities) in simulations of $N = 10000$ 
particles. When we choose only those cavities for which the packing fraction
is close to the bulk value, the number of cavities used in the averaging reduces to  
approximately 32, which results in greater uncertainty in the values of the overlap. 
In Fig.\ref{polPTS} we show  data for the estimates of the PTS length scale. In the top panel we  show
the results obtained by taking all cavities irrespective of the packing fraction
and the observed correlation seems  not to grow much. In the
bottom panel we  show the data for the case where only those
cavities are used in the analysis for which the packing fraction is close 
to the bulk value. The correlations indeed increase substantially 
in this case. In this case we have chosen $M = 4 \times 4 $ boxes 
of linear size $l = 0.35$. We find that A=0.50 fits the data 
better for this model system.

\bibliography{si}